\documentclass[12pt]{article}
\usepackage{axodraw}
\usepackage{pstricks}
\usepackage{color}
\usepackage{amssymb,amsmath,bm,bbm}
\usepackage{epsf}
\usepackage{epsfig}
\usepackage{afterpage}
\usepackage{longtable}
\usepackage{cite}
\usepackage{latexsym, mathrsfs}
\usepackage{dsfont}
\usepackage{graphics}
\usepackage{url}
\usepackage{paralist}

\DeclareMathOperator{\diag}{diag}

\def\slash#1{\setbox0=\hbox{$#1$}\dimen0=\wd0
      \setbox1=\hbox{/} \dimen1=\wd1 \ifdim\dimen0>\dimen1
      \rlap{\hbox to \dimen0{\hfil/\hfil}} #1                        \else
      \rlap{\hbox to \dimen1{\hfil$#1$\hfil}}
      /   \fi}

\newcommand{\lsim}{
\mathrel{\hbox{\rlap{\hbox{\lower4pt\hbox{$\sim$}}}\hbox{$<$}}}}

\newcommand{\gsim}{
\mathrel{\hbox{\rlap{\hbox{\lower4pt\hbox{$\sim$}}}\hbox{$>$}}}}

\newcommand{\vcb}{|V_{cb}|}
\newcommand{\vtd}{|V_{td}|}
\newcommand{\vub}{|V_{ub}|}
\newcommand{\vts}{|V_{ts}|}
\newcommand{\vus}{|V_{us}|}

\def\eps{\varepsilon}

\newcommand{\tev}{\, {\rm TeV}}
\newcommand{\gev}{\, {\rm GeV}}
\newcommand{\mev}{\, {\rm MeV}}

\newcommand{\Heff}{{\cal H}_\text{eff}}

\newcommand{\re}{{\rm Re}}
\newcommand{\im}{{\rm Im}}

\def\beq{\begin{equation}}
\def\eeq{\end{equation}}

\newcommand{\be}{\begin{equation}}
\newcommand{\ee}{\end{equation}}
\newcommand{\bea}{\begin{eqnarray}}
\newcommand{\eea}{\end{eqnarray}}
\newcommand{\nn}{\nonumber}
\newcommand{\no}{\nonumber}
\newcommand{\bi}{\begin{itemize}}
\newcommand{\ei}{\end{itemize}}
\newcommand{\ord}{{\cal O}}

\def\kpn{K^+\rightarrow\pi^+\nu\bar\nu}

\def\klpn{K_{L}\rightarrow\pi^0\nu\bar\nu}

\newcommand{\gm}{\gamma^\mu}

\medskipamount 

\newlength{\textlength}
\newlength{\overlinelength}
\newcommand{\ovl}[2][.55]{\settowidth{\textlength}{$#2$}
  \setlength{\overlinelength}{0.1pt}
  \addtolength{\overlinelength}{0.75\textlength}
  \makebox[\textlength][s]{$#2$} \hspace{-.55\textlength}
  \hspace{-\overlinelength}\hspace{#1\overlinelength}
  \overline{\makebox[\overlinelength][s]{\vphantom{$#2$}}}
  \hspace{-#1\overlinelength}\hspace{.55\textlength}}

\newcommand{\kkm}{$K\!-\!\ovl{K}{}\,$\ mixing}

\newcommand{\TeV}{\rm TeV}
\begin{document}
\begin{titlepage}
\vspace*{-0.5truecm}


\begin{flushright}
{FLAVOUR(267104)-ERC-30}
\end{flushright}

\vfill

\begin{center}
\boldmath
{\Large\textbf{Particle-Antiparticle Mixing, CP Violation and Rare $K$ and $B$ Decays in a Minimal Theory of Fermion Masses}}
\unboldmath
\end{center}

\begin{center}
{\bf  Andrzej J.~Buras, Jennifer Girrbach and Robert Ziegler }
\vspace{0.3truecm}

{\footnotesize
 {\sl Physik Department, Technische Universit\"at M\"unchen, D-85748 Garching, Germany}

{\sl TUM Institute for Advanced Study, Technische Universit\"at M\"unchen, \\ D-85748 Garching, Germany
}

}

\end{center}

\begin{abstract}
{\small\noindent
We present a detailed study of $\Delta F=2$ observables and of rare $K^+(K_L)$ 
and $B_{s,d}$ meson decays in a ``Minimal Theory of Fermion Masses" (MTFM). 
In this theory
Yukawa couplings are generated through the mixing with heavy vectorlike (VF)
fermions. This implies corrections to the SM quark 
couplings to $W^\pm$, 
$Z^0$ and Higgs so that FCNC processes receive contributions from tree level 
$Z^0$ and Higgs exchanges and  $W^\pm$ bosons couple to right-handed quarks.
In a particular version of this model in which  the Yukawa matrix $\lambda^D$
in the heavy {\it down} fermion sector is {\it unitary}, 
$\lambda^U = \mathds{1}$ and  $M = M_{\rm VF}$ 
is fixed,
only three real and positive definite parameters describe New Physics (NP) 
contributions to all $\Delta F=2$ and $\Delta F=1$ observables  in 
$K$ and $B_{s,d}$ systems  once the known quark masses and 
the CKM matrix are correctly reproduced.  For $M\ge 1\tev$ NP 
 contributions to $B_{s,d}^0-\bar B_{s,d}^0$ mixings are found to be very small.
 While in principle NP contributions to $\varepsilon_K$ and $\Delta F=1$ processes 
could be large, the correlation between $\varepsilon_K$ and $K_L\to\mu^+\mu^-$
eliminates basically NP contributions to $\varepsilon_K$  and  right-handed 
current contributions to $\Delta F=1$ FCNC observables.
We find CMFV structure in $B_{s,d}$ decays with 
$\mathcal{B}(B_{s,d}\to\mu^+\mu^-)$ uniquely enhanced for  $M=3~\tev$ 
by at least $35\%$ and almost
up to a factor of two over 
their SM values. Also $\mathcal{B}(\kpn)$ and $\mathcal{B}(\klpn)$ are 
uniquely enhanced 
by similar amount but the correlation between them differs from the CMFV one.
We emphasize various correlations between $K$ and $B_{s,d}$ decays that could 
test this scenario. The model favours $\gamma\approx 68^\circ$, $\vub\approx 0.0037$, $S_{\psi K_S}\approx 0.72$, 
$S_{\psi\phi}\approx0.04$ and $ 4.2 \times 10^{-9}\le\mathcal{B}(B_s\to\mu^+\mu^-)\le 5.0 \times 10^{-9}$ for $M=3-4\tev$. }
\end{abstract}

%
%
%
\end{titlepage}

\setcounter{page}{1}
\pagenumbering{arabic}

\section{Introduction}\label{sec:int}
\setcounter{equation}{0}

 In a ``minimal theory of fermion masses" (MTFM) \cite{Buras:2011ph}
the fermionic content of the SM is extended by heavy vectorlike 
fermions, with flavor-anarchical Yukawa couplings, that mix with SM chiral 
quarks. Small SM Yukawa couplings arise then from small mixing angles 
between the heavy and light sectors. Although the hierarchical structure of the mixing is put in by hand, this model can be regarded as an effective 
description of the fermionic sector of a large class of existing flavour models 
and thus might serve as a useful reference frame for a further understanding of flavour hierarchies in the SM. 

This mechanism of quark mass generation implies corrections to the SM quark 
couplings to $W^\pm$, 
$Z^0$ and Higgs so that FCNC processes receive contributions from tree level 
$Z^0$ and Higgs exchanges and  $W^\pm$ couple to right-handed quarks implying 
the presence of right-handed currents in a number of observables.
The first question then arises whether this framework, while generating 
measured quark masses and CKM parameters is consistent with 
the constraints from electroweak precision observables, 
tree-level decays and in particular FCNC transitions for heavy fermion 
masses in the reach of the LHC. The second question is whether there is 
a particular pattern of deviations from SM predictions for FCNC processes in 
this model that could further test it in the flavour precision era.

In \cite{Buras:2011ph} we have derived general formulae for the modified couplings of 
the SM heavy gauge bosons $(W^\pm,Z^0,H^0)$ to quarks and we have discussed 
the bounds on heavy fermion masses from present data finding globally 
consistency with the data for heavy fermion masses in a few TeV range. 

One of the central formulae in \cite{Buras:2011ph} was the leading order expression for the 
SM quark masses
\be\label{masses}
m_{ij}^X=v\varepsilon_i^Q\varepsilon_j^X \lambda^X_{ij}, \qquad (X=U,D),
\ee 
where 
\be\label{equ:eps}
\varepsilon^{Q,U,D}_i= \frac{m_i^{Q,U,D}}{M_i^{Q,U,D}}  
\ee
with $m_i$ describing the mixing between the light and heavy sectors and $M_i$ 
standing for the heavy fermion masses (see Section 2). The matrices $\lambda^{U,D}$ 
are the heavy Yukawa couplings which in \cite{Buras:2011ph} have been assumed to 
be anarchical $\ord(1)$ real numbers. While such an approach allowed 
a first look at the phenomenological implications of this model
it did not allow a study of CP violation and a meaningful identification of
correlations between various flavour observables.

The goal of the present paper is to present a more general study with 
$\lambda^{U,D}$ being first arbitrary complex matrices for which we 
will perform approximate but analytic diagonalization. Subsequently 
we will consider a specific structure for  $\lambda^{U,D}$ requiring 
them to be {\it unitary} matrices. This will allow us to simplify the expressions 
of interest and significantly reduce the number of free parameters. In 
this manner a rather transparent phenomenology will follow. In particular we 
will find a number of correlations between different observables that can be 
tested in the future. Moreover it will turn out that in the simplest version 
of the {\it unitary model} (UM) in which $\lambda^U=1$, to be termed the 
trivially
UM (TUM), very definite predictions for particle-antiparticle mixing and the 
 deviations from the SM expectations for rare 
$K$ and $B_{s,d}$ decays are found.

Our paper is organized as follows. In Section \ref{sec:mod} we 
summarize briefly 
those ingredients of the MTFM that we will need in our analysis. In particular 
we discuss the case of unitary heavy Yukawa matrices which specify the UM.
In Section~\ref{sec:Diag} we will perform the diagonalization of the 
induced quark mass matrices that will allow us very general expressions for 
quark masses, mixing angles and the couplings of SM heavy bosons to 
quarks in terms of the fundamental parameters of the model. We will then 
apply these formulae in the case of the UM and TUM 
in which we set $\lambda^U= \mathds{1}$. The TUM has only three 
free real and positive definite parameters after all the CKM parameters and quark masses 
have been determined and a common  mass $M$ for heavy fermions has been fixed.
This makes TUM very predictive.

In Section \ref{sec:trans} we present the formulae for the effective
Hamiltonians governing particle-antiparticle mixings 
$K^0-\bar K^0$ and $B_{d,s}-\bar B_{d,s}$ that  in addition to the 
usual SM box diagrams receive tree-level contributions 
from $Z^0$ and $H^0$ exchanges. Only induced tree-level $Z^0$ exchanges are 
relevant in MTFM.
We also give a compendium of formulae relevant 
for numerical analysis of $\Delta F=2$ observables. In Section \ref{sec:Heff}
the effective Hamiltonians for
$s\to d\nu\bar\nu$, $b\to q \nu\bar\nu$ and $b\to q \ell^+\ell^-$ ($q=d,s$) transitions are given.   There we also comment on  the 
contributions of $W^\pm$, $Z^0$ and $H^0$ to $B\to X_s\gamma$ decay which  in the TUM for $M$ of order few TeV turn out to be negligible.
In Section \ref{sec:rare} we calculate  {the}
most interesting rare decay
 branching ratios in {the} $K$ and $B$ meson systems, including {those for the processes} $\kpn$,
$\klpn$, $K_L\to\pi^0 \ell^+\ell^-$, $B\to K^{(*)}\nu\bar\nu$, $B\to X_{s,d}\nu\bar\nu$, $B_{s,d}\to\mu^+\mu^-$, $B^+\to\tau^+\nu_\tau$ and  $K_L \to\mu^+\mu^-$.
 Finally 
 constraints from $b\to s \ell^+\ell^-$ transitions like 
$B\to K^*(K)\ell^+\ell^-$ decays are discussed.

In Section \ref{sec:anatomy} we will search for a global structure of NP 
contributions in the TUM in order to understand better the numerical results 
presented in the subsequent section.
 This search  is accompanied  by the following questions:
\begin{itemize}
\item
Can the so-called 
$\varepsilon_K$--$S_{\psi K_S}$ tension present in the SM be 
removed in the TUM?
\item
Which values of $\vub$ and of the angle $\gamma$ in the unitarity triangle are favoured by the TUM on the basis of FCNC processes 
and what are the implications for $B^+\to\tau^+\nu_\tau$?
\item
What are the implications for
the mixing induced CP-asymmetry $S_{\psi\phi}$ and the branching ratios 
for $B_{s,d}\to\mu^+\mu^-$ and how this model faces new results from  
LHCb \cite{LHCbBsmumu} ?
\item
What are the implications for $\kpn$,  $\klpn$ and $K_L\to \mu^+\mu^-$ decays that can probe 
high energy scales beyond the reach of the LHC \cite{Buras:2012jb}.
\end{itemize}
In Section \ref{sec:numerics} a detailed
numerical analysis is presented.
In particular we study the correlations not only between various $\Delta F=1$
observables but also between $\Delta F=1$ and $\Delta F=2$ observables.
 We summarize
our results in Section \ref{sec:summary}. Few technicalities are relegated to appendices.

\section{The Model}\label{sec:mod}
\setcounter{equation}{0}

\subsection{Basic Ingredients}

The aim of this section is to briefly review the most important ingredients of 
MTFM. A detailed theoretical discussion is presented in \cite{Buras:2011ph}.

In this minimal model we add heavy vectorlike fermions that mix with chiral fermions. The hierarchical structure of SM fermion
masses can
then be  explained through mass hierarchies entering the mixing pattern. We reduce the number of parameters such that it is still possible
to reproduce the SM Yukawa couplings and that at the same time  flavour violation is suppressed. In this way we can identify the minimal
FCNC effects. The Higgs couples only to vectorlike but not to chiral fermions, so that  SM Yukawas arise solely through mixing. However
this mixing induces flavour violation already at the tree level proportional to $v^2/M^2$,
where $M$ is the mass of the vectorlike fermion. This is due to the fact that
\begin{compactitem}
 \item SU(2)$_L$ doublets mix with SU(2)$_L$ singlets and
\item the Higgs couplings are no longer aligned with SM fermion masses.
\end{compactitem}
Now we specify the field content of the model focusing on the quark sector: we have three generations of chiral quarks
\begin{equation}
u_{Ri},d_{Ri},q_{Li} = \binom{u_{Li}}{d_{Li}} \hspace{1cm} i=1,2,3
\end{equation}
and add for each of them a vectorlike pair of heavy quarks\footnote{As for scales above the heavy quark masses the model contains 18
dynamical quarks, our model is not asymptotically free but the one-loop beta function at this scales 
is almost zero. We anticipate that the model being embedded in a larger gauge 
group will eventually be asymptotically free.}
\begin{equation}
U_{Ri},U_{Li},D_{Ri},D_{Li},Q_{Ri}= \binom{U_{Ri}^Q}{D_{Ri}^Q},Q_{Li}= \binom{U_{Li}^Q}{D_{Li}^Q} \hspace{1cm}  i=1,2,3. 
\end{equation}
The Lagrangian is then of the form (omitting kinetic terms)
\begin{align}
\begin{split}
\label{Lorigin}
- {\cal L} & = \tilde{h} \lambda^U_{ij} \bar{Q}_{Li} U_{Rj} + h \lambda^D_{ij} \bar{Q}_{Li} D_{Rj}  + M^U_{i}  \bar{U}_{Li} U_{Ri} +
M^D_{i}\bar{D}_{Li}  D_{Ri}  + M^Q_{i} \bar{Q}_{Ri} Q_{Li} \\
& + m^U_{i}  \bar{U}_{Li} u_{Ri} + m^D_{i}  \bar{D}_{Li} d_{Ri} + m^Q_{i}  \bar{Q}_{Ri} q_{Li} + {\rm h.c.}
\end{split}
\end{align}
where we assumed that $m^X$ and $M^X$ ($X = U,\,D,\,Q$) can be diagonalized simultaneously. Then   $M^X$ and $m^X$ are diagonal matrices
with positive entries. Instead of $m_i^X$ we will use $\eps_i^{X}$ defined in
(\ref{equ:eps}) which are also real and positive.
In order to get the low-energy effective Lagrangian that contains only SM fields we proceed in two steps: First we integrate out the heavy
states
using their equations of motions
in unbroken $SU(2)_L\times U(1)_Y$. If we relax the assumption of aligned $m^X$ and $M^X$ we additionally have to diagonalize the mass
matrix. As eigenstates we then get three heavy vectorlike quarks and three massless states which can be identified as SM quarks. Second we
include EWSB and redefine light fields to get canonical kinetic terms.
The relevant part of the effective Lagrangian in the mass eigenstate basis for gauge bosons 
and $H^0$ but in flavour basis for quarks reads \cite{Buras:2011ph}\footnote{The masses $m_{ij}^{U,D} $ in ${\cal L}_\text{eff}$ should not
be 
confused with  $m^{U,D,Q}$ in ${\cal L}$, but as we traded the latter masses 
for $\varepsilon_i^{U,D,Q}$ no problems of this sort should arise in what follows.} {($v=174\gev$)}
\begin{align}
\begin{split}
\label{equ:Leff}
- {\cal L}_{\rm eff} & \supset \frac{g}{\sqrt2} \left( W^+_\mu j^{\mu -}_{\rm charged} + {\rm h.c.} \right) +  \frac{g}{2 c_{\rm w}} Z_\mu
j^\mu_{\rm neutral} \\
& + \overline{u}_{Li}m^U_{ij} u_{Rj} +  \overline{d}_{Li} m^D_{ij} d_{Rj}  + \frac{H}{\sqrt2} \left( \overline{u}_{Li} y^U_{ij} u_{Rj} + 
\overline{d}_{Li} y^D_{ij} d_{Rj} \right) + {\rm h.c.}  
\end{split} 
\end{align}
where the quark masses and Yukawa couplings are given as 
(summation over $k=1,2,3$)
\begin{subequations}
 \begin{align}
\label{effm}
 m^X_{ij}  & = v {\bar \eps}^Q_i {\bar \eps}^X_j \lambda^X_{ij}  -\frac{v}{2} \left( A^X_L \right)_{ik}  {\bar \eps}^Q_k {\bar \eps}^X_j
 \lambda^X_{kj} -\frac{v}{2} \left( A^X_R \right)_{kj}  {\bar \eps}^Q_i {\bar \eps}^X_k \lambda^X_{ik} \,,\\
 \label{effy}
 y^X_{ij} & = \frac{m^{X}_{i j}}{v} - \left( A^X_L \right)_{ik}  {\bar \eps}^Q_k {\bar \eps}^X_j \lambda^X_{kj} -\left( A^X_R \right)_{kj}
 {\bar \eps}^Q_i {\bar \eps}^X_k \lambda^X_{ik}\,.
 \end{align}
\end{subequations}
The charged and neutral current read
\begin{subequations}
 \begin{align}\label{equ:jcharged}
 j^{\mu -}_{\rm charged} & = \overline{u}_{Li} \left[ \delta_{ij} - \frac{1}{2} \left(A^U_L\right)_{ij} - \frac{1}{2}
\left(A^D_L\right)_{ij}  \right] \gm d_{Lj} + \overline{u}_{Ri}  \left( A^{UD}_R\right)_{ij} \gm d_{Rj}  \,,\\
\begin{split}\label{equ:jneutral}
  j^\mu_{\rm neutral} & = \overline{u}_{Li} \left[ \delta_{ij} - \left(A^U_L\right)_{ij}  \right] \gm u_{Lj} + \overline{u}_{Ri}  \left(
  A^U_R\right)_{ij}  \gm u_{Rj}    \\
  & - \overline{d}_{Li} \left[ \delta_{ij} - \left(A^D_L\right)_{ij}  \right] \gm d_{Lj} -  \overline{d}_{Ri} \left( A^D_R\right)_{ij} \gm
  d_{Rj} - 2 s_{\rm w}^2 j^\mu_{\rm elmag} \,
\end{split}
 \end{align}
\end{subequations}
with ($X=U,D$)
\begin{subequations}
 \begin{align}
 \left( A^X_L \right)_{ij} & = \frac{v^2}{{\bar M}_k^X {\bar M}_k^X} {\bar \eps}^Q_i {\bar \eps}^Q_j  \lambda_{ik}^{X}  \lambda_{jk}^{X*}
\,, &
 \left( A^X_R \right)_{ij} & = \frac{v^2}{{\bar M}_k^Q {\bar M}_k^Q} {\bar \eps}^X_i {\bar \eps}^X_j  \lambda_{kj}^{X}  \lambda_{ki}^{X*}\,,
 \label{ALXARX}\\
 \left( A^{UD}_R\right)_{ij} & = \frac{v^2}{{\bar M}_k^Q {\bar M}_k^Q} {\bar \eps}^U_i {\bar \eps}^D_j  \lambda_{kj}^{D} 
\lambda_{ki}^{U*}\,,
 \\
 \bar{\eps}_i^X & = \frac{\eps_i^X}{\sqrt{1+ \eps^X_i \eps^X_i}} \,,& \bar{M}_k^X & = M_k^X (1 + \eps^X_k \eps^X_k)\,. \label{Qeqn}
 \end{align}
\end{subequations}
{We note that $A_L^{U,D}$ enter both the charged and neutral currents.}
Explicit Feynman rules in mass eigenstate basis for both gauge bosons and fermions will be derived in Sec.~\ref{sec:Diag}.

\subsection{The Unitary Model (UM)}\label{sec:unitarymodel}
\subsubsection{Basic Assumptions}
We will now discuss an explicit model based on the Lagrangians 
(\ref{Lorigin}) and (\ref{equ:Leff}) and the 
following two assumptions:
\begin{itemize}
\item
{\it Universality} of heavy masses:
\be\label{UNI}
M_i^Q=M_i^U=M_i^D =M
\ee
for all $i$. The crucial assumption is that the matrices above are proportional 
to a unit matrix. The assumption that the masses in these three matrices are 
equal to each other is less important but helps in reducing the number of 
free parameters.
\item
{\it Unitarity} of the Yukawa matrices $\lambda^U$ and $\lambda^D$. While 
other forms, like symmetric or hermitian matrices could be considered,  the 
unitarity of Yukawa matrix once assumed is valid in any basis and
 simplifies in a profound 
manner the charged and in particular neutral current interactions written 
in terms of fundamental parameters of the model. 
As we will see below, the resulting structure of interactions 
resembles the one in Randall-Sundrum (RS) scenarios.
\end{itemize}

\subsubsection{Parameter Counting and Explicit Parametrizations 
of \boldmath{$\lambda^U$} and \boldmath{$\lambda^D$}}

We have the following parameters:
\begin{itemize}
\item
10 real parameters: $M$ and $\varepsilon^{Q,U,D}_i$ for $i=1,2,3$.
\item
$\lambda^U$ and $\lambda^D$  being unitary matrices 
have each 3 real parameters and 6 phases.
\item 
 Phase redefinitions of $U_R$, $D_R$ and $Q_L$ allow, as clearly seen from 
the first two terms in the basic Lagrangian, to remove 8 phases. We are 
thus left with 
\be\label{parameters}
{\rm 16~~ real~~ parameters~~ and ~~ 4 ~~phases.}
\ee
\end{itemize}
Subtracting six quark masses and four parameters of the CKM matrix
our model has the following number of new parameters
\be
{\rm 7~~ new ~~real~~ parameters~~ and ~~ 3 ~~new~~phases.}
\ee

The parameters in (\ref{parameters}) can be distributed as follows
\begin{itemize}
\item
10 real parameters: $M$ and $\varepsilon^{Q,U,D}_i$ for $i=1,2,3$.
\item 
As phase redefinitions of $Q_L$ can be made only once,  
$\lambda^U$ and $\lambda^D$ will have different number of phases but the 
same number (3) of real parameters.
\end{itemize}

For $\lambda^U$ and $\lambda^D$ one can use explicit parametrizations of 
unitarity matrices known from the Standard Model (CKM matrix) and the 
corresponding matrix in the Littlest Higgs model with T-parity (LHT)  describing the interactions of quarks 
with mirror quarks \cite{Blanke:2006sb}. 

In the minimal version of MTFM we use for  $\lambda^D$ the CKM-like 
parametrization
\begin{equation}\label{2.72}
 \lambda^D=
\left(\begin{array}{ccc}
c_{12}^dc_{13}^d&s_{12}^dc_{13}^d&s_{13}^de^{-i\delta^d}\\ -s_{12}^dc_{23}^d
-c_{12}^ds_{23}^ds_{13}^de^{i\delta^d}&c_{12}^dc_{23}^d-s_{12}^ds_{23}^ds_{13}^de^{i\delta^d}&
s_{23}^dc_{13}^d\\ s_{12}^ds_{23}^d-c_{12}^dc_{23}^ds_{13}^de^{i\delta^d}&-s_{23}^dc_{12}^d
-s_{12}^dc_{23}^ds_{13}^de^{i\delta^d}&c_{23}^dc_{13}^d
\end{array}\right)\,,
\end{equation}
with
$c_{ij}^d=\cos\theta_{ij}^d$ and $s_{ij}^d=\sin\theta_{ij}^d$. 
$c_{ij}^d$ and
$s_{ij}^d$ can all be chosen to be positive
and  $\delta^d$ may vary in the
range $0\le\delta^d\le 2\pi$.

Following \cite{Blanke:2006sb} the mixing matrix $\lambda^U$ can be conveniently parameterized, generalizing
the usual CKM parameterization, as a product of three rotations, and
introducing a complex phase in each of them, thus obtaining ($c_{ij}^u=\cos\theta_{ij}^u$, $s_{ij}^u=\sin\theta_{ij}^u$)
\be
\addtolength{\arraycolsep}{3pt}
\lambda^U= \begin{pmatrix}
1 & 0 & 0\\
0 & c_{23}^u & s_{23}^u e^{- i\delta^u_{23}}\\
0 & -s_{23}^u e^{i\delta^u_{23}} & c_{23}^u\\
\end{pmatrix}\,\cdot
 \begin{pmatrix}
c_{13}^u & 0 & s_{13}^u e^{- i\delta^u_{13}}\\
0 & 1 & 0\\
-s_{13}^u e^{ i\delta^u_{13}} & 0 & c_{13}^u\\
\end{pmatrix}\,\cdot
 \begin{pmatrix}
c_{12}^u & s_{12}^u e^{- i\delta^u_{12}} & 0\\
-s_{12}^u e^{i\delta^u_{12}} & c_{12}^u & 0\\
0 & 0 & 1\\
\end{pmatrix}
\ee
Performing the product one obtains the expression
\be\label{lambdaumatrix}
\addtolength{\arraycolsep}{3pt}
\lambda^U= \begin{pmatrix}
c_{12}^u c_{13}^u & s_{12}^u c_{13}^u e^{-i\delta^u_{12}}& s_{13}^u e^{-i\delta^u_{13}}\\
-s_{12}^u c_{23}^u e^{i\delta^u_{12}}-c_{12}^u s_{23}^us_{13}^u e^{i(\delta^u_{13}-\delta^u_{23})} &
c_{12}^u c_{23}^u-s_{12}^u s_{23}^us_{13}^u e^{i(\delta^u_{13}-\delta^u_{12}-\delta^u_{23})} &
s_{23}^uc_{13}^u e^{-i\delta^u_{23}}\\
s_{12}^u s_{23}^u e^{i(\delta^u_{12}+\delta^u_{23})}-c_{12}^u c_{23}^us_{13}^u e^{i\delta^u_{13}} &
-c_{12}^u s_{23}^u e^{i\delta^u_{23}}-s_{12}^u c_{23}^us_{13}^u e^{i(\delta^u_{13}-\delta^u_{12})} &
c_{23}^uc_{13}^u\\
\end{pmatrix}
\ee
 which completes the explicit parametrization of the model.

However, even without looking at details the property of the unitarity of 
$\lambda^U$ and $\lambda^D$  combined with the {\it universality} of the heavy 
fermion masses implies a very transparent structure of corrections to weak 
charged and neutral currents before one rotates to the mass eigenstates 
for SM quarks.

\subsubsection{Implications}

The general expressions for the gauge couplings simplify considerably. 
 Indeed we find
\begin{subequations}
\begin{align}
\left( A^U_L \right)_{ij} & = \frac{v^2}{\bar M^2} {\bar \eps}^Q_i 
{\bar \eps}^Q_j  \delta_{ij} &
\left( A^D_L \right)_{ij} & = \frac{v^2}{\bar M^2} {\bar \eps}^Q_i {\bar \eps}^Q_j  \delta_{ij}  \\
\left( A^{UD}_R\right)_{ij} & = \frac{v^2}{\bar M^2} {\bar \eps}^U_i {\bar \eps}^D_j  \lambda_{kj}^{D}  \lambda_{ki}^{*U} \\
\left( A^U_R \right)_{ij} & = \frac{v^2}{\bar M^2} {\bar \eps}^U_i {\bar \eps}^U_j  \delta_{ij}&
\left( A^D_R \right)_{ij} & = \frac{v^2}{\bar M^2}{\bar \eps}^D_i {\bar \eps}^D_j  \delta_{ij}
\end{align}
\end{subequations}

In order to obtain these expressions, without loosing the generality,
 we have modified the universality assumption in (\ref{UNI}) so that it is 
 valid for $\bar M_i$ and not $M_i$. As we will see in our phenomenological analysis $\bar M_{1,2}=M_{1,2}$ to an excellent
accuracy but $ \bar M_3\not=M_3$ and this redefinition 
has to be made only in the case of $M_3$.

Inserting these expressions into neutral and charged currents we note that
\begin{itemize}
\item
 The neutral current is diagonal in the flavour basis but the universality 
of interactions is broken by different values of $\varepsilon^{Q,U,D}_i$ 
 for $i=1,2,3$ that are necessary  in order to explain the mass spectrum of quarks.
 Such universality breakdown is known from the RS scenarios where the 
universality is broken by the bulk masses. After rotation to mass eigenstates 
we will obtain FCNC currents at the tree level but they will be partly 
protected by the difference in masses. This structure is analogous to 
the RS-GIM \cite{Agashe:2004cp}. 
\item
The left-handed charged currents are also diagonal within generations but 
the right-handed ones not. This will generally imply that the RH mixing 
matrices in the mass eigenstate basis will differ from LH-ones.

\item 
An interesting structure arises in this model also in the case of mass matrices  and Yukawa 
couplings. Defining 
\begin{equation}\label{equ:lamdaprime}
\lambda^{U^\prime}_{i j}  ={\bar \eps}^Q_i {\bar \eps}^U_j \lambda^U_{ij} \qquad  \lambda^{D^\prime}_{i j}   = {\bar \eps}^Q_i {\bar
\eps}^D_j \lambda^D_{ij}. 
\end{equation}
we find 
\begin{subequations}
\begin{align}
\begin{split}
m^U_{ij}  & = v \left[ \lambda^{U^\prime}_{i j}  -\frac{1}{2} \frac{v^2}{M^2}
(\bar\eps_i^Q)^2\lambda^{U^\prime}_{i j} -\frac{1}{2}\frac{v^2}{\bar M^2}
(\bar\eps_j^U)^2  \lambda^{U^\prime}_{ij} \right] + \ord({v^5/M^4})  \\
m^D_{ij}  & = v \left[ \lambda^{D^\prime}_{i j} -\frac{1}{2}\frac{v^2}{\bar M^2} 
(\bar\eps_i^Q)^2  \lambda^{D^\prime}_{i j} -\frac{1}{2}\frac{v^2}{\bar M^2}
(\bar\eps_j^D)^2 \lambda^{D^\prime}_{ij} \right] + \ord({v^5/M^4}) 
\end{split}\\
\begin{split} 
y^U_{ij} & = \frac{m^{U}_{i j}}{v} - \frac{v^2}{\bar M^2}(\bar\eps_i^Q)^2   \lambda^{U^\prime}_{i j}
-\frac{v^2}{M^2}(\bar\eps_j^U)^2   \lambda^{U^\prime}_{ij}  + \ord({v^4/\bar M^4})   \\
y^D_{ij} & = \frac{m^{D}_{i j}}{v} - \frac{v^2}{\bar M^2}(\bar\eps_i^Q)^2   \lambda^{D^\prime}_{i j} - \frac{v^2}{\bar
M^2}(\bar\eps_j^D)^2   \lambda^{D^\prime}_{ij} + \ord({v^4/M^4})\,. \end{split}
\end{align}
\end{subequations}
Note that again if the $\eps_i^{Q,U,D}$ where independent of the index $i$ 
the Higgs couplings would be aligned with quark masses and we would not have 
tree level Higgs FCNC in the mass eigenbasis.
\item
The matrices $\lambda^{U^\prime}$ and $\lambda^{D^\prime}$ in 
Eq.~(\ref{equ:lamdaprime}) are not unitary 
with the departure from unitarity given by the hierarchy of $\varepsilon_i$ 
parameters. 
\end{itemize}

In order to simplify the notation in the rest of the paper we will 
suppress the {\it bars} and work with $\eps_i^X$ and $M$ which really 
represent  $\bar\eps_i^X$ and $\bar M$. In this context it is important to 
keep in mind that 
$\bar\eps_i^X\le 1$ (see \ref{Qeqn}) so that with this notation in the 
subsequent formulae
\be\label{epsbound}
\eps_i^X\le 1.
\ee

\subsubsection{Trivially Unitary Model (TUM)}
In our first phenomenological analysis it will be useful to consider 
a special case of the Unitary Model in which the Yukawa matrix $\lambda^U$ 
is trivial:
\be
\lambda^U=\mathds{1}.
\ee

In this model, to be called TUM in what follows, we have then
\be\label{parametersTUM}
{\rm 13~~ real~~ parameters~~ and ~~ 1 ~~phase.}
\ee

 This means 
that TUM has only four new parameters
\be
{\rm 4~~ new ~~real~~ parameters~~ and ~~ 0 ~~new~~phases.}
\ee

Thus in the TUM CP violation is governed by a single CP phase $\delta^d$ in 
the Yukawa matrix $\lambda^D$, which as we will show is equal to the CKM phase.
Yet the phenomenology of this model will differ from the SM one due 
to new contributions from tree-level diagrams that contain new 
flavour violating parameters absent in the SM. In fact in this 
model it is very easy to express most of the parameters of the model in 
terms of the SM quark masses  and the mixing angles of the CKM 
matrix. The remaining four free parameters can be chosen to be
\be\label{newTUM}
M, \quad  \eps_3^Q, \quad s_{13}^d, \quad s_{23}^d.
\ee
Note that all these parameters are positive definite and 
 $s_{13}^d$ and $s_{23}^d$ are smaller than unity due to the unitarity 
of $\lambda^D$.  We will also see that in order to obtain the 
correct top quark mass for $M$ of order few $\tev$ one has
$0.80\le\eps_3^Q\le 1.0$. Further restrictions  on new parameters come from FCNC transitions.

In order to perform phenomenology we have to rotate the quark fields to 
the mass eigenbasis and express ten fundamental parameters of the model 
in terms of six quark masses and four parameters of the CKM matrix. 
This we will do in the next Section.

\section{Diagonalization of the Quark Mass Matrices}\label{sec:Diag}
\setcounter{equation}{0}

\subsection{Preliminaries}

In this section we give general analytic expressions for the leading order rotation matrices $V_{L,R}^{X}$ ($X = U,D$) that diagonalize the mass
matrices $ m^X_{ij} = v \lambda^X_{ij} \eps_i^Q \eps_j^X$
via
\be
V_L^{X\dagger} m^X V_R^X = m^X_\text{diag}\,,\quad X = U,D\,.
\ee
The CKM matrix is then given as 
\be\label{CKM}
V_\text{CKM} = (V^U_L)^\dagger V^D_L.
\ee
With this at hand we can derive the flavour changing $Z,\,W$ and $H$
couplings in the mass eigenstate basis.

In what follows we do not  yet specify the parametrization of Yukawa matrices 
$\lambda^U$ and $\lambda^D$  so that the formulae in this section unless 
explicitly mentioned apply to MTFM at large.

Introducing 
\be
\eps^M_{ij} \equiv \frac{{\eps}_i^M}{\eps^M_j} \quad {\rm for} \quad i<j, \quad M=Q,U,D
\ee
 one can rewrite the mass matrices as
\be
m^X_{ij} = v \eps^Q_3 \eps^X_3 \lambda_{ij}^X \eps^Q_{i3} \eps^X_{j3}\,.
\ee
The diagonalization is done as an
expansion in powers of a small  auxiliary parameter $\epsilon$ that first can be set equal to the Cabibbo angle $\epsilon= 0.23$
\footnote{We emphasize that this value is indicative in order to perform perturbative  expansion but has not 
been used in our numerical analysis.}.
 To this end it is useful to have a rough estimate for the squared ratios $\eps^M_{ij}$. We find 

\begin{align}
(\eps^{Q}_{13})^2 & \sim V_{ub}^2 \sim \epsilon^6 & (\eps^{Q}_{23})^2 & \sim V_{cb}^2 \sim \epsilon^4  \label{equ:epsQscaling}\\
(\eps^{U}_{13})^2 & \sim \frac{m_u^2}{m_t^2} \frac{1}{V_{ub}^2} \sim \epsilon^9 & (\eps^{U}_{23})^2 & \sim  \frac{m_c^2}{m_t^2}
\frac{1}{V_{cb}^2} 
\sim \epsilon^{3.6}  \label{equ:epsQscaling1}  \\
(\eps^{D}_{13})^2 & \sim  \frac{m_d^2}{m_b^2} \frac{1}{V_{ub}^2}  \sim \epsilon^{2.8 \div 4} &  (\eps^{D}_{23})^2 & \sim 
\frac{m_s^2}{m_b^2}
\frac{1}{V_{cb}^2}  \sim \epsilon^{0.8 \div 1.6}\,.  \label{equ:epsQscaling2}
\end{align}
We conclude that except for $(\eps^{D}_{23})^2 $, which appears too large 
to be considered as expansion parameter, we can expand in the remaining ratios.

In our results for the mass eigenvalues, CKM matrix and rotation matrix we use the shorthand notation defined in 
Appendix~\ref{app:notation}.

 \subsection{ Quark Masses}
The  mass eigenvalues in the up and down sector are given as
\begin{subequations}\label{equ:Quarkmasses}
\begin{align}
m_b & = v \eps_3^Q\eps_3^D \sqrt{\hat{\lambda}^D_{33}}\equiv
 v \eps_3^Q\eps_3^D \kappa_b
\,, & m_t & =v \eps_3^Q\eps_3^U \sqrt{\hat{\lambda}^U_{33}}\equiv
v \eps_3^Q\eps_3^U \kappa_t
\,,  \\
m_s &  = v\eps_2^Q\eps_2^D  \frac{|\tilde{\lambda}_{22}^D|}{\sqrt{\hat{\lambda}^D_{33}}}\equiv v\eps_2^Q\eps_2^D \kappa_s
\,, &
m_c &  =
v\eps_2^Q\eps_2^U
 \frac{|\tilde{\lambda}_{22}^U| }{\sqrt{\hat{\lambda}^U_{33} }}\equiv 
v\eps_2^Q\eps_2^U\kappa_c
\,, \\
m_d &  =  v\eps^Q_1\eps^D_1\frac{ |\det{\lambda^D}|}{|\tilde{\lambda}^D_{22}| }
\equiv  v\eps^Q_1\eps^D_1\kappa_d
\,,
& m_u &  = v\eps_1^Q\eps_1^U \frac{|\det{\lambda^U}|}{|\tilde{\lambda}^U_{22}| }
\equiv  v\eps_1^Q\eps_1^U\kappa_u \,,
\end{align}
\end{subequations}
where for later convenience we have introduced parameters $\kappa_i$ that collect the dependence 
on Yukawa couplings.
This structure is very similar to the RS scenario (see \cite{Agashe:2004cp,Blanke:2008zb} and references therein). The fermion shape functions $f_i^{Q,u,d}$
correspond to our $\eps_i^{Q,U,D}$ but we have here no  exponential hierarchy 
but rather a powerlike one. Moreover, our calculation is a bit more general as we
did not expand in $\eps_{23}^D$. 
 
\subsection{ Rotations to Mass Eigenbasis} 
We parametrize the rotation matrices in the {\it up sector} in the following manner:
\begin{equation}\label{equ:VUL}
V^U_{L} = 
\begin{pmatrix}
1 & \eps^Q_{12} u_1^{L} & \eps^Q_{13} u_2^{L} \\
-  \eps^Q_{12} (u_1^{L})^* & 1 & \eps^Q_{23} u_3^{L} \\
\eps^Q_{13} u_4^{L} & -  \eps^Q_{23} (u_3^{L})^* & 1
\end{pmatrix}\cdot
\begin{pmatrix}
1 & 0 & 0 \\
0 & e^{i b_U} & 0\\
0 & 0 & e^{i c_U} 
\end{pmatrix}
\end{equation}
\begin{equation}\label{equ:VUR}
V^U_{R} = 
\begin{pmatrix}
1 & \eps^U_{12} u_1^{R} &  \eps^U_{13}  u_2^{R} \\
-   \eps^U_{12}  (u_1^{R})^* & 1 &  \eps^U_{23}  u_3^{R} \\
 \eps^U_{13}  u_4^{R} & -   \eps^U_{23}  (u_3^{R})^* & 1
\end{pmatrix}\cdot
\begin{pmatrix}
1 & 0 & 0 \\
0 & e^{i b_U} & 0\\
0 & 0 & e^{i c_U} 
\end{pmatrix}
\end{equation}
with the $\ord(1)$ coefficients $u_i^{L,R}$ that are listed in Appendix~\ref{app:notation}
and phases that depend on the $z_i$ defined below in (\ref{equ:zi}) according to
\begin{equation}\label{zphases}
e^{i b_U} = \frac{z_1^*}{|z_1|} \quad \quad e^{i c_U} =  \frac{z_1^*}{|z_{1}|} \frac{z_{3}^*}{|z_3|}.
\end{equation}
In the unitary model the coefficients $u_i^{L,R}$ can be written as  functions of angles and phases. 

In the {\it down sector} rotations are parametrized as
\begin{equation}\label{equ:VDL}
V^D_L = 
\begin{pmatrix}
1 & \eps^Q_{12} d_1^L &  \eps^Q_{13} d_2^L \\
-   \eps^Q_{12} (d_1^L)^* & 1 &  \eps^Q_{23} d_3^L \\
 \eps^Q_{13} d_4^L & -  \eps^Q_{23}  (d_3^L)^* & 1
\end{pmatrix}\cdot
\begin{pmatrix}
1 & 0 & 0 \\
0 & e^{i b_D} & 0\\
0 & 0 & e^{i c_D} 
\end{pmatrix}
\end{equation}
\begin{equation}\label{equ:VDR}
V^D_R = 
\begin{pmatrix}
1 &  \eps^D_{12} d_1^R &  \eps^D_{13} d_2^R \\
 \eps^D_{12} d_3^R & d_4^R &  \eps^D_{23}  d_5^R \\
 \eps^D_{13}  d_6^R &  -   \eps^D_{23}  (d_5^R)^* & d_4^R
\end{pmatrix}\cdot
\begin{pmatrix}
1 & 0 & 0 \\
0 & e^{i b_D} & 0\\
0 & 0 & e^{i c_D} 
\end{pmatrix}
\end{equation}
with the $\ord(1)$ coefficients $d_i^{L,R}$ (see Appendix~\ref{app:notation})
and phases  according to
\begin{equation}
e^{i b_D} = e^{i b_U} = \frac{z_1^*}{|z_1|} \quad \quad e^{i c_D} = e^{i c_U} = \frac{z_1^*}{|z_{1}|}
\frac{z_{3}^*}{|z_3|}.
\end{equation}
The translation to the notation used in \cite{Blanke:2008zb} can be found in the Appendix~\ref{app:notation}. However  our
$V_R^D$ is more general.

\subsection{CKM Matrix} 
Using the rotation matrices derived above we can calculate the CKM matrix 
in (\ref{CKM}). 
Expanding again in
$\epsilon$  and guided by (\ref{equ:epsQscaling})-(\ref{equ:epsQscaling2}) we get
\begin{align}
s_{12}c_{13} = \lambda & = |z_1|\,, & s_{23} c_{13} =  A\lambda^2 & = |z_{3}|\,, &s_{13} e^{-i\delta} = A\lambda^3\left( \rho - i
\eta\right) = 
z_2\,,
\end{align}
with
\begin{subequations}\label{equ:zi}
\begin{align}
z_1 & = \eps^Q_{12} \left(  \frac{\tilde{\lambda}^D_{12} }{\tilde{\lambda}^D_{22} } -  \frac{\tilde{\lambda}^U_{12} }{\tilde{\lambda}^U_{22}
}  \right) =  \varepsilon_{12}^Q \left(d_1^L - u_1^L\right)\equiv  
\varepsilon_{12}^Q\alpha_{12} \,,\\
z_2  &= \eps^Q_{13} \left(  \frac{\hat{\lambda}^D_{13}  }{\hat{\lambda}^D_{33}  } +  \frac{\tilde{\lambda}^U_{13} }{\tilde{\lambda}^U_{22} }
-  \frac{\tilde{\lambda}^U_{12} }{\tilde{\lambda}^U_{22} } \frac{\hat{\lambda}^D_{23}  }{\hat{\lambda}^D_{33} }\right)= \varepsilon_{13}^Q
\left(d_2^L + u_4^{L\star} - d_3^L u_1^{L}\right)\equiv \varepsilon_{13}^Q\alpha_{13}\,, \\
z_{3} & = \eps^Q_{23} \left( \frac{\hat{\lambda}^D_{23}  }{\hat{\lambda}^D_{33} } -  \frac{\lambda_{23}^U }{\lambda_{33}^{U} } \right) =
\varepsilon_{23}^Q \left(d_3^L - u_3^L \right)\equiv \varepsilon_{23}^Q\alpha_{23}\,,
\end{align}
\end{subequations}
where for later convenience we have introduced parameters $\alpha_{ij}$ that collect the dependence 
on Yukawa couplings.

Analogous formulae can also be found in \cite{Blanke:2008zb} in Eq.~(3.9), (3.10). The relation between our notation and the one used
in \cite{Blanke:2008zb} is summarized in Appendix~\ref{app:notation}. 
From these three elements  we can extract all four parameters of $V_\text{CKM}$ and then use the usual PDG convention
\begin{align}\label{equ:VCKMWinkel}
 V_\text{CKM} = \begin{pmatrix}
                c_{12} c_{13} & s_{12} c_{13} & s_{13} e^{-i\delta} \\
		-s_{12}c_{23}-c_{12} s_{23} s_{13} e^{i\delta} & c_{12} c_{23} - s_{12} s_{23} s_{13} e^{i\delta} & s_{23} c_{13}\\
		s_{12} s_{23}- c_{12} c_{23} s_{13} e^{i\delta} & -c_{12} s_{23} - s_{12} c_{23} s_{13} e^{i\delta} & c_{23} c_{13}
                \end{pmatrix}.
\end{align}
The three parameters $s_{12} = |V_{us}|/c_{13}$, $s_{13} = |V_{ub}|$ and $s_{23} = |V_{cb}|/c_{13}$   can be extracted from tree level
decays and should
not be sensitive to NP unless the charged currents are significantly 
modified by NP contributions. As we discuss in subsection~\ref{Wcouplings} this turns out to be 
not the case in the TUM.

\subsection{Modified $Z$ and $W$ couplings}
While in the quark flavour basis the $Z$ couplings were 
determined by the matrices $A_{L,R}^{U,D}$, in the mass eigenstate basis 
they are governed by the matrices $ \tilde A_L^X $ ($X=U,D$) given by
\cite{Buras:2011ph}
\begin{align}
 \tilde A_L^X & = m_\text{diag}^X \tilde B_L^X m_\text{diag}^X\,,\qquad \qquad\tilde A_R^X  = m_\text{diag}^X \tilde B_R^X
m_\text{diag}^X\,,\\
\tilde B_L^X & = V_R^{X\dagger} B_L^X V_R^X\,,\qquad \qquad\quad\tilde B_R^X  = V_L^{X\dagger} B_R V_L^X\,.
\end{align}
Here
\begin{align}
 B_L^X =
\tfrac{1}{M_X^2}\diag\left(\tfrac{1}{\epsilon_1^X\epsilon_1^X},\tfrac{1}{\epsilon_2^X\epsilon_2^X},\tfrac{1}{\epsilon_3^X\epsilon_3^X}
\right)\,,\qquad B_R =
\tfrac{1}{M_Q^2}\diag\left(\tfrac{1}{\epsilon_1^Q\epsilon_1^Q},\tfrac{1}{\epsilon_2^Q\epsilon_2^Q},\tfrac{1}{\epsilon_3^Q\epsilon_3^Q}
\right)\,.
\end{align}
As seen in (\ref{equ:jcharged}) and~(\ref{equ:jneutral}) in quark 
flavour basis 
the same matrices $A_{L}^{U,D}$ appear in left-handed charged and neutral currents. However the
rotation to mass eigenstate basis affects charged and neutral currents 
in a different manner. The left handed $W$ couplings are then governed by\footnote{The CKM matrix comes from the usual SM $W$ couplings, but
the second
and third flavour violating coupling is new.} 
\begin{align}
&  V_\text{CKM}\,,\qquad V_L^{U\dagger} A_L^U V_L^D\equiv \tilde{A}_L^U V_\text{CKM}\,,\qquad V_L^{U\dagger} A_L^D
V_L^D\equiv  V_\text{CKM}\tilde{A}_L^D\,.
\end{align}
Induced right handed couplings of $W$ are described in the quark flavour 
basis by $A^{UD}_R$.
In the mass eigenstate basis these couplings are given by
\begin{align}
 \tilde A_R^{UD} & = m_\text{diag}^U \tilde B_R^{UD} m_\text{diag}^D\,,\qquad \tilde B_R^{UD} = V_L^{U\dagger} B_R V_L^D\,.
\end{align}
As already stated in \cite{Buras:2011ph} the couplings scale according to
\begin{align}
 \left(\tilde A_L^X\right)_{ij} \sim \frac{v^2}{M_X^2}\eps_i^Q\eps_j^Q\,,\qquad\qquad  \left(\tilde A_R^X\right)_{ij}
\sim\frac{v^2}{M_Q^2} \eps_i^X \eps_j^X\,
\end{align}
and for the right handed $W$ coupling we get 
\begin{align}
 \left(\tilde A_R^{UD}\right)_{ij} \sim \frac{v^2}{M_Q^2}\eps_i^U\eps^D_j\,.
\end{align}
Inserting the rotation matrices $V_{L,R}^{U,D}$ presented above and expanding in $\epsilon$ we can calculate explicitly these flavour
violating $Z$ and $W$ couplings. 
The analytic expressions for these couplings 
are rather complicated and can be found in Appendix~\ref{app:Deltas}. 

Now comes an important point. As seen  in Appendix~\ref{app:Deltas} 
in the MTFM all 
these flavour changing tree-level
couplings (neutral and charged) depend on the same parameters that determine the SM quark masses and CKM parameters. However, generally the 
number of 
fundamental parameters is larger than ten and even after the correct quark 
masses and CKM parameters have been reproduced in this model, a number of 
free parameters will remain implying in principle non-MFV interactions. 
This point has been in particular stressed in a general context in 
\cite{Lalak:2010bk}.   We will investigate the size of these effects in TUM in
Section~\ref{sec:numerics}.

In Sec.~\ref{sec:unitarymodel} we showed that in the unitary model the gauge couplings~-- except $A_R^{UD}$
for right handed $W$ couplings~-- are diagonal in the flavour basis but different for the three generations. However, rotating to the mass
eigenstate basis off-diagonal elements are generated:
\begin{align}
 &\tilde A_L^X = \frac{v^2}{M_X^2} V_L^{X\dagger}\diag\left(\eps_1^{Q2}, \eps_2^{Q2},  \eps_3^{Q2} \right) V_L^X\,,\\
&\tilde A_R^X = \frac{v^2}{M_Q^2} V_R^{X\dagger}\diag\left(\eps_1^{X2}, \eps_2^{X2},  \eps_3^{X2} \right) V_R^X\,.
\end{align}
Explicit formulas are listed in Appendix~\ref{app:Deltas}. In order 
to simplify the notation we use in what follows shorthand notation:
\be
\eps_i^{Q2}\equiv (\eps_i^Q)^2, \qquad \eps_i^{X2}\equiv (\eps_i^X)^2, \qquad
\eps_i^{Q4}\equiv (\eps_i^Q)^4.
\ee

\boldmath
\subsection{Final Results for $\Delta_{L,R}^{ij}(Z)$,  $\Delta_{L,R}^{ij}(W)$}
\label{firstdeltas}
\unboldmath
In the expressions for effective Hamiltonians presented in the subsequent sections, the fundamental 
role is played by the couplings  $\Delta_{L,R}^{ij}(Z)$ that are defined 
as follows
\be\label{eq:3.14}
 \mathcal{L}_\text{FCNC}(Z)\equiv\left[ \mathcal{L}_{L}(Z)+ \mathcal{L}_{R}(Z)\right]\,,
\ee
where
\bea\label{eq:3.15}
 \mathcal{L}_{L}(Z) &=& \left[ \Delta_L^{sd}(Z)(\bar s \gamma_\mu P_L d)+\Delta_L^{bd}(Z)(\bar b \gamma_\mu P_L d)
+\Delta_L^{bs}(Z)(\bar b \gamma_\mu P_L s)\right] Z^{\mu} \,,\\
\label{eq:3.16}
 \mathcal{L}_{R}(Z)&=& \left[ \Delta_R^{sd}(Z)(\bar s \gamma_\mu P_R d)+\Delta_R^{bd}(Z)(\bar b \gamma_\mu P_R d)
+\Delta_R^{bs}(Z)(\bar b \gamma_\mu P_R s)\right] Z^{\mu} \,,
\eea
and $\Delta_{L,R}^{ij}(Z)$ are the elements of the matrices $\hat\Delta_{L,R}(Z)$. 

The modifications of the $W^\pm$  couplings to the SM quarks, $\Delta_{L,R}^{ij}(W)$,  can be similarly
summarized by the  
Lagrangian:
\be\label{eq:W3.14}
 \Delta\mathcal{L}(W)\equiv\left[ \mathcal{L}_{L}(W)+ \mathcal{L}_{R}(W)\right]\,,
\ee
where
\bea\label{eq:W3.15}
 \mathcal{L}_{L}(W) &=& \left[ \Delta_L^{td}(W)(\bar t \gamma_\mu P_L  d)+\Delta_L^{ts}(W)(\bar t \gamma_\mu P_L s)
+\Delta_L^{tb}(W)(\bar t \gamma_\mu P_L  b)\right] W^{\mu} \,,\\
\label{eq:W3.16}
 \mathcal{L}_{R}(W)&=& \left[ \Delta_R^{td}(W)(\bar t \gamma_\mu P_R  d)+\Delta_R^{ts}(W)(\bar t \gamma_\mu P_R s)
+\Delta_R^{tb}(W)(\bar t \gamma_\mu P_R b)\right] W^{\mu} \,,
\eea
and $\Delta_{L,R}^{ij}(W)$ are the elements of the matrices $\hat\Delta_{L,R}(W)$.

In the mass eigenstate basis these couplings are given in terms of $\tilde A^X_{L,R}$ and $\tilde A_R^{UD}$ 
as follows 
\begin{align}
 \Delta_L^{ij}(Z) & = -\frac{g}{2 c_W} (\tilde A_L^D)_{ij}\,,\\
\Delta_R^{ij}(Z) & = \frac{g}{2 c_W} (\tilde A_R^D)_{ij}\,.
\end{align}
\begin{align}\label{DLW}
&\Delta_L^{ij}(W) = \frac{g}{2\sqrt{2}}\left[\left(\tilde A_L^U V_\text{CKM}\right)_{ij}+\left(V_\text{CKM}\tilde A_L^D
\right)_{ij} \right]\,,\\
&\Delta_R^{ij}(W) =
-\frac{g}{\sqrt{2}}\left(\tilde A_R^{UD}\right)_{ij}\,\label{DRW}.
\end{align}
 Note that 
\be
 \Delta_{L,R}^{ji}(Z)=(\Delta_{L,R}^{ij}(Z))^*, \qquad  \Delta_{L,R}^{ji}(W)=(\Delta_{L,R}^{ij}(W))^*~.
\ee

\boldmath
\subsection{Results for $\Delta_{L,R}^{ij}(H)$}
\label{firstdeltasH}
\unboldmath
For completeness we list the flavour violating  $H$ couplings even if 
 in the minimal 
version of our model  the Higgs contributions to all processes considered 
by us are as expected negligible \cite{Buras:2011ph}. 

In analogy with the flavour violating couplings of $Z^0$ we introduce
\be\label{eq:3.14H}
 \mathcal{L}_\text{FCNC}(H)\equiv\left[ \mathcal{L}_{L}(H)+ \mathcal{L}_{R}(H)\right]\,,
\ee
where
\bea\label{eq:3.15H}
 \mathcal{L}_{L}(H) &=& \left[ \Delta_L^{sd}(H)(\bar s  P_L d)+\Delta_L^{bd}(H)(\bar b P_L d)
+\Delta_L^{bs}(H)(\bar b P_L s)\right] H \,,\\
\label{eq:3.16H}
 \mathcal{L}_{R}(H)&=& \left[ \Delta_R^{sd}(H)(\bar s P_R d)+\Delta_R^{bd}(H)(\bar b  P_R d)
+\Delta_R^{bs}(H)(\bar b  P_R s)\right] H \,.
\eea
The Higgs couplings to light down-type quarks are given as ($i,j = d,s,b$): 
\begin{align}
 \Delta_L^{ij}(H) & = \frac{1}{\sqrt{2}v}\left( m_i^{D}( \tilde A_L^D)^\star_{ji} + ( \tilde A_R^D)^\star_{ji}m_j^{D}
\right)\,,\\
 \Delta_R^{ij}(H) & = \frac{1}{\sqrt{2}v}\left(( \tilde A_L^D)_{ij} m_j^{D} +  m_i^{D}( \tilde A_R^D)_{ij}
\right)\,.\\
\end{align}
We observe that relative to flavour violating $Z^0$ couplings, these 
      couplings are suppressed for all SM quarks but top quark by roughly $m_i/v$ where $m_i$ denote the masses of SM quarks. Consequently
$H^0$ contributions  do not play any role in the processes considered by us.

\subsection{Expressing Model Parameters in terms of Quark Masses and CKM Parameters}
For our numerical analysis it is useful to express analytically as much as 
possible the fundamental model parameters in terms of the measured quark masses and CKM 
parameters so that the number of free parameters relevant for the analysis 
of various decays will be significantly reduced. Simultaneously 
correct values of quark masses and CKM parameters will be incorporated 
in our analysis automatically. We consider first the general case and 
subsequently show the case of the TUM where the analysis becomes very 
simple and transparent.

\subsubsection{General Case}

First from (\ref{equ:Quarkmasses}) we can determine
$\eps_i^{D,U}$ in terms of quark masses, $\eps_i^{Q}$ and the parameters 
in the Yukawa matrices that are hidden in $\kappa_i$. We find
\begin{subequations}\label{equ:epsiDU}
\begin{align}
\eps_3^D & = \frac{1}{\eps_3^Q}\frac{m_b}{v \kappa_b}
\,, & \eps_3^U & = \frac{1}{\eps_3^Q}\frac{m_t}{v \kappa_t}
\,,  \\
\eps_2^D & = \frac{1}{\eps_2^Q}\frac{m_s}{v \kappa_s}
\,, & \eps_2^U & = \frac{1}{\eps_2^Q}\frac{m_c}{v \kappa_c}
\,,  \\
\eps_1^D & = \frac{1}{\eps_1^Q}\frac{m_d}{v \kappa_d}
\,, & \eps_1^U & = \frac{1}{\eps_1^Q}\frac{m_u}{v \kappa_u}
\,.
\end{align}
\end{subequations}
Next using (\ref{equ:zi}) we can determine $\varepsilon_1^Q$ and $\varepsilon_2^Q$ 
in terms of $s_{13}$, $s_{23}$, $\varepsilon_3^Q$  and the parameters 
in the Yukawa matrices that are hidden in $\alpha_{ij}$. We find 
\be\label{equ:epsiQ}
\varepsilon_1^Q= \varepsilon^Q_3\frac{s_{13}}{|\alpha_{13}|}, 
\qquad 
\varepsilon_2^Q= \varepsilon^Q_3\frac{s_{23}c_{13}}{|\alpha_{23}|}
\ee

Finally writing 
\be
\alpha_{ij}=|\alpha_{ij}| e^{-i\phi_{ij}}
\ee
we find the following two conditions
\be\label{twoconditions}
\frac{|\alpha_{13}|}{|\alpha_{23}||\alpha_{12}|}=\frac{s_{13}}{s_{23}s_{12}c_{13}^2}, \qquad 
\phi_{13}=\delta_{\rm CKM}=\gamma,
\ee
with $\gamma$ being one angle of the unitarity triangle that up to the sign 
equals the phase of $V_{ub}$.

In this manner we could directly express analytically the fundamental parameters of the model in terms of six quark masses and three mixing angles that can be 
determined from high energy collider experiments and tree-level decays, 
respectively. With two conditions in (\ref{twoconditions}) we can fix one 
phase and one mixing parameter in the Yukawa matrices in terms of the 
remaining five mixing parameters in these matrices and three phases. The 
additional two real free parameters are $\epsilon_3^Q$ and the common mass 
$M$ of the heavy fermions.

At this stage one comment should be made. This procedure can be straightforwardly 
executed numerically if one is allowed to set $\epsilon_{23}^D=0$ in 
the expression for $\hat\lambda_{33}^D$, which appears to be a good approximation if the diagonal terms in $\lambda^D$ are largest. Otherwise these formulae 
could be used iteratively setting first  $\epsilon_{23}^D=0$.

\subsubsection{Trivially Unitary Model}

In this case the number of free parameters is decreased to four 
that are listed in (\ref{newTUM}). Even more importantly the two conditions 
in (\ref{twoconditions}) simplify considerably 
\begin{align}\label{newtwoconditions}
\begin{split}
&\frac{s^d_{13}}{s^d_{23}\left|e^{i\delta} c_{23}^d t_{12}^d +
s_{23}^d s_{13}^d\right|}=\frac{s_{13}}{s_{23}s_{12}c_{13}^2}, \qquad 
\delta^d=\delta_{\rm CKM}=\gamma.\\
&\Rightarrow \quad t_{12}^d = -t_{23}^d s_{13}^d\cos\gamma  + t_{23}^d s_{13}^d\sqrt{-\sin^2\gamma + \frac{1}{s_{23}^{d4}}\frac{s_{12}^2
s_{23}^2c_{13}^4}{s_{13}^2}}\,.
\end{split}
\end{align}

\begin{figure}[!tb]
\begin{center}
\includegraphics[width = 0.5\textwidth]{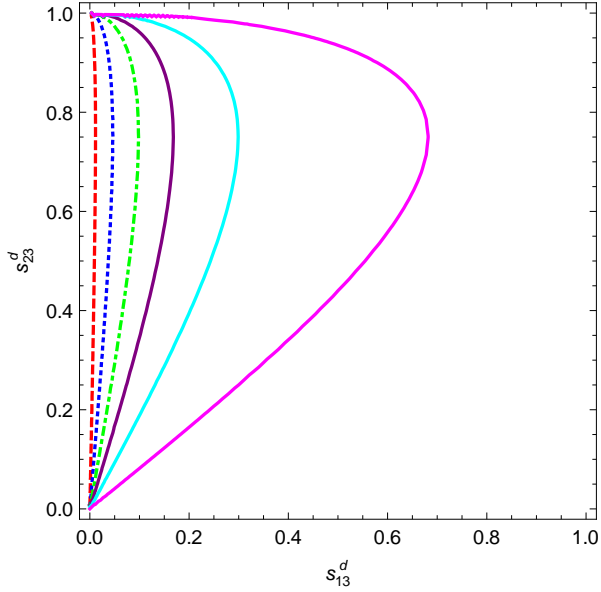}
\end{center}
\caption{\it Correlation between  $s_{23}^d$ and $s_{13}^d$  in the TUM for  $|V_{ub}| = 0.0037$ for different $s_{12}^d$: 0.05 (dashed
red), 0.2 (dotted blue), 0.4 (dot dashed green), 0.6 (solid purple), 0.8
(solid cyan), 0.95 (solid magenta). }\label{fig:angle}~\\[-2mm]\hrule
\end{figure}

In this manner all conditions from quark masses and the measured CKM matrix 
from tree-level decays can be automatically satisfied and once these 
parameters have been fixed all the observables analyzed by us are given 
entirely in terms of the four real parameters listed in  (\ref{newTUM}).  In Fig.~\ref{fig:angle}  we show the
correlation between $s_{13}^d$,  $s_{23}^d$ and $s_{12}^d$ implied by the 
condition in (\ref{newtwoconditions})
using  $\vub=0.0037$.
As we will include in our analysis many observables we expect a number of correlations between them in this scenario.  We also
emphasize that in 
this model CP-violation is governed entirely by the CKM phase but as in addition to CKM parameters new flavour parameters are present  we
will find some deviations 
from the correlations present in the SM.

 Moreover in TUM  $\kappa_t=\kappa_c=\kappa_u=1$ and we find in particular 
an important relation 
\be
\eps_3^U\eps_3^Q=\frac{m_t(M)}{v}.
\ee
For our nominal value $M=3\tev$ one finds $m_t(M)=139\gev$. Taking into account
that $\eps_3^X\le 1$ and $v=174\gev$ we find the allowed range for $\eps_3^Q$
\be\label{epsrange}
0.80\le \eps_3^Q\le 1.0.
\ee
As this parameter plays a prominent role in our phenomenology, this range, 
following from the desire to obtain correct top-quark mass, has significant 
impact on possible deviations of FCNC observables from SM expectations.
As $m_t(M)$ decreases with increasing $M$, the lower limit in (\ref{epsrange}) 
also decreases with increasing $M$ but this decrease is only logarithmic and in the range of $M$ in the reach of the LHC, this dependence can be neglected.

The beauty
of this simple 
scenario lies in the fact that at the end, when the data improve and the 
theoretical uncertainties in FCNC observables will be reduced,
 we will be able to fix completely 
the Yukawa interactions in the heavy vectorial quark sector and the mixing 
parameters $\varepsilon_i^{Q,U,D}$ that connect this sector with SM quarks.

\boldmath
\subsection{Couplings  $\tilde A^X_{L,R}$  in the TUM}
\unboldmath
 The general expressions for  $\tilde A^X_{L,R}$ and $\tilde A_R^{UD}$  even in 
the unitary model are rather lengthy and have been presented 
in Appendix~\ref{app:Deltas}. However, they are simpler in the TUM and we 
list them here.

In the TUM the up-type couplings simplify considerably, since $u_i^{L,R} = 0$ and $\kappa_{u,c,t} = 1$. Consequently we get diagonal up-type
couplings:
\begin{align}
 \tilde A_L^U & =
\frac{1}{M_U^2}\text{diag}\left(\frac{m_u^2}{\eps_1^{U2}},\,\frac{m_c^2}{\eps_2^{U2}},\,\frac{m_t^2}{\eps_3^{U2}}\right) =
\frac{v^2}{M_U^2}\text{diag}\left(\eps_1^{Q2},\,\eps_2^{Q2},\,\eps_3^{Q2}\right)\,,\\
 \tilde A_R^U & =
\frac{1}{M_Q^2}\text{diag}\left(\frac{m_u^2}{\eps_1^{Q2}},\,\frac{m_c^2}{\eps_2^{Q2}},\,\frac{m_t^2}{\eps_3^{Q2}}\right) =
\frac{v^2}{M_Q^2}\text{diag}\left(\eps_1^{U2},\,\eps_2^{U2},\,\eps_3^{U2}\right)\,.
\end{align}
This means that in the TUM there are no-tree level contributions to FCNC 
processes in the up-quark sector, e.g. in $D^0-\bar D^0$ system. If one day 
 NP effects in this system will be convincingly shown to be present, the 
condition $\lambda^U=1$ will have to be removed.

The down-type couplings are non-diagonal and given as
\begin{align}
 \tilde A_L^D &= \frac{v^2}{M_D^2}V_\text{CKM}^\dagger\text{diag}\left(\eps_1^{Q2},\,\eps_2^{Q2},\,\eps_3^{Q2}\right)V_\text{CKM}\,,\\
\tilde A_R^D &=
\frac{1}{M_Q^2}m^D_\text{diag}V_\text{CKM}^\dagger\text{diag}\left(\frac{1}{\eps_1^{Q2}},\,\frac{1}{\eps_2^{Q2}},\,\frac{1}{\eps_3^{Q2}}
\right)V_\text { CKM}m^D_\text{diag}\,,
\end{align}

These formulae are very transparent. Indeed 
all phase matrices drop out. This means that {\it all} information about FCNCs in the TUM are encoded in the $\eps_i^Q$. We again observe
that for 
\be\label{epsrel}
\eps_1^{Q}=\eps_2^{Q}=\eps_3^{Q}
\ee
the coupling matrices $\tilde A_L^D$ and $\tilde A_R^D$ are diagonal and there 
are no FCNC transitions mediated by $Z^0$ at the tree level. However, the hierarchical structure of the CKM matrix breaks this relation as in MTFM 
\be\label{Xij}
\frac{\eps_1^{Q}}{\eps_2^{Q}}=\vus X_{12}, \qquad
\frac{\eps_1^{Q}}{\eps_3^{Q}}=\vub X_{13}, \qquad
\frac{\eps_2^{Q}}{\eps_3^{Q}}=\vcb X_{23}, 
\ee
where we have introduced quantities $X_{ij}$ that will be useful 
later on. In \cite{Buras:2011ph} $X_{ij}$
have been assumed to be close to unity.  In the case of 
arbitrary $\lambda^D$ and $\lambda^U$  they can differ significantly from 
unity and from each other. But when all constraints are 
taken into account the equality (\ref{epsrel}) remains  badly broken 
and tree-level FCNC transitions mediated by $Z^0$ are present.

\boldmath
\subsection{Modification of $W^\pm$ Couplings  in the TUM}\label{Wcouplings}
\unboldmath
We analyse next the modification of the charged couplings in the  TUM. Using 
the general formula for the corrections to left-handed couplings in 
(\ref{DLW}) we find the {\it effective} CKM matrix 
\be
\left[V_{ij}^{\rm CKM}\right]_{\rm eff}= V_{ij}^{\rm CKM}(1-\frac{v^2}{M^2}\eps^{Q2}_i).
\ee
The effective CKM matrix is clearly non-unitary, a property know from other 
studies in which heavy fermions mix with the SM quarks \cite{delAguila:2000kb,delAguila:2000aa,delAguila:2000rc,Buras:2009ka}. See for
instance
\cite{Buras:2009ka}, where a detailed study of this effect in the context 
of a RS scenario has been presented. With $v=174\gev$, $M=3.0\tev$ and $\eps^Q_i\le 1$ these corrections are at most at the level of $0.5\%$ and thus negligible.
Therefore, in our numerical analysis it is legitimate to use the unitarity of the CKM matrix and neglect the corrections to left-handed couplings of $W^\pm$.

For the right-handed $W^\pm$ couplings we get
\begin{align}
 \tilde A_R^{UD} & = \frac{1}{M_Q^2}m^U_\text{diag}\text{diag}\left(\frac{1}{\eps_1^{Q2}},\,\frac{1}{\eps_2^{Q2}},\,\frac{1}{\eps_3^{Q2}}
\right)V_\text { CKM}m^D_\text{diag}\,
\end{align}
and consequently using (\ref{DRW})
\be
\Delta_R^{ij}(W) =
-\frac{g}{\sqrt{2}} \frac{m_i^Um_j^D}{M^2}\frac{V_{ij}^{\rm CKM}}{\eps_i^{Q2}},
\ee
where $m_i^U$ and $m_j^D$ are SM quark masses normalized at $M$.

Using this formula we can check, whether the presence of right-handed currents 
could help in explaining the differences between the determination of 
$\vub$ in exclusive semi-leptonic decays, inclusive $B$ decays and 
$B^+\to\tau^+\nu_\tau$ as proposed in \cite{Crivellin:2009sd,Chen:2008se} and analysed in detail 
in \cite{Buras:2010pz,Crivellin:2011ba}.

Following \cite{Buras:2010pz} we find for exclusive semileptonic decays the effective $V_{ub}$
\be 
V_{ub}^{\rm excl}=C_+ V_{ub},
\ee
while in the case of $B^+\to\tau^+\nu_\tau$
\be 
V_{ub}^{\tau}=C_-V_{ub},
\ee
where 
\be
C_\pm=1\pm\frac{m_u m_b}{M^2}\frac{1}{\eps_1^{Q2}}.
\ee

We find that the sign of the correction in $C_+$ is  opposite to the one required for the explanation of the difference between 
inclusive and exclusive $\vub$ determination
but the correction turns out to be so small that it can be neglected. 
Similar comment applies to $C_-$.
Indeed using (\ref{Xij}) we have  for $M=3.0\tev$
\be
C_\pm=1\pm \frac{m_u m_b}{M^2}\frac{1}{(\eps_3^{Q2} X_{13}^2\vub^2)}\approx 
1\pm 2\frac{1}{\eps_3^{Q2} X_{13}^2}10^{-5}
\ee
However, our analysis of FCNC processes will imply $X_{13}\ge2$
rendering 
these effects to be  totally negligible. 
Recent decrease of the experimental 
branching ratio for $B^+\to\tau^+\nu_\tau$  \cite{BelleICHEP,Tarantino:2012mq} makes the case for RH couplings 
of $W^\pm$ much weaker than two years ago anyway.

With all this information at hand we are now ready to turn our attention 
to FCNC transitions in this model. We should emphasize that the formulae 
given in the next two sections apply to MTFM at large and only when the 
couplings $\Delta^{ij}_{L,R}(Z)$ are specified to the TUM, one obtains 
the formulae specific to this model.

\boldmath
\section{$\Delta F=2$ Transitions}\label{sec:trans}
\unboldmath
\setcounter{equation}{0}

\subsection{Standard Model Results}
The dominant contributions to the off-diagonal elements $M^i_{12}$ in the neutral $K$and $B_{q}$ meson mass matrices come from SM box-diagrams. They 
are given as follows
\bea\label{eq:3.4}
\left(M_{12}^K\right)^*_\text{SM}&=&\frac{G_F^2}{12\pi^2}F_K^2\hat
B_K m_K M_{W}^2\left[
\lambda_c^{2}\eta_1S_0(x_c)+\lambda_t^{2}\eta_2S_0(x_t)+
2\lambda_c\lambda_t\eta_3S_0(x_c,x_t)
\right]
\,,\\
\left(M_{12}^q\right)^*_\text{SM}&=&{\frac{G_F^2}{12\pi^2}F_{B_d}^2\hat
B_{B_d}m_{B_d}M_{W}^2
\left[
\left(\lambda_t^{(q)}\right)^2\eta_B S_0(x_t)
\right]}\,,\label{eq:3.6}
\eea
where $x_i=m_i^2/M_W^2$ and
\be\label{CKMV}
\lambda_i=V_{is}^*V_{id},\qquad \lambda_i^{(q)}=V_{ib}^*V_{iq}
\ee
 with $V_{ij}$ being the elements of the CKM matrix. 
 Here, $S_0(x_i)$ and
$S_0(x_c,x_t)$
are one-loop box functions for which explicit expressions are given e.\,g.~in \cite{Blanke:2006sb}.
The
factors $\eta_i$ are QCD {corrections} evaluated at the NLO level in
\cite{Herrlich:1993yv,Herrlich:1995hh,Herrlich:1996vf,Buras:1990fn,Urban:1997gw}. For $\eta_1$ and $\eta_3$ also NNLO corrections
are known \cite{Brod:2010mj,Brod:2011ty}. Finally $\hat B_K$ and $\hat B_{B_q}$ are the well-known
non-perturbative factors.

It should be emphasized that in the SM only a single operator
\be
{Q}_1^\text{VLL}(K)=\left(\bar s\gamma_\mu P_L d\right)\left(\bar s\gamma^\mu P_L d\right)\,
\ee
and
\be
{Q}_1^\text{VLL}(B_q)=\left(\bar b\gamma_\mu P_L q\right)\left(\bar b\gamma^\mu P_L q\right)\,
\ee
contributes to $M_{12}^K$ and $M_{12}^q\;(q=d,s)$, respectively. Moreover complex phases are only present in the CKM factors.

Our next goal is to generalize these formulae to include the new tree level contributions from $Z$ exchanges  as shown in
Fig.~\ref{fig:Figure Z}. We will see that three distinct new features will characterize these new contributions:
\begin{enumerate}
 \item The flavour structure will differ from the CKM one.
\item
FCNC transitions will appear already {at the tree level} as opposed to the one-loop SM contributions in \eqref{eq:3.4} and \eqref{eq:3.6}.
\item In addition to ${Q}_1^\text{VLL}(K)$ and ${Q}_1^\text{VLL}(B_q)$ (with $q=d,s$) new operators will be present in the effective Hamiltonians in question.
\end{enumerate}

\boldmath
\subsection{Tree Level $Z$ Contributions}
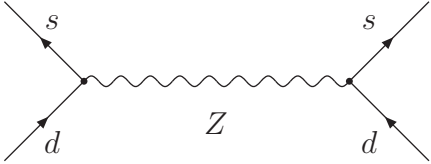
\begin{figure}
\begin{center}
\begin{picture}(150,100)(0,0)
\ArrowLine(10,10)(40,40)
\ArrowLine(40,40)(10,70)
\Photon(40,40)(140,40){2}{9}
\Vertex(40,40){1.3}
\Vertex(140,40){1.3}
\ArrowLine(140,40)(170,70)
\ArrowLine(170,10)(140,40)
\Text(90,20)[cb]{{\Black{$Z$}}}
\Text(28,60)[cb]{{\Black{$s$}}}
\Text(28,13)[cb]{{\Black{$d$}}}
\Text(148,60)[cb]{{\Black{$s$}}}
\Text(148,13)[cb]{{\Black{$d$}}}
\end{picture}
\end{center}
\caption{Tree-level flavour changing $Z^0$ contribution to \kkm\ (the diagram rotated with $90^\circ$ also exists). }\label{fig:Figure Z}
\end{figure}

\unboldmath
We begin our discussion with the tree level $Z$ exchanges contributing 
to $\Delta S=2$ transitions in Fig.~\ref{fig:Figure Z}. Analogous diagrams contribute to $B_{d,s}^0-\bar B_{d,s}^0$ mixings. 

The FCNC Lagrangian is given in (\ref{eq:3.14})--(\ref{eq:3.16})
in terms of
$\Delta_{L,R}^{ij}(Z)$. These couplings are complex quantities
and introduce new flavour and CP-violating interactions that can have a pattern very different from the CKM one. 
Explicit expressions for  $\Delta_{L,R}^{ij}(Z)$ in MTFM are given in 
 Section~\ref{firstdeltas} and
Appendix~\ref{app:Deltas}. 

The diagrams in Fig.~\ref{fig:Figure Z} lead to the following effective Hamiltonian for $\Delta S=2$ transitions mediated by $Z$:
\bea
\left[\Heff^{\Delta S=2}\right]_\text{Z}&=&\frac{1}{2M_\text{Z}^2}\left[\big[\Delta_L^{sd}(Z)\big]^2\left(\bar s\gamma_\mu P_L
d\right)\left(\bar s\gamma^\mu P_L d\right)
+\big[\Delta_R^{sd}(Z)\big]^2\left(\bar s\gamma_\mu P_R d\right)\left(\bar s\gamma^\mu P_R d\right)\right.\nonumber\\
&& \hspace{1cm} {} +\left.2\Delta_L^{sd}(Z)\Delta_R^{sd}(Z)\left(\bar s\gamma_\mu P_L d\right)\left(\bar s\gamma^\mu P_R d\right)\right.\Big]\,.
\label{eq:3.17}
\eea
For the $B_{d,s}^0-\bar B_{d,s}^0$ Hamiltonians one has to replace ``$sd$'' by ``$bd$'' and ``$bs$'', respectively.

The Hamiltonian in (\ref{eq:3.17}) is valid at scales $\mathcal{O}(M_\text{Z})$ and has to be evolved to low energy scales
$\mu=\mathcal{O}(3\gev)$, $\mu=\mathcal{O}(m_b)$ at which the hadronic matrix elements of the operators in question can be evaluated by
lattice methods. The relevant anomalous dimension matrices necessary for this renormalization group evolution have been calculated
at two-loop level in \cite{Ciuchini:1997bw,Buras:2000if} and analytic formulae for the relevant QCD factors analogous to $\eta_i$
in (\ref{eq:3.4}) and (\ref{eq:3.6}) can be found in~\cite{Buras:2001ra}.  The latter formulae do not include the $\ord(\alpha_s)$
corrections to the Wilson coefficients of the relevant new operators at 
$\mu_Z = \ord(M_\text{Z})$. Recently such corrections to $\Delta
F=2$ observables originating in tree level 
contributions of gauge bosons and scalars have been calculated in 
\cite{Buras:2012fs}
and we 
will include them in our analysis.

In order to use the renormalization group analysis of \cite{Buras:2001ra,Buras:2012fs} we
recall the operator basis used there:
\begin{subequations}\label{equ:operatorsZ}
\bea
{Q}_1^\text{VLL}&=&\left(\bar s\gamma_\mu P_L d\right)\left(\bar s\gamma^\mu P_L d\right)\,,\\
{Q}_1^\text{VRR}&=&\left(\bar s\gamma_\mu P_R d\right)\left(\bar s\gamma^\mu P_R d\right)\,,\\
{Q}_1^\text{LR}&=&\left(\bar s\gamma_\mu P_L d\right)\left(\bar s\gamma^\mu P_R d\right)\,,\\
{Q}_2^\text{LR}&=&\left(\bar s P_L d\right)\left(\bar s P_R d\right)\,,
\eea
\end{subequations}
where we suppressed colour indices as they are summed up in each factor. For instance $\bar s\gamma_\mu P_L d$ stands for $\bar s_\alpha\gamma_\mu P_L d_\alpha$ and similarly for other factors.

A straightforward calculation gives us the effective Hamiltonian
for $\Delta S=2$ transitions in the basis (\ref{equ:operatorsZ}) with the Wilson coefficients corresponding to
$\mu_Z=\mathcal{O}(M_\text{Z})$ \cite{Buras:2012fs}
\begin{align}\begin{split}\label{eq:3.19}
 \left[\mathcal{H}_\text{eff}^{\Delta S = 2}\right]_\text{Z}  =&
\frac{(\Delta_L^{sd}(Z))^2}{2M_Z^2}C_1^\text{VLL}(\mu_Z)Q_1^\text{VLL}+\frac{(\Delta_R^{sd}(Z))^2}{2M_Z^2}
C_1^\text{VRR}(\mu_Z)Q_1^\text{VRR} \\
&+\frac{\Delta_L^{sd}(Z)\Delta_R^{sd}(Z)}{
M_Z^2} \left [ C_1^\text{LR}(\mu_Z) Q_1^\text{LR} +C_2^\text{LR}(\mu_Z) Q_2^\text{LR} \right]\,.\end{split}
\end{align}
where including NLO QCD corrections \cite{Buras:2012fs}
\begin{align}\label{equ:WilsonZ}
\begin{split}
C_1^\text{VLL}(\mu)=C_1^\text{VRR}(\mu) 
& = 1+\frac{\alpha_s}{4\pi}\left(-2\log\frac{M_Z^2}{\mu^2}+\frac{11}{3}\right)\,,\end{split}\\
\begin{split}
 C_1^\text{LR}(\mu) 
& =1+\frac{\alpha_s}{4\pi}
\left(-\log\frac{M_Z^2}{\mu^2}-\frac{1}{6}\right)\,,\end{split}\\
C_2^\text{LR}(\mu) &=\frac{\alpha_s}{4\pi}\left(-6\log\frac{M_Z^2}{\mu^2}-1\right)\,.
\end{align}

The latter coefficients are also valid 
for $B_d^0-\bar B_d^0$ and $B_s^0-\bar B_s^0$ systems, where in the 
rest of the formulae  $sd$ should be replaced by $bd$ and $bs$, respectively. 

 We should remark that also tree-level Higgs contributions to $\Delta F=2$ 
processes are present. However, as we have seen the corresponding flavour 
violating Higgs couplings to light quarks are very strongly suppressed and 
these 
contributions are totally negligible.  
We will not discuss them further. The general structure of such contributions 
including NLO QCD corrections can be found in \cite{Buras:2012fs}.

\boldmath
\subsection{Hadronic Matrix Elements}
\unboldmath

In order to complete the analysis of $\Delta F=2$ processes we have to 
include renormalization group QCD evolution from the high energy scales 
down to scales at which hadronic matrix elements are evaluated. In what 
follows we will summarize the expressions for
tree-level contributions to the off-diagonal 
elements $M_{12}$ for the $\Delta S=2$ transition. Analogous expressions for 
$B_{s,d}$ systems can be easily obtained in the same manner.

In presenting our results we will use
the so-called $P_i^a$ QCD factors of \cite{Buras:2001ra} that include both hadronic matrix
elements of contributing operators and renormalization group evolution from high energy to low energy scales. These factors depend on the
system considered, on the high energy matching scale and on the renormalization 
scheme used to renormalize the operators. The formulae for these factors
have been given in \cite{Buras:2001ra} in the  $\overline{\rm MS}$-NDR  renormalization scheme. This scheme dependence is canceled by the non-logarithmic
$\ord(\alpha_s)$ corrections calculated in \cite{Buras:2012fs} and 
given in the previous subsection. The 
logarithmic corrections calculated also there 
cancel the scale dependence of $P^a_i$ as demonstrated in that paper. 

The formulae for various contributions to $M_{12}$ defined in the case of 
$\Delta S=2$ through
\be
\left(M_{12}^K\right)^\ast=\langle\bar K^0|\Heff^{\Delta S=2}|K^0\rangle\,,
\label{eq:3.23}
\ee
are easily obtained from 
the Hamiltonians presented above by replacing the operators by their 
hadronic matrix elements
\be\label{matrixelements}
\langle Q^a_i(\mu_Z)\rangle \equiv \frac{m_K F_K^2}{3} P^a_i(\mu_Z)
\ee
with analogous definition for the $B_{s,d}$ systems. We list now the resulting 
NLO expressions.
For the $Z$ contribution we find \cite{Buras:2012fs}
\begin{align}\begin{split}\label{M12Z}
 \left(M_{12}^K\right)^\star_Z   =&
 \frac{(\Delta_L^{sd}(Z))^2}{2M_Z^2}C_1^\text{VLL}(\mu_Z)\langle Q_1^\text{VLL}(\mu_Z)\rangle
 +\frac{(\Delta_R^{sd}(Z))^2}{2M_Z^2}
 C_1^\text{VRR}(\mu_Z)\langle Q_1^\text{VLL}(\mu_Z)\rangle \\
 &+\frac{\Delta_L^{sd}(Z)\Delta_R^{sd}(Z)}{
 M_Z^2} \left [ C_1^\text{LR}(\mu_Z) \langle Q_1^\text{LR}(\mu_Z)\rangle +
 C_2^\text{LR}(\mu_Z) \langle Q_2^\text{LR}(\mu_Z)\rangle \right]\,,\end{split}
 \end{align}
The relevant Wilson coefficients for $\Delta S = 2$ are given in Eq.~(\ref{equ:WilsonZ}) with obvious replacements for $B_{s,d}$ systems.

 In order to find numerical values of $P_i^a$ and consequently 
$\langle Q^a_i(\mu_Z)\rangle $
one needs the values of 
the corresponding non-perturbative parameters $B^a_i$ defined in  
\cite{Buras:2001ra}. These are given in terms of the parameters $B_i$
used in \cite{Boyle:2012qb,Bertone:2012cu,Bouchard:2011xj} as follows:
\begin{subequations}
\begin{align}
& B_1^\text{VLL}(\mu_0) = B_1^\text{VRR}(\mu_0) = B_1(\mu_0)\,,\\
& B_1^\text{LR}(\mu_0) = B_5(\mu_0)\,,\\
& B_2^\text{LR}(\mu_0) = B_4(\mu_0)\,.
\end{align}
\end{subequations}

In the case of the $K^0-\bar K^0$ system, 
the values for $B_i$ in the $\overline{\rm MS}$-NDR scheme have been provided 
in \cite{Boyle:2012qb,Bertone:2012cu}. We have just
used the average of the results in \cite{Boyle:2012qb,Bertone:2012cu} that
are consistent with each other.
On the other hand the most recent results for 
the $B_{d,s}^0-\bar B^0_{d,s}$ systems \cite{Bouchard:2011xj} in the same 
renormalization scheme are given not for $B_i$ but for $B_i^\text{eff}F^2_{B_q}$ as this 
reduces the errors in $\langle Q^a_i(\mu_H)\rangle$. 
In Table~\ref{tab:B_i} we collect the values of 
\begin{align}\label{equ:defDi}
& D_i^\text{eff}(K)\equiv B_i^\text{eff}(\mu_L) F^2_K\,\text{ for}\quad i = 4,5\,,\\
& D_i^\text{eff}(B_q)\equiv B_i^\text{eff}(\mu_b) F^2_{B_q}\,\text{ for}\quad i = 4,5\,,\quad q = d,s\,,
\end{align}
  at the relevant scale  $\mu_0$ given in the last column.

The effective parameters
$B^\text{eff}_i(\mu_L)$ are defined in the case of $K^0-\bar K^0$ 
mixing ($\mu_L=3\gev$) by 
\be
\label{Balat}
B^\text{eff}_i(\mu_L)\equiv
\left(\frac{m_K}{m_s(\mu_L)+m_d(\mu_L)}\right)^2 B_i(\mu_L)
\ee
with the most recent values for $B_{4,5}$ at $\mu_L=3\gev$ in 
the  $\overline{\text{MS}}$-NDR scheme given by \cite{Boyle:2012qb,Bertone:2012cu} 
\be
B_4=0.76\pm0.07, \qquad B_5=0.56\pm0.06~.
\ee

In the case of $B_{d,s}^0-\bar B_{d,s}^0$ mixings one has to make the replacements
$\mu_L\to \mu_b=4.2\gev$ and  change appropriately flavours
\be
B^\text{eff}_i(\mu_b)\equiv
\left(\frac{m_B}{m_b(\mu_b)+m_d(\mu_b)}\right)^2 B_i(\mu_b)
\ee
with an analogous formula for the $B_s^0-\bar B^0_s$ system.
The values of weak decay constants and of meson masses required to obtain the hadronic matrix elements by means of
(\ref{matrixelements}) in the case of  $K^0-\bar K^0$ system and analogous 
expressions for $B^0_{d,s}-\bar B^0_{d,s}$ systems
are given in Table~\ref{tab:input}. For the SM contributions  we use the 
necessary input from the latter table. 

\begin{table}[!ht]
{\renewcommand{\arraystretch}{1.3}
\begin{center}
\begin{tabular}{|c||c|c|c|}
\hline
&$D_4^\text{eff}$&$D_5^\text{eff}$&$\mu_0$\\
\hline
\hline
$K^0$-$\bar K^0$ &0.675 &0.517 &3.0\gev\\
\hline
$B_d^0$-$\bar B_d^0$&0.093&0.127&4.2\gev\\
\hline
$B_s^0$-$\bar B_s^0$&0.135 &0.178&4.2\gev\\
\hline
\end{tabular}
\end{center}}
\caption{\it Central values of the parameters $D_i(M)$ in units of $\gev^2$
in the $\overline{\text{MS}}$-NDR scheme based on~\cite{Boyle:2012qb,Bertone:2012cu} for $K^0-\bar K^0$ system and  \cite{Bouchard:2011xj} for 
$B_{d,s}^0-\bar B^0_{d,s}$ systems.
The scale $\mu_0$ at which $D_i(M)$ are evaluated is given in the last column.  
\label{tab:B_i}}
\end{table}

Finally, we collect 
the values of  $\langle Q^a_i(\mu_Z)\rangle $ contributing to
$(M_{12})_Z$ for $\mu_Z=m_t$
in Table~\ref{tab:Qi}. As the SM Wilson coefficients in the SM are 
also evaluated at $m_t$ we have also chosen this scale for the 
Wilson coefficients of new operators. As we include NLO corrections 
to the Wilson coefficients the same final results for the mixing 
amplitudes  up to higher order corrections would be obtained if we 
used $\mu_Z=M_Z$ or any value $\ord(M_W,m_t)$.

\begin{table}[!ht]
{\renewcommand{\arraystretch}{1.3}
\begin{center}
\begin{tabular}{|c||c|c|}
\hline
&$\langle Q_1^\text{LR}(m_t)\rangle$& $\langle Q_2^\text{LR}(m_t)\rangle$\\
\hline
\hline
$K^0$-$\bar K^0$ &$-0.11$ &0.18  \\
\hline
$B_d^0$-$\bar B_d^0$& $-0.21$ &0.27  \\
\hline
$B_s^0$-$\bar B_s^0$& $-0.30$ &0.40 \\
\hline
\end{tabular}
\end{center}}
\caption{\it Hadronic matrix elements $\langle Q_i^a(m_t)\rangle$  in units of GeV$^3$ at $m_t=163\gev$.
\label{tab:Qi}}
\end{table}

Concerning the new contributions to the Wilson coefficients of 
${Q}_1^\text{VLL}$  and the contributions from ${Q}_1^\text{VRR}$ 
operators they can be included effectively by replacing the 
flavour independent
$S_0(x_t)$ in the SM formulae by the functions $S(M)$
($M=K,B_d,B_s$):
\begin{equation}\label{sum}
S(M)=S_0(x_t)+ [\Delta S(M)]_{\rm VLL}+[\Delta S(M)]_{\rm VRR}.
\end{equation}
One finds \cite{Buras:2012jb}
\be\label{Zprime1}
[\Delta S(B_q)]_{\rm VLL}=
\left[\frac{\Delta_L^{bq}(Z)}{\lambda_t^{(q)}}\right]^2
\frac{4\tilde r}{M^2_{Z}g_{\text{SM}}^2}, \qquad
[\Delta S(K)]_{\rm VLL}=
\left[\frac{\Delta_L^{sd}(Z)}{\lambda_t^{(K)}}\right]^2
\frac{4\tilde r}{M^2_{Z}g_{\text{SM}}^2},
\ee
where $g_{\text{SM}}$ is defined in (\ref{gsm}) and 
$\tilde r=1.068$.  $[\Delta S(M)]_{\rm VRR}$  is then
found from the formulae above by simply replacing L by R. 
The important new property is the flavour dependence in these functions and
the fact that they carry new complex phases.

\subsection{Combining SM and Tree Contributions}
The final results for $M_{12}^K$, $M_{12}^d$ and $M_{12}^s$, that govern the analysis of $\Delta F=2$ transitions in the MTFM in question, are then given by
\be
M_{12}^i=\left(M_{12}^i\right)_\text{SM}+\left(M_{12}^i\right)_\text{Z}
\equiv \left(M_{12}^i\right)_\text{\rm SM}+\left(M_{12}^i\right)_\text{NP},
\qquad(i=K,d,s)
\,,
\label{eq:3.33}
\ee
with $\left(M_{12}^i\right)_\text{SM}$ given in (\ref{eq:3.4})--(\ref{eq:3.6}) and $\left(M_{12}^i\right)_\text{Z}$ in (\ref{M12Z}).

\subsection[Basic formulae for $\Delta F=2$ observables]
{\boldmath Basic formulae for $\Delta F=2$ observables}

Having the mixing amplitudes $M^i_{12}$ at hand we can calculate all relevant 
$\Delta F=2$ observables. To this end we collect below those formulae that we used in our numerical analysis.

The $K_L-K_S$ mass difference is given by
\be
\Delta M_K=2\left[\re\left(M_{12}^K\right)_\text{\rm SM}+\re\left(M_{12}^K\right)_\text{NP}\right]\,,
\label{eq:3.34}
\ee
and the CP-violating parameter $\varepsilon_K$ by
\be
\varepsilon_K=\frac{\kappa_\eps e^{i\varphi_\eps}}{\sqrt{2}(\Delta M_K)_\text{exp}}\left[\im\left(M_{12}^K\right)_\text{\rm
SM}+\im\left(M_{12}^K\right)_\text{NP}\right]\,,
\label{eq:3.35}
\ee
where $\varphi_\eps = (43.51\pm0.05)^\circ$ and $\kappa_\eps=0.94\pm0.02$ \cite{Buras:2008nn,Buras:2010pza} takes into account 
that $\varphi_\eps\ne \tfrac{\pi}{4}$ and includes long distance effects in $\im( \Gamma_{12})$ and $\im (M_{12})$. 

For the mass differences in the $B_{d,s}^0-\bar B_{d,s}^0$ systems we have
\be
\Delta M_q=2\left|\left(M_{12}^q\right)_\text{\rm SM}+\left(M_{12}^q\right)_\text{NP}\right|\qquad (q=d,s)\,.
\label{eq:3.36}
\ee
Let us then write \cite{Bona:2005eu}
\be
M_{12}^q=\left(M_{12}^q\right)_\text{\rm SM}+\left(M_{12}^q\right)_\text{NP}=\left(M_{12}^q\right)_\text{\rm SM}C_{B_q}e^{2i\varphi_{B_q}}\,,
\label{eq:3.37}
\ee
where
\be
\left(M_{12}^d\right)_\text{\rm SM}=\left|\left(M_{12}^d\right)_\text{\rm SM}\right|e^{2i\beta}\,,\qquad\beta\approx 22^\circ\,,
\label{eq:3.38}
\ee
\be
\left(M_{12}^s\right)_\text{\rm SM}=\left|\left(M_{12}^s\right)_\text{\rm SM}\right|e^{2i\beta_s}\,,\qquad\beta_s\simeq -1^\circ\,.
\label{eq:3.39}
\ee
Here the phases $\beta$ and $\beta_s$ are defined through
\be
V_{td}=|V_{td}|e^{-i\beta}\quad\textrm{and}\quad V_{ts}=-|V_{ts}|e^{-i\beta_s}\,.
\label{eq:3.40}
\ee
We find then
\be
\Delta M_q=(\Delta M_q)_\text{\rm SM}C_{B_q}\,,
\label{eq:3.41}
\ee
and
\bea
S_{\psi K_S} &=& \sin(2\beta+2\varphi_{B_d})\,,
\label{eq:3.42} \\
S_{\psi\phi} &= & \sin(2|\beta_s|-2\varphi_{B_s})\,,
\label{eq:3.43}
\eea
with the latter two observables being the coefficients of $\sin(\Delta M_d t)$ and $\sin(\Delta M_s t)$ in the 
time dependent asymmetries in $B_d^0\to\psi K_S$ and $B_s^0\to\psi\phi$, respectively. 
At this stage a few comments on the assumptions leading to expressions (\ref{eq:3.42}) and (\ref{eq:3.43}) are 
in order. These simple formulae follow only if there are no weak phases in the decay amplitudes for 
$B_d^0\to\psi K_S$ and $B_s^0\to\psi\phi$ as is the case in the SM and also in the LHT model, where due to 
T-parity there are no new contributions to decay amplitudes at tree level so that these amplitudes are
dominated by SM contributions \cite{Blanke:2006sb}. In the model discussed in the present paper new contributions 
to decay amplitudes with non-vanishing weak phases are in principle present at tree level. However, as we demonstrated previously in the
case of TUM 
these contribution can be totally neglected when calculating $S_{\psi K_S}$ and 
$S_{\psi\phi}$. 

\section{Effective Hamiltonians for \boldmath{$\Delta F=1$} Decays}\label{sec:Heff}
\setcounter{equation}{0}

\subsection{Preliminaries}
The goal of the present section is to give formulae for the effective
Hamiltonians relevant for rare $K$ and $B$ decays that in addition to
SM one-loop contributions  could generally include tree level contributions from the SM
$Z$ gauge boson,  the tree level neutral Higgs $H^0$ and the corrections 
to SM one-loop contributions due to the modification of the $W^\pm$ couplings.
 In the absence of $\tan\beta$ 
enhancement present in supersymmetric models and generally multi-Higgs 
models, the $H^0$ contributions can be neglected as their flavour violating 
couplings to quarks are strongly suppressed in our model and the 
couplings to leptons are strongly suppressed by small lepton masses. Similarly we find that in TUM
the corrections from modified $W^\pm$ couplings in loop diagrams governing in 
the SM rare $K$ and $B$ decays are much 
smaller than the tree-level $Z^0$ amplitudes. This can be expected on 
the basis of our estimates of such corrections in subsection~\ref{Wcouplings}.
 Therefore in what follows 
we will present only the effective Hamiltonians based on the SM loop 
contributions and the induced tree level $Z^0$ exchanges.

 The case of radiative decay $B\to X_{s}\gamma$ is special as here there 
are no contributions in the SM and MTFM at the tree level and one has to 
check whether 
the modifications of $W^\pm$ couplings and $Z^0$ exchanges in loop 
diagrams absent in the SM can have a visible impact on the branching 
ratio. In particular the presence of right-handed $W^\pm$ 
couplings in this decay can lead to enhancement factors $m_t/m_b$. 
 Moreover as now  $W^\pm$, $Z^0$ and $H^0$ couplings between light and heavy quarks are involved one has to check whether heavy quark
contributions to $B\to X_s\gamma$ are 
significant. Following formalism for $W^\pm$ and $Z^0$ contributions with both 
left-handed and right-handed couplings developed in \cite{Buras:2011zb} in 
the context of gauged flavour models and in \cite{Blanke:2011ry} in the context of  $SU(2)_L \times SU(2)_R \times U(1)$ model, we have verified that the NP contributions from  $W^\pm$ and  $Z^0$
exchanges to  $B\to X_{s}\gamma$ even in the presence of heavy quarks  are negligible in the TUM. Extending 
this formalism to $H^0$ internal exchanges we have found that also these 
contributions are very small. As this analysis, including QCD corrections, 
is rather involved and the NP corrections in the TUM are negligible 
anyway, we will present it elsewhere in the context of a more 
general version of the MTFM.

In \cite{Blanke:2008yr} 
 general formulae for effective Hamiltonians resulting from tree level neutral
gauge boson exchanges with arbitrary masses and arbitrary left-handed and right-handed couplings have been presented in the context of the RS scenario.
We could in principle
apply them directly to our case. However, as in our case we have 
only one neutral gauge boson instead of three, we can write the relevant formulae in a simpler form, as done already 
in \cite{Buras:2012jb}, than it was done in \cite{Blanke:2008yr}.

\boldmath
\subsection{Effective Hamiltonian for $\bar s\to \bar d\nu\bar\nu$}\label{sec:sdnn}
\unboldmath
The effective Hamiltonian for $\bar s \rightarrow \bar d\nu\bar\nu$ transitions 
 resulting from $Z$--penguin and box diagrams 
is given in the SM as follows
\be\label{Heffpr}
\left[\Heff^{\nu\bar\nu}\right]^K_\text{SM}=g_{\text{SM}}^2\sum_{\ell=e,\mu,\tau}{\left[\lambda_c {X_\text{NNL}^\ell(x_c)}+\lambda_t X(x_t)\right]}
(\bar s\gamma_\mu P_L d)(\bar\nu_\ell\gamma_\mu P_L \nu_\ell)+h.c.\,,
\ee
where $x_i=m_i^2/M_W^2$, $\lambda_i =V_{is}^*V_{id}^{}$ and $V_{ij}$ are the elements of the CKM matrix. {$X_\text{NNL}^\ell(x_c)$} and $X(x_t)$ comprise 
internal charm and top quark contributions, respectively. They are known to high accuracy including QCD corrections
\cite{Buchalla:1998ba,Buras:2005gr,Buras:2006gb} and electroweak corrections \cite{Brod:2008ss,Brod:2010hi}. For convenience we have
introduced
\be\label{gsm}
g_{\text{SM}}^2=4\frac{G_F}{\sqrt 2}\frac{\alpha}{2\pi\sin^2\theta_W}\,.
\ee

\begin{figure}
\begin{center}
\begin{picture}(100,100)(0,0)
\ArrowLine(0,10)(30,40)
\ArrowLine(30,40)(0,70)
\Photon(30,40)(100,40){2}{6}
\Vertex(30,40){1.3}
\Vertex(100,40){1.3}
\ArrowLine(100,40)(130,70)
\ArrowLine(130,10)(100,40)
\Text(65,21)[cb]{{\Black{$Z$}}}
\Text(20,60)[cb]{{\Black{$s$}}}
\Text(20,13)[cb]{{\Black{$d$}}}
\Text(110,60)[cb]{{\Black{$\nu$}}}
\Text(110,13)[cb]{{\Black{$\nu$}}}
\end{picture}
\end{center}
\caption{\it Tree level contributions of $Z$ to the $s\to d\nu\bar\nu$ effective Hamiltonian.\label{KplusZ_H}}
\end{figure}
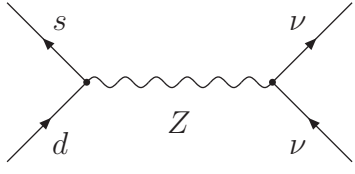

 In the MTFM (\ref{Heffpr}) is modified by tree-level diagrams in 
 Fig.\ \ref{KplusZ_H}.
The FCNC $Z\bar sd$ vertex has been given in (\ref{eq:3.14})-(\ref{eq:3.16})
with explicit expressions for $\Delta_L^{sd}(Z)$ and $\Delta_R^{sd}(Z)$ given 
in Section~\ref{firstdeltas} and 
in Appendix \ref{app:Deltas}. 

For the $Z\nu\bar\nu$ coupling we analogously write 
\be\label{Lnunu}
\mathcal L_{\nu\bar\nu}(Z)=  \Delta_L^{\nu\nu}(Z)(\bar\nu\gamma_\mu P_L\nu)Z^\mu,
\qquad
\Delta_L^{\nu\nu}(Z)=-\frac{g}{2 c_W}.
\ee

A straightforward calculation of the diagram in Fig.\ \ref{KplusZ_H} results in a new contribution to $\left[\Heff^{\nu\bar\nu}\right]^K$
\be
\left[\Heff^{\nu\bar\nu}\right]_{Z}^K=\frac{\Delta_L^{\nu\nu}(Z)}{M_{Z}^2}
\left[\Delta_L^{sd}(Z)(\bar s\gamma^\mu P_L d)+\Delta_R^{sd}(Z)
(\bar s\gamma^\mu P_R d)\right]
\left(\bar\nu\gamma_\mu P_L\nu\right)+h.c.\,.\label{eq:HeffKnunuZ1}
\ee

Combining this contribution with the SM contribution in  \eqref{Heffpr},
\be
\left[\Heff^{\nu\bar\nu}\right]^K = \left[\Heff^{\nu\bar\nu}\right]^K_\text{SM}+
\left[\Heff^{\nu\bar\nu}\right]_{Z}^K\,,
\ee
we find
\bea
\left[\Heff^{\nu\bar\nu}\right]^K &=&
g_{\text{SM}}^2\sum_{\ell=e,\mu,\tau}
{\left[\lambda_c {X_\text{NNL}^\ell(x_c)}+\lambda_t X_L(K)\right]}
(\bar s\gamma^\mu P_L d)(\bar\nu_\ell\gamma_\mu P_L\nu_\ell)\nn\\
&& {}
+ g_{\text{SM}}^2\sum_{\ell=e,\mu,\tau}{\left[\lambda_t^{(K)}X_R(K)\right]}
(\bar s\gamma^\mu P_Rd)(\bar\nu_\ell\gamma_\mu P_L\nu_\ell)
+h.c.\,.
\eea
Here we have introduced the functions $X_L(K)$ and $X_R(K)$
\be\label{XLK}
X_L(K)=X(x_t)+\frac{\Delta_L^{\nu\bar\nu}(Z)}{g^2_{\rm SM}M_Z^2}
                                       \frac{\Delta_L^{sd}(Z)}{V_{ts}^* V_{td}},
\ee
\be\label{XRK}
X_R(K)=\frac{\Delta_L^{\nu\bar\nu}(Z)}{g^2_{\rm SM}M_Z^2}
                                       \frac{\Delta_R^{sd}(Z)}{V_{ts}^* V_{td}}.
\ee
Finally the SM loop function $X(x_t)$, resulting from $Z$-penguin and box diagrams  is given as follows
\be
X(x_t)=\eta_X~{\frac{x_t}{8}}\;\left[{\frac{x_t+2}{x_t-1}}
+ {\frac{3 x_t-6}{(x_t -1)^2}}\; \ln x_t\right], \qquad \eta_X=0.994
\ee
$\eta_X$ is
QCD correction to these diagrams \cite{Buchalla:1998ba,Misiak:1999yg}
 when $m_t\equiv m_t(m_t)$.

Some comments are in order:
\bi
\item
In the SM only a single operator 
$(\bar s\gamma_\mu P_L d)(\bar\nu\gamma_\mu P_L\nu)$ is present. This is due to the purely left-handed structure of $SU(2)_L$ gauge couplings.
\item
In the MTFM in question also the operator 
$(\bar s\gamma_\mu P_R d)(\bar\nu\gamma_\mu P_L\nu)$
is present, as $\hat\Delta_R(Z)$ coupling matrices have
non-diagonal entries. 
\item
On the other hand we will keep the neutrino couplings SM-like, that is 
 purely left-handed.
\item
{As all NP contributions have been collected in the term proportional to $\lambda_t^{(K)}$, {$X_\text{NNL}^\ell(x_c)$} contains only the SM contributions} that are known including QCD corrections at the NNLO level \cite{Buras:2005gr,Buras:2006gb}.
\ei

\boldmath
\subsection{Effective Hamiltonian for $b\to d\nu\bar\nu$ and $b\to s\nu\bar\nu$}
\label{sec:bqnn}\unboldmath

Let us now generalize the result obtained in the previous section to the case of $b\to d\nu\bar\nu$ and $b\to s\nu\bar\nu$
transitions. Basically only two steps have to be performed:
\begin{enumerate}
\item
All flavour indices have to be adjusted appropriately.
\item
The charm quark contribution can be safely neglected in $B$ physics.
\end{enumerate}
The effective Hamiltonian for $b\to q\nu\bar\nu$ ($q=d,s$) is then given as follows:
\bea
\left[\Heff^{\nu\bar\nu}\right]^{B_q} &=&
g_{\text{SM}}^2\sum_{\ell=e,\mu,\tau}
{\left[V_{tq}^\ast V_{tb} X_L(B_q)\right]}
(\bar q\gamma^\mu P_L b)(\bar\nu_\ell\gamma_\mu P_L\nu_\ell)\nn\\
&& + g_{\text{SM}}^2\sum_{\ell=e,\mu,\tau}{\left[V_{tq}^\ast V_{tb} X_R(B_q)\right]}
(\bar q\gamma^\mu P_R b)(\bar\nu_\ell\gamma_\mu P_L\nu_\ell)
+h.c.\,,
\eea

with

\be\label{XLRB}
 X_{\rm L}(B_q)=X(x_t)+\left[\frac{\Delta_{L}^{\nu\nu}(Z)}{M_Z^2g^2_{\rm SM}}\right]
\frac{\Delta_{L}^{qb}(Z)}{ V_{tq}^\ast V_{tb}}
 \qquad 
 X_{\rm R}(B_q)=\left[\frac{\Delta_{L}^{\nu\nu}(Z)}{M_Z^2g^2_{\rm SM}}\right]
\frac{\Delta_{R}^{qb}(Z)}{ V_{tq}^\ast V_{tb}}.
\ee

Again all relevant {$\Delta_{L,R}^{bq}$ entries} in the MTFM can be found in 
Section~\ref{firstdeltas} and Appendix~\ref{app:Deltas}. 

Note that the functions $X_{L,R}(K)$ and $X_{L,R}(B_q)$ presented above 
depend on the quark flavours involved, through the flavour indices in the $\Delta_{L,R}^{ij}(Z)$
{$(i,j=s,d,b)$} couplings and through the CKM elements that have been 
factored out. 
 While in principle $\Delta_{L,R}^{ij}(Z)$ could be aligned with the corresponding CKM factors, this is generally not the case and the
functions in question 
become complex quantities that are flavour dependent.
 This should be contrasted {with} the case of the SM and CMFV 
models where the decays in question in the $K$, $B_d$ and $B_s$ systems are governed by a \emph{flavour-universal} loop 
function $X(x_t)$ and the only flavour dependence enters through 
the CKM factors. 
Consequently certain SM-relations and more generally 
CMFV-relations can be violated in MTFM. However, as we will see below in TUM 
they are satisfied with high precision in the $B_{s,d}$ systems once all existing constraints on FCNC processes are taken into account. The case of $\kpn$ is 
different.

\boldmath
\subsection{Effective Hamiltonian for $b\to d\ell^+\ell^-$ and $b\to s\ell^+\ell^-$}\label{sec:bqll}
\unboldmath
The effective Hamiltonian for $b\to q\ell^+\ell^-$ ($q=d,s$) can straightforwardly be obtained following the derivation of the effective Hamiltonian for 
  $s\to d\ell^+\ell^-$ transition, presented in \cite{Blanke:2008yr} and 
properly adjusting all flavour indices. In addition, in contrast to the $s\to d\ell^+\ell^-$ transition, now also the dipole operator {contributions} mediating the decay $b\to s\gamma$ become relevant. 
Explicit formulae
for these contributions will be presented below. In the following we will
denote the total contribution of the dipole operators to the effective Hamiltonian in question simply by $\Heff(b\to s\gamma)$.  As already
mentioned 
previously in the TUM, this Hamiltonian is governed by SM contributions 
and its explicit form will not be given here.

\begin{figure}
\begin{center}
\begin{picture}(150,100)(0,0)
\ArrowLine(10,10)(40,40)
\ArrowLine(40,40)(10,70)
\Photon(40,40)(140,40){2}{9}
\Vertex(40,40){1.3}
\Vertex(140,40){1.3}
\ArrowLine(140,40)(170,70)
\ArrowLine(170,10)(140,40)
\Text(90,20)[cb]{{\Black{$Z$}}}
\Text(28,60)[cb]{{\Black{$b$}}}
\Text(28,13)[cb]{{\Black{$s$}}}
\Text(148,60)[cb]{{\Black{$\ell$}}}
\Text(148,13)[cb]{{\Black{$\ell$}}}
\end{picture}
\end{center}
\caption{\it Tree level contributions of $Z$ to the $b\to s\ell^+\ell^-$ effective Hamiltonian.}\label{Bleptons}
\end{figure}
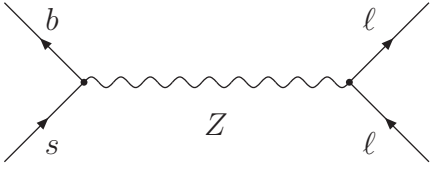

 The relevant Feynman diagram for the $Z$ contribution,  
shown in Fig.\ \ref{Bleptons},
contains on the l.\,h.\,s.\ the vertex, which we already encountered in the 
case of {the} $b\to s \nu \bar \nu$ decay. The relevant FCNC Lagrangian for
 $Z\bar bs$ couplings has been given in (\ref{eq:3.14})-(\ref{eq:3.16}).
For the $\ell^+ \ell^-$ vertex we write in analogy to (\ref{Lnunu}) 
\begin{equation}
\mathcal{L}_{\ell \bar \ell}(Z)= \left[\Delta_L^{\ell\ell}(Z)
(\bar\ell\gamma_\mu P_L\ell)
+\Delta_R^{\ell\ell}(Z)(\bar\ell\gamma_\mu P_R\ell)\right]
Z^\mu\,,
\end{equation}
where with definitions in (\ref{DeltasVA})
\be
\Delta_A^{\nu\nu}(Z)=-\frac{g}{2 c_W}, \quad 
\Delta_R^{\mu\mu}(Z)=-\frac{g}{2 c_W}2 s_W^2~.
\ee

We find ($q=d,s$) then
\be\label{eq:Heffqll}
 \Heff(b\to s \ell\bar\ell)
= \Heff(b\to s\gamma)
-  \frac{4 G_{\rm F}}{\sqrt{2}} \frac{\alpha}{4\pi}V_{ts}^* V_{tb} \sum_{i = 9,10} [C_i(\mu)Q_i(\mu)+C^\prime_i(\mu)Q^\prime_i(\mu)]
\end{equation}
where
\be\label{QAQVL}
Q_9  = (\bar s\gamma_\mu P_L b)(\bar \ell\gamma^\mu\ell),\qquad
Q_{10}  = (\bar s\gamma_\mu P_L b)(\bar \ell\gamma^\mu\gamma_5\ell),
\ee
\be\label{QAQVR}
Q^\prime_9  = (\bar s\gamma_\mu P_R b)(\bar \ell\gamma^\mu\ell), \qquad
Q^\prime_{10}  = (\bar s\gamma_\mu P_R b)(\bar \ell\gamma^\mu\gamma_5\ell)\,.
\ee
For  the  Wilson coefficients we find
\begin{align}
 \sin^2\theta_W C_9 &=[\eta_Y Y_0(x_t)-4\sin^2\theta_W Z_0(x_t)]
-\frac{1}{g_{\text{SM}}^2}\frac{1}{M_{Z}^2}
\frac{\Delta_L^{sb}(Z)\Delta_V^{\mu\bar\mu}(Z)} {V_{ts}^* V_{tb}} ,\\
   \sin^2\theta_W C_{10} &= -\eta_Y Y_0(x_t) -\frac{1}{g_{\text{SM}}^2}\frac{1}{M_{Z}^2}
\frac{\Delta_L^{sb}(Z)\Delta_A^{\mu\bar\mu}(Z)}{V_{ts}^* V_{tb}}, \label{C10}\\
  \sin^2\theta_W C^\prime_9         &=-\frac{1}{g_{\text{SM}}^2}\frac{1}{M_{Z}^2}
\frac{\Delta_R^{sb}(Z)\Delta_V^{\mu\bar\mu}(Z)}{V_{ts}^* V_{tb}},\\
  \sin^2\theta_W C_{10}^\prime   &= -\frac{1}{g_{\text{SM}}^2}\frac{1}{M_{Z}^2}
\frac{\Delta_R^{sb}(Z)\Delta_A^{\mu\bar\mu}(Z)}{V_{ts}^* V_{tb}},\label{C10P}
 \end{align}
where we have defined
\begin{align}\label{DeltasVA}
\begin{split}
 &\Delta_V^{\mu\bar\mu}(Z)= \Delta_R^{\mu\bar\mu}(Z)+\Delta_L^{\mu\bar\mu}(Z),\\
&\Delta_A^{\mu\bar\mu}(Z)= \Delta_R^{\mu\bar\mu}(Z)-\Delta_L^{\mu\bar\mu}(Z).
\end{split}
\end{align}

Here $Y_0(x_t)$ and $Z_0(x_t)$ are one-loop functions, analogous to $X_0(x_t)$, that
result from various penguin and box diagrams and given as follows
\be
Y_0(x_t)=\frac{x_t}{8}\left(\frac{x_t-4}{x_t-1}+ \frac{3 x_t\log x_t}{(x_t-1)^2}\right)
\ee
\be
Z_0(x_t) = -\frac{1}{9}\log x_t + \frac{18 x_t^4-163 x_t^3+ 259 x_t^2 -108 x_t}{144(x_t-1)^3}+ \frac{32 x_t^4-38 x_t^3-15
x_t^2+18x_t}{72(x_t-1)^4}\log x_t~.
\ee
$\eta_Y=1.012$ is QCD correction evaluated for  $m_t=m_t(m_t)$ 
\cite{Buchalla:1998ba,Misiak:1999yg}.

Defining 
\begin{align}
 Y_q & = \eta_Y Y_0(x_t) + \frac{\Delta_L^{qb}(Z)\Delta_A^{\mu\bar\mu}(Z)}{g_\text{SM}^2 M_Z^2 V_{tq}^\star V_{tb}}\,,\label{equ:Yq}\\
Y_q^\prime & = \frac{\Delta_R^{qb}(Z)\Delta_A^{\mu\bar\mu}(Z)}{g_\text{SM}^2 M_Z^2 V_{tq}^\star V_{tb}}\,,\label{equ:Yqp}\\
Z_q & = Z_0(x_t) + \frac{1}{4\sin^2\theta_W}\frac{2\Delta_R^{\mu\bar\mu}(Z)\Delta_L^{qb}(Z)}{g_\text{SM}^2 M_Z^2 V_{tq}^\star V_{tb}}\,,\\
Z_q^\prime & = \frac{1}{4\sin^2\theta_W}\frac{2\Delta_R^{\mu\bar\mu}(Z)\Delta_R^{qb}(Z)}{g_\text{SM}^2 M_Z^2 V_{tq}^\star V_{tb}}\,,
\end{align}
we can write the Wilson coefficients as
\begin{align}
 \sin^2\theta_W C_9 & = Y_q-4\sin^2\theta_W Z_q\,,\\
  \sin^2\theta_W C^\prime_9   & = Y_q^\prime-4\sin^2\theta_W Z_q^\prime\,,\\
\sin^2\theta_W C_{10}  &  = - Y_q\,,\\
 \sin^2\theta_W C_{10}^\prime  & = - Y_q^\prime\,.\\
\end{align}

The effective Hamiltonian for  $s\to d\ell^+\ell^-$ transition can be obtained 
directly from \cite{Blanke:2008yr} or from formulae given above by replacing 
$q$ by $K$ and appropriately changing the flavour indices.

 \section{Rare Decays}\label{sec:rare}
\setcounter{equation}{0}

\boldmath
\subsection{$K^+ \rightarrow \pi^+\nu\bar\nu$ and $K_L \rightarrow \pi^0\nu\bar\nu$}
\unboldmath

Having at hand the effective Hamiltonian for $\bar s\to \bar d\nu\bar\nu$ transitions
derived in Section \ref{sec:sdnn} it is now straightforward to obtain
explicit expressions for the branching ratios $\mathcal{B}(\kpn)$ and $\mathcal{B}(\klpn)$.
 Reviews of these two decays can be found in 
\cite{Buras:2004uu,Isidori:2006yx,Smith:2006qg}.

The 
branching ratios for the two  
$K \to \pi \nu\bar \nu$ modes that follow from the Hamiltonian  
in Section \ref{sec:sdnn}
 can be written generally as
\begin{gather} \label{eq:BRSMKp}   
  \mathcal{B} (K^+\to \pi^+ \nu\bar\nu) = \kappa_+ \left [ \left ( \frac{{\rm Im} X_{\rm eff} }{\lambda^5}
  \right )^2 + \left ( \frac{{\rm Re} X_{\rm eff} }{\lambda^5} 
  - P_c(X)  \right )^2 \right ] \, , \\
\label{eq:BRSMKL} \mathcal{B}( K_L \to \pi^0 \nu\bar\nu) = \kappa_L \left ( \frac{{\rm Im} 
    X_{\rm eff} }{\lambda^5} \right )^2 \, ,
\end{gather}
where \cite{Mescia:2007kn}
\begin{equation}\label{kapp}
\kappa_+=(5.36\pm0.026)\cdot 10^{-11}\,, \quad \kappa_{\rm L}=(2.31\pm0.01)\cdot 10^{-10}
\ee
and \cite{Buras:2005gr,Buras:2006gb,Brod:2008ss,Isidori:2005xm,Mescia:2007kn}.
\be
P_c(X)=0.42\pm0.03.
\end{equation}
The short distance contributions are described by
\be\label{Xeff}
X_{\rm eff} = V_{ts}^* V_{td} (X_{L}(K) + X_{R}(K)),
\ee
where $X_{L,R}(K)$ are given in (\ref{XLK}) and (\ref{XRK}).
They include both
the SM contributions from $Z$-penguin and box diagrams and the tree-level 
$Z$ contributions calculated here.
 The numerical analysis of both decays is presented in Section \ref{sec:numerics}.

 Experimentally we have \cite{Artamonov:2008qb}
\be\label{EXP1}
\mathcal{B}(\kpn)_\text{exp}=(17.3^{+11.5}_{-10.5})\cdot 10^{-11}\,,
\ee
and the $90\%$ C.L. upper bound \cite{Ahn:2009gb}
\be\label{EXP2}
\mathcal{B}(\klpn)_\text{exp}\le 2.6\cdot 10^{-8}\,.
\ee

In the SM one finds
\cite{Brod:2008ss,Brod:2010hi}
\be
\mathcal{B}(\kpn)_\text{SM} =(8.5\pm 0.7)\cdot 10^{-11}\,,
\ee
\be
\mathcal{B}(\klpn)_\text{SM} =(2.6\pm 0.4)\cdot 10^{-11}\,,
\ee
where the errors are dominated by CKM uncertainties. This should be compared
with the experimental values given in (\ref{EXP1}) and (\ref{EXP2}). Clearly we have to wait for improved data.

It should be emphasized that $\mathcal{B}(\klpn)_\text{SM}$ 
depends sensitively on CKM parameters, in particular not only on $\vcb$, as is also the case of 
$\mathcal{B}(\kpn)_\text{SM}$, but also on 
$\vub$ as it is a CP-violating observable. In our case the optimal value of $\vub$ will be $0.0037$ implying the central value 
for this branching ratio at $2.9\cdot 10^{-11}$, consistent with 
the value quoted above but higher. On the other hand our central value for 
$\mathcal{B}(\kpn)_\text{SM}$ agrees well with the one given in 
\cite{Brod:2010hi}.

\boldmath
\subsubsection{$K_L\to\pi^0\ell^+\ell^-$}
\unboldmath

The rare decays $K_L\to\pi^0e^+e^-$ and $K_L\to\pi^0\mu^+\mu^-$ are dominated
by CP-violating contributions. The indirect CP-violating
contributions are determined by the measured decays 
$K_S\to\pi^0 \ell^+\ell^-$ and the parameter $\varepsilon_K$ in 
a model independent manner. It is the dominant contribution within the SM 
where one finds
\cite{Mescia:2006jd}
\begin{gather}
\mathcal{B}(K_L\to\pi^0e^+e^-)_\text{SM}=
3.54^{+0.98}_{-0.85}\left(1.56^{+0.62}_{-0.49}\right)\cdot 10^{-11}\,,\label{eq:KLpee}\\
\mathcal{B}(K_L\to\pi^0\mu^+\mu^-)_\text{SM}= 1.41^{+0.28}_{-0.26}\left(0.95^{+0.22}_{-0.21}\right)\cdot
10^{-11}\label{eq:KLpmm}\,,
\end{gather}
with the values in parentheses corresponding to the destructive interference
between directly and indirectly CP-violating contributions. 
The last discussion  of the theoretical status of this interference
sign can be found in \cite{Prades:2007ud} where the results of \cite{Isidori:2004rb,Friot:2004yr,Bruno:1992za} are
critically analysed. From this discussion, constructive interference
seems to be  favoured though more work is necessary. In view of significant
uncertainties in the SM prediction we will mostly use these decays 
to test whether the correlations of them with $\klpn$ and $\kpn$ decays 
can have an impact on the latter. To this end we will confine our analysis 
to the case of the constructive interference between the directly and 
indirectly CP-violating contributions.

The present experimental bounds
\be
\mathcal{B}(K_L\to\pi^0e^+e^-)_\text{exp} <28\cdot10^{-11}\quad\text{\cite{AlaviHarati:2003mr}}\,,\qquad
\mathcal{B}(K_L\to\pi^0\mu^+\mu^-)_\text{exp} <38\cdot10^{-11}\quad\text{\cite{AlaviHarati:2000hs}}\,,
\ee
are still by one order of magnitude larger than the SM predictions, leaving 
thereby large room for NP contributions. 
In the LHT model the branching
ratios for both decays can be enhanced at most 
by a factor of 1.5 \cite{Blanke:2006eb,Blanke:2008ac}. Slightly larger 
effects are still allowed in RSc \cite{Blanke:2008yr}. Much larger effects 
have been found in general $Z'$ models \cite{Buras:2012jb}.
However our numerical analysis 
demonstrates that NP effects in these decays in the TUM are even smaller than 
in LHT.

In the LHT model, where only SM operators are present
the effects 
of NP can be compactly summarized by generalization of the 
real SM functions $Y_0(x_t)$ and $Z_0(x_t)$ to two complex functions $Y_K$ and 
$Z_K$, respectively. As demonstrated in the context of the corresponding 
analysis within RSc  \cite{Blanke:2008yr}, also in the presence of RH 
currents two complex functions $Y_K$ and $Z_K$
are sufficient to describe jointly the SM and NP contributions.
Consequently the LHT formulae (8.1)--(8.8) of \cite{Blanke:2006eb} with 
$Y_K$ and $Z_K$ given below can be used to study these decays
in the context of tree-level $Z$ exchanges. Application of these 
formulae for general $Z$ and $Z'$ exchanges can be found in  
\cite{Buras:2012jb}.
The original papers behind these formulae can 
be found in 
\cite{Buchalla:2003sj,Isidori:2004rb,Friot:2004yr,Mescia:2006jd,Buras:1994qa}.

Using the formulae of \cite{Blanke:2008yr} we find
\be\label{YK}
Y_K=\eta_Y Y_0(x_t)+ \left[\frac{\Delta_{A}^{\mu\bar\mu}(Z)}{M_{Z}^2g^2_{\rm SM}}\right]
\frac{\Delta_{V}^{sd}(Z)}{ V_{ts}^\ast V_{td}},
\ee
\be
Z_K=Z_0(x_t)+\frac{1}{4\sin^2\theta_W}\left[\frac{2\Delta_{R}^{\mu\bar\mu}(Z)}{M_{Z'}^2g^2_{\rm SM}}\right]
\frac{\Delta_{V}^{sd}(Z)}{ V_{ts}^\ast V_{td}},
\ee
where $\Delta_V^{sd}$ is defined in (\ref{DeltasVA}). 

 \boldmath
\subsection{$K_L\to\mu^+\mu^-$}\label{sec:KLmumu}
\unboldmath
As discussed in  \cite{Blanke:2008yr} in models with tree-level gauge boson 
exchanges in this decay 
the real function $Y_0(x_t)$ is replaced by the complex
function 
\be\label{YAK}
Y_{\rm A}(K)= \eta _Y Y_0(x_t)
+\frac{\left[\Delta_A^{\mu\bar\mu}(Z)\right]}{M_{Z}^2g_\text{SM}^2}
\left[\frac{\Delta_L^{sd}(Z)-\Delta_R^{sd}(Z)}{V_{ts}^\star V_{td}}\right]\,
\equiv |Y_A(K)|e^{i\theta_Y^K}.
\ee

Only the so-called short distance (SD)
part to a dispersive contribution
to $K_L\to\mu^+\mu^-$ can be reliably calculated. We have then 
following \cite{Buras:2004ub} 
($\lambda=0.2252$)
\be
\mathcal{B}(K_L\to\mu^+\mu^-)_{\rm SD} =
 2.03\cdot 10^{-9} \left[\bar P_c\left(Y_K\right)+
A^2 R_t\left|Y_A(K)\right|\cos\bar\beta_{Y}^K\right]^2\,,
\ee
where $R_t $ and $A$ are defined through $\vtd=\vus\vcb R_t$ and $\vcb=A\lambda^2$, respectively.  Moreover
\be
\bar\beta_{Y}^K \equiv \beta-\beta_s-\theta^K_Y\,,
\qquad
\bar P_c\left(Y_K\right) \equiv \left(1-\frac{\lambda^2}{2}\right)P_c\left(Y_K\right)\,,
\ee
with $P_c\left(Y_K\right)=0.113\pm 0.017$
\cite{Gorbahn:2006bm}. Here
$\beta$ and $\beta_s$ are the phases of $V_{td}$ and $V_{ts}$ defined in
(\ref{eq:3.40}).

The extraction of the short distance
part from the data is subject to considerable uncertainties. The most recent
estimate gives \cite{Isidori:2003ts}
\be\label{eq:KLmm-bound}
\mathcal{B}(K_L\to\mu^+\mu^-)_{\rm SD} \le 2.5 \cdot 10^{-9}\,,
\ee
to be compared with $(0.8\pm0.1)\cdot 10^{-9}$ in the SM 
\cite{Gorbahn:2006bm}.
The numerical results are discussed in Section~\ref{sec:numerics}.  In 
fact we will find that this decay plays a very significant role in our analysis 
as was already signaled by the rough estimates presented in 
 \cite{Buras:2011ph}.

\boldmath
\subsection{$B_{d,s} \to \mu^+ \mu^-$}
\unboldmath
We will next consider the important 
 decays $B_{d,s} \to \mu^+ \mu^-$, that suffer from helicity suppression in the 
SM. This suppression cannot be removed through the tree level exchange of $Z$ 
boson  but in principle could be removed through tree level 
exchanges of the Higgs boson. However the flavour conserving $H\mu\bar\mu$
vertex is proportional to the muon mass and in contrast to SUSY and general
two Higgs doublet models this suppression cannot be canceled by a large
$\tan\beta$ enhancement. 
Therefore {in what follows} we restrict our
attention to the contributions of the $Z^0$ boson both through SM 
penguin and box contributions and its generated tree-level exchanges,
calculated in Section \ref{sec:bqll}.

Following \cite{Blanke:2008yr} and assuming that the CKM parameters 
have been determined independently 
of NP and are universal we find 
\be\label{GB/SM}
\frac{\mathcal{B}(B_q\to\mu^+\mu^-)}{\mathcal{B}(B_q\to\mu^+\mu^-)^{\rm SM}}
=\left|\frac{Y_A(B_q)}{\eta_Y Y_0(x_t)}\right|^2,
\ee
where $Y_A(B_q)$ is given by
\be\label{YAB}
Y_{\rm A}(B_q)= \eta_Y Y_0(x_t)
+\frac{\left[\Delta_A^{\mu\bar\mu}(Z)\right]}{M_{Z}^2g_\text{SM}^2}
\left[\frac{\Delta_L^{qb}(Z)-\Delta_R^{qb}(Z)}{V_{tq}^\star V_{tb}}\right]\,
\equiv |Y_A(B_q)|e^{i\theta_Y^{B_q}}.
\ee

As stressed in 
\cite{DescotesGenon:2011pb,deBruyn:2012wj,deBruyn:2012wk} \footnote{We follow here presentation and notations of \cite{deBruyn:2012wj,deBruyn:2012wk}.},
when comparing
the theoretical branching 
ratio $\mathcal{B}(B_s\to\mu^+\mu^-)$ with experimental data quoted by LHCb, ATLAS and CMS,
a correction factor has to be included which takes care of $\Delta\Gamma_s$
effects
that influence the extraction of this branching ratio from the data:
\be
\label{Fleischer1}
\mathcal{B}(B_{s}\to\mu^+\mu^-)_{\rm th} =
r(y_s)~\mathcal{B}(B_{s}\to\mu^+\mu^-)_{\rm exp}, \quad r(0)=1.
\ee
Here
\be
r(y_s)\equiv\frac{1-y_s^2}{1+\mathcal{A}^\lambda_{\Delta\Gamma} y_s}
\approx 1 - \mathcal{A}^\lambda_{\Delta\Gamma} y_s
\ee
with
\be
y_s\equiv\tau_{B_s}\frac{\Delta\Gamma_s}{2}=0.088\pm0.014.
\ee
The quantity $\mathcal{A}^\lambda_{\Delta\Gamma}$ is discussed below.

The branching ratios $\mathcal{B}(B_q\to\mu^+\mu^-)$ are only sensitive to
the absolute value of $Y_A(B_q)$. However,
as pointed out in \cite{deBruyn:2012wj,deBruyn:2012wk}  in
the flavour precision era these decays could allow to get also some information
on the phase of $Y_A(B_q)$. 
The authors of \cite{deBruyn:2012wk,Fleischer:2012fy} provide general expressions for
$\mathcal{A}^\lambda_{\Delta\Gamma}$ and
$S_{\mu^+\mu^-}^s$ as functions of Wilson coefficients involved. Using
these formulae we find in our model very simple formulae that
reflect the fact that $Z^0$ and not scalar operators dominate NP 
contributions:
\be\label{Smumus}
\mathcal{A}^\lambda_{\Delta\Gamma}=\cos (2\theta^{B_s}_Y-2\varphi_{B_s}), \quad
S_{\mu^+\mu^-}^s=\sin (2\theta^{B_s}_Y-2\varphi_{B_s}).
\ee
Both $\mathcal{A}^\lambda_{\Delta\Gamma}$ and
$S_{\mu^+\mu^-}^s$ are theoretically clean observables. 

While $\Delta\Gamma_d$ is very small and $y_d$ can be set to zero,
in the case of $B_d\to\mu^+\mu^-$ one can still consider the CP asymmetry
$S_{\mu^+\mu^-}^d$ \cite{Fleischer:2012fy}, for which we simply find
\be\label{Smumud}
S_{\mu^+\mu^-}^d=\sin(2\theta^{B_d}_Y-2\varphi_{B_d}).
\ee

As demonstrated in Section~\ref{sec:numerics} the TUM model considered by us turns out to give values for
$\mathcal{B}(B_{s,d}\to\mu^+\mu^-)$ that uniquely 
differ from SM predictions. However, similarly to 
 the SM and CMFV models we find in TUM to a very good accuracy
\be\label{ADG}
\mathcal{A}^\lambda_{\Delta\Gamma}=1, \quad S_{\mu^+\mu^-}^s=0,
\quad r(y_s)=0.912\pm0.014
\ee
basically independent of NP parameters considered.
These are definite predictions of TUM which will be tested one day.

In our numerical analysis in Section~\ref{sec:numerics} we will discuss then only the branching ratios 
for these decays. Now, the most recent  results from LHCb read \cite{Aaij:2012ac,LHCbBsmumu}
\be\label{LHCb2}
\mathcal{B}(B_{s}\to\mu^+\mu^-) = (3.2^{+1.5}_{-1.2}) \times 10^{-9}, \quad
\mathcal{B}(B_{s}\to\mu^+\mu^-)^{\rm SM}=(3.23\pm0.27)\times 10^{-9},
\ee
\be\label{LHCb3}
\mathcal{B}(B_{d}\to\mu^+\mu^-) \le  9.4\times 10^{-10}, \quad
\mathcal{B}(B_{d}\to\mu^+\mu^-)^{\rm SM}=(1.07\pm0.10)\times 10^{-10}.
\ee
We have shown also SM predictions for these
observables \cite{Buras:2012ru} that do not include the correction
${r(y_s)}$. If this factor is included one finds \cite{deBruyn:2012wj,deBruyn:2012wk}
\be
\label{FleischerSM}
\mathcal{B}(B_{s}\to\mu^+\mu^-)^{\rm SM}_{\rm corr}= (3.5\pm0.3)\cdot 10^{-9}.
\ee
It is this branching that should be compared in such a case
with the results of LHCb given above. 
For the latest discussions of these issues see 
\cite{deBruyn:2012wj,deBruyn:2012wk,Buras:2012ru,Fleischer:2012fy}.

As in TUM $\mathcal{A}^\lambda_{\Delta\Gamma}=1$  independently of NP parameters 
we will include the correction in question
in the
experimental branching ratio using the values in (\ref{ADG}).
If this is done the experimental results in (\ref{LHCb2}) is reduced by $9\%$
 and 
we find 
\be\label{LHCb2corr}
\mathcal{B}(B_{s}\to\mu^+\mu^-)_{\rm corr} =(2.9^{+1.4}_{-1.1}) \times 10^{-9}, 
\ee
that should be compared with the SM result in (\ref{LHCb2}). While the central 
theoretical value agrees very well with experiment, the large experimental 
error still allows for  NP contributions. In our plots we will show 
the result in~(\ref{LHCb2corr}).

This completes the analytic analysis of the $B_{s,d} \to \mu^+ \mu^-$ decays. 
The numerical results are discussed in Section~\ref{sec:numerics}.

\subsection{\boldmath $B \to \{X_s,K, K^*\} \nu\bar \nu$}
Following the analysis of \cite{Altmannshofer:2009ma}, 
the branching ratios of the $B \to \{X_s,K, K^*\}\nu\bar \nu$  
modes in the presence of RH currents can be written as follows
 \bea
 \mathcal{B}(B\to K \nu \bar \nu) &=& 
 \mathcal{B}(B\to K \nu \bar \nu)_{\rm SM} \times\left[1 -2\eta \right] \epsilon^2~, \label{eq:BKnn}\\
 \mathcal{B}(B\to K^* \nu \bar \nu) &=& 
 \mathcal{B}(B\to K^* \nu \bar \nu)_{\rm SM}\times\left[1 +1.31\eta \right] \epsilon^2~, \\
 \mathcal{B}(B\to X_s \nu \bar \nu) &=& 
 \mathcal{B}(B\to X_s \nu \bar \nu)_{\rm SM} \times\left[1 + 0.09\eta \right] \epsilon^2~,\label{eq:Xsnn}
 \eea
 where 
\be\label{etaepsilon}
 \epsilon^2 = \frac{ |X_{\rm L}(B_s)|^2 + |X_{\rm R}(B_s)|^2 }{
 |\eta_X X_0(x_t)|^2 }~,  \qquad
 \eta = \frac{ - {\rm Re} \left( X_{\rm L}(B_s) X_{\rm R}^*(B_s)\right) }
{ |X_{\rm L}(B_s)|^2 + |X_{\rm R}(B_s)|^2 }~,
 \ee
with $X_{L,R}(B_s)$ defined in (\ref{XLRB}). The issue of long-distance 
contributions to these short-distance formulae is discussed in 
\cite{Kamenik:2009kc}.

The predictions for the SM branching  ratios 
are~\cite{Bartsch:2009qp,Kamenik:2009kc,Altmannshofer:2009ma}
\bea
\mathcal{B}(B\to K \nu \bar \nu)_{\rm SM}   &=& (3.64 \pm 0.47)\times 10^{-6}~, \no \\
\mathcal{B}(B\to K^* \nu \bar \nu)_{\rm SM} &=& (7.2 \pm 1.1)\times 10^{-6}~, \no \\
\mathcal{B}(B\to X_s \nu \bar \nu)_{\rm SM} &=& (2.7 \pm 0.2)\times 10^{-5}~, 
\label{eq:BKnnSM}
\eea
are respectively by factors of four, eleven and twenty below
the experimental bounds~\cite{Barate:2000rc,:2007zk,:2008fr}.

We would like already announce at this point that the TUM model satisfies easily the present experimental bounds and makes 
definite predictions for $\epsilon$ and $\eta$ defined in 
(\ref{etaepsilon}):
\be
\epsilon > 1, \qquad \eta\approx 0.
\ee
The last result is the consequence of the strong suppression of right-handed contributions
when all constraints are taken into account. More details will be given in 
Section~\ref{sec:numerics}.

\boldmath
\subsection{$B^+\to \tau^+\nu$}\label{btaunu}                   
\unboldmath
\subsubsection{Standard Model Results}
We now look at the tree-level decay $B^+ \to \tau^+ \nu$ which in the SM is 
mediated by the $W^\pm$  exchange with the resulting branching ratio  given by
\begin{equation} \label{eq:Btaunu}
\mathcal{B}(B^+ \to \tau^+ \nu)_{\rm SM} = \frac{G_F^2 m_{B^+} m_\tau^2}{8\pi} \left(1-\frac{m_\tau^2}{m^2_{B^+}} \right)^2 F_{B^+}^2
|V_{ub}|^2 \tau_{B^+}~.
\end{equation}
Evidently this result is subject to significant parametric uncertainties induced in (\ref{eq:Btaunu}) by $F_{B^+}$ and $V_{ub}$. However, it is expected 
that these uncertainties will be eliminated in this decade and 
a precise prediction will be possible. Anticipating this we will present 
the results for fixed values of these parameters.

In the literature
in order to find the SM prediction for this branching ratio one eliminates 
these uncertainties by using $\Delta M_d$,  $\Delta M_d/\Delta M_s$ and 
$S_{\psi K_S}$ \cite{Bona:2009cj,Altmannshofer:2009ne} and taking experimental 
values for these three quantities. This strategy has a weak point as 
 the experimental 
values of $\Delta M_{d,s}$ used 
in this strategy may not be the one corresponding to the true value of 
the SM. However, proceeding in this manner one finds 
\cite{Altmannshofer:2009ne}
\begin{equation}\label{eq:BtaunuSM1}
\mathcal{B}(B^+ \to \tau^+ \nu)_{\rm SM}= (0.80 \pm 0.12)\times 10^{-4},
\end{equation}
with a similar result obtained 
by the UTfit collaboration
\cite{Bona:2009cj} and CKM-fitters.

For quite some time this result was by a factor of two below the data from 
Belle and BaBar.
However this disagreement
of the data with the SM softened significantly with the new
data from Belle Collaboration \cite{BelleICHEP}. The
new world average provided by the UTfit collaboration \cite{Tarantino:2012mq}
\be\label{Belle}
\mathcal{B}(B^+ \to \tau^+ \nu)_{\rm exp} = (0.99 \pm 0.25) \times 10^{-4}~
\ee
is in perfect agreement with the SM, even if NP providing a slight enhancement 
of this branching ratio is presently favoured.

The full clarification of the left room for NP in this decay
will have to wait for the
data from Super-B machine at KEK. In the meantime hopefully 
improved values for $F_{B^+}$ from lattice 
and $\vub$ from tree level decays will allow us to make a precise prediction 
for this decay without using the experimental value for $\Delta M_d$. In 
TUM it will turn out that the favoured value of $\vub=0.0037$ in this model
implies  through  (\ref{eq:Btaunu}) central value for $\mathcal{B}(B^+ \to \tau^+ \nu)\approx 0.88 \times
10^{-4}$, that is very close to the data and the question arises whether modification of $W^\pm$ couplings in TUM still allows to keep this
agreement.

\subsubsection{Effect of Modified \boldmath{$W^\pm$} Couplings}
In the presence of modified $W^\pm$ couplings we find 
\begin{equation} \label{eq:Btaunu-general}
\mathcal{B}(B^+ \to \tau^+ \nu) = 
\frac{1}{64\pi M_W^4}m_{B^+} m_\tau^2 \left(1-\frac{m_\tau^2}{m^2_{B^+}} \right)^2 F_{B^+}^2
\tau_{B^+}
\left|\Delta_L^{\nu\tau}(\Delta_L^{ub\star}-\Delta_R^{ub\star})\right|^2,
\end{equation}
with
\begin{align}
 &\Delta_L^{ub}=-\frac{g}{\sqrt{2}}V_{ub} + \Delta_L^{13}(W)\,,\qquad \Delta_R^{ub} = \Delta_R^{13}(W)\,,\qquad
\Delta_L^{\nu\tau}=-\frac{g}{\sqrt{2}}\,
\end{align}
with 
$\Delta_{L,R}^{13}(W)$ given in Section~\ref{firstdeltas}.
For $\Delta_L^{13}=\Delta_R^{13}=0$ this expression reduces to the SM expression in (\ref{eq:Btaunu}). Using our estimates of the corrections to $W$ couplings 
in subsection~\ref{Wcouplings}
we find 
\be
\sqrt{2}\left|\frac{\Delta_L^{13}-\Delta_R^{13}}{g V_{ub}}\right|\le 10^{-2}
\ee
and consequently NP corrections to $\mathcal{B}(B^+ \to \tau^+ \nu)$ can 
be safely neglected.

\boldmath
\section{Basic Structure of New Physics Contributions in the TUM}\label{sec:anatomy}
\unboldmath
\setcounter{equation}{0}
\subsection{Preliminaries}
The spirit of our analysis presented in this paper differs significantly 
from the one of our first paper \cite{Buras:2011ph} and many papers 
found in the literature in which the main goal is to find out whether 
a given model is roughly consistent with the present experimental bounds. 
In doing this quite often one imposes the constraint that NP contributions 
are at most as large as the SM contributions. 

In view of increased precision in experimental data and in the theory to be 
expected in this decade such a passive approach to the tests of new model 
constructions is not satisfactory. Below we will follow a more aggressive 
approach by identifying correlations between various observables predicted 
by the TUM and proposing various tests of this scenario.

In many NP extensions of the SM, like LHT model, RS scenarios of various sort 
and supersymmetric models the identification of specific correlations between 
various observables is challenged by the multitude of new free parameters 
present in these models. However,
in the case of TUM  we are in a comfortable situation that our model does not 
have many free parameters to describe FCNC processes as several of 
its fundamental parameters have been already used to describe
successfully  the spectrum of 
quark masses and the values of the CKM parameters. In fact as we already 
discussed previously once the common mass of the vectorial heavy fermions 
has been fixed, only three real and positive definite parameters are left 
to our disposal and NP contributions to all FCNC processes in the down quark 
sector are entirely given in terms of these parameters and the parameters 
of the SM. Very important is the fact that the sole CP-violating phase in 
the TUM equals the KM phase and is equal to the angle $\gamma$ in the unitarity 
triangle.

\boldmath
\subsection{Facing the Anomalies in $\Delta F=2$ Data}\label{Anomalies}
\unboldmath

Before entering the details let us first ask the question how TUM faces 
the anomalies in the data for $\Delta F=2$ observables 
identified first in 
\cite{Lunghi:2008aa,Buras:2008nn}. Indeed 
the SM does not offer a fully satisfactory 
description of these data. Here the prominent role is played by the
$\epsilon_K-S_{\psi K_S}$ tension within the SM. 
In this context it should be emphasized that because of this 
tension the pattern of deviations from SM expectations for other 
observables depends often on 
whether $\epsilon_K$ or $S_{\psi K_S}$ is used as a basic observable to 
fit the CKM parameters. As both observables can in principle receive important 
contributions from NP, none of them is optimal for this goal. The solution 
to this problem will be  solved one day by measuring the CKM parameters 
with the help 
of tree-level decays. Unfortunately, the tension between the inclusive 
and exclusive determinations of $\vub$ and the poor knowledge of the angle 
$\gamma$ 
from tree-level decays preclude this solution at present. However a good 
agreement of the SM value for the ratio $\Delta M_s/\Delta M_d$ with the data 
and the fact that the present determinations of $\gamma$ are in the ballpark of 
$70^\circ$,  imply two basic scenarios for the three observables 
$\vub$, $\epsilon_K$ and $S_{\psi K_S}$. Moreover for fixed $\gamma$ and 
$\Delta M_s/\Delta M_d$  the latter three 
observables are strongly correlated within the SM with each other. We have:
 \begin{itemize}
\item
{\bf Exclusive (small) $\vub$ Scenario 1:}
$|\varepsilon_K|$ is smaller than its experimental determination,
while $S_{\psi K_S}$ is very close to the central experimental value.
\item
{\bf Inclusive (large) $\vub$ Scenario 2:}
$|\varepsilon_K|$ is consistent with its experimental determination,
while $S_{\psi K_S}$ is significantly higher than its  experimental value.
\end{itemize}

Thus dependently which scenario is considered we need either  
{\it constructive} NP contributions to $|\varepsilon_K|$ (Scenario 1) 
or {\it destructive} NP contributions to  $S_{\psi K_S}$ (Scenario 2). 
However this  NP should not spoil the agreement with the data 
for $S_{\psi K_S}$ (Scenario 1) and for $|\varepsilon_K|$ (Scenario 2).

While introducing these two scenarios, one should emphasize the following difference between them. 
In Scenario 1, the central value of $|\varepsilon_K|$ is visibly smaller than 
the very precise data  but the still  significant parametric uncertainty 
due to $\vcb^4$ dependence in $|\varepsilon_K|$ and a large uncertainty 
in the charm contribution found at the NNLO level in \cite{Brod:2011ty} 
does not make this problem as pronounced as this is the case of 
Scenario 2, where large $\vub$ implies definitely a value of $S_{\psi K_S}$ 
that is by $3\sigma$ above the data.

Now models with many new parameters can face successfully both scenarios 
removing the deviations from the data for certain range of their parameters 
but in simpler models often 
only one scenario can be admitted. For instance in models with constrained 
MFV there are no NP contributions to  $S_{\psi K_S}$ but NP contributions to 
$\varepsilon_K$ are possible so that Scenario 1 is selected in that framework. 
On the other hand in ${\rm 2HDM_{\overline{MFV}}}$ \cite{Buras:2010mh}  there are no relevant 
NP contributions to $\varepsilon_K$ but the presence of flavour blind phases 
can have a significant impact on $S_{\psi K_S}$ so that Scenario 2 is selected.
On the other hand in models with $U(2)^3$ flavour symmetry both 
scenarios can be accommodated implying an interesting triple correlation 
between the values of $S_{\psi K_S}$, $S_{\psi\phi}$ and $\vub$ \cite{Buras:2012sd}.

As we will demonstrate in the next subsection first at a semi-quantitative 
level, the TUM  cannot improve the description of  $\Delta F=2$ data relative 
to the SM for the following reasons:
\begin{itemize}
\item
NP contributions to $B_{s,d}^0-\bar B_{s,d}^0$ mixings are very small so that
\be
S_{\psi K_S}, \qquad S_{\psi\phi},\qquad \Delta M_d, \qquad \Delta M_s
\ee
remain SM-like, although the predicted value of $S_{\psi K_S}$ depends 
sensitively on the chosen value of $\vub$. We simply find that within 
$1\%$ accuracy
\be
C_{B_d}=C_{B_s}=1, \qquad \varphi_{B_d}=\varphi_{B_s}=0.
\ee
\item
NP effects in $\varepsilon_K$ can in principle be large so that at first 
sight one could choose exclusive scenario for $\vub$ and enhance $\varepsilon_K$
through NP contributions. However in TUM the LR contributions uniquely 
suppress $\varepsilon_K$ relatively to its SM value and have to be compensated 
by LL  contributions. But the increase of these contributions is bounded 
from above by the upper bound on $K_L\to\mu^+\mu^-$ and we find that in 
order not to violate this bound LL contributions can just compensate LR 
contributions so that $\varepsilon_K$ is very close to its SM value.
\end{itemize}

In view of this situation, within the  TUM 
the optimal choice for $\vub$ is between the 
Scenarios 1 and 2. Inspecting then the dependence on $\gamma$ we are then 
lead to the following optimal central values for the four CKM parameters that 
we will use in our analysis:
\be\label{fixed}
\vus=0.2252, \qquad \vcb=0.0406, \qquad \vub=0.0037,\qquad \gamma=68^\circ, 
\ee
where the values for
 $|V_{us}|$ and  $|V_{cb}|$ have been measured
 in tree level
decays. Moreover the value for $\gamma$ is consistent with the
ratio $\Delta M_d/\Delta M_s$ in the model considered: 
\be\label{Ratio}
\left(\frac{\Delta M_s}{\Delta M_d}\right)_{\rm TUM}=
\left(\frac{\Delta M_s}{\Delta M_d}\right)_{\rm SM}= 34.5\pm 3.0\qquad {\rm exp:~~ 35.0\pm 0.3}.
\ee

In Table~\ref{tab:SMpred} we 
summarize for completeness the results for various observables in the TUM model 
that for the input in (\ref{fixed}) and in Table~\ref{tab:input}.
are the same as in the SM.

\begin{table}[!tb]
\centering
\begin{tabular}{|c||c|c|}
\hline
 & TUM   & Experiment\\
\hline
\hline
  \parbox[0pt][1.6em][c]{0cm}{} $|\varepsilon_K|$ & $2.00(22)  \cdot 10^{-3}$  &     $ 2.228(11)\times 10^{-3}$ \\
 \parbox[0pt][1.6em][c]{0cm}{}$S_{\psi K_S}$& 0.725(25)  & $0.679(20)$\\
\parbox[0pt][1.6em][c]{0cm}{}$S_{\psi \phi}$& 0.039  & $0.002(9)$\\
 \parbox[0pt][1.6em][c]{0cm}{}$\Delta M_s\, [\text{ps}^{-1}]$ &19.0(21) &$17.73(5)$ \\
 \parbox[0pt][1.6em][c]{0cm}{} $\Delta M_d\, [\text{ps}^{-1}]$ &0.55(6)  &  $0.507(4)$\\
\parbox[0pt][1.6em][c]{0cm}{}$\mathcal{B}(B^+\to \tau^+\nu_\tau)$&  $0.88(14) \cdot 10^{-4}$ & $0.99(25) \times
10^{-4}$\\
\hline
\end{tabular}
\caption{\it TUM predictions for various observables for  $|V_{ub}|=3.7\cdot 10^{-3}$  and $\gamma =
68^\circ$ compared to experiment. 
}\label{tab:SMpred}~\\[-2mm]\hrule
\end{table}

We note that:
\begin{itemize}
\item
$|\varepsilon_K|$ is lower than the central experimental value but still 
within $1\sigma$ when hadronic and parametric uncertainties are taken into account. 
\item
$S_{\psi K_S}$ is by $2\sigma$ larger than its central experimental value.
\item
$S_{\psi\phi}$ is very close to the experimental value.
\item
$\Delta M_s$  and $\Delta M_d$, although 
slightly above the data,  are both in  good agreement with the latter 
when hadronic uncertainties are taken into account. In particular their 
ratio is in an excellent agreement with data.
\item
Choosing higher value of $\gamma$ would bring $|\varepsilon_K|$ closer 
to the data. But then also $\Delta M_d$ would be larger implying worse 
agreement with the data for $\Delta M_d$ and $\Delta M_s/\Delta M_d$.
\end{itemize}
These results depend on the lattice input and in the case 
of $\Delta M_d$ on the value of $\gamma$. Therefore to get a better insight 
both lattice input and the tree level determination of $\gamma$ 
have to improve.

Thus the description of $\Delta F=2$ observables in TUM, similarly to the SM,
is not fully satisfactory but sufficiently good that we can continue our 
analysis and investigate  
the predictions in this model for
\be
B_{s,d}\to\mu^+\mu^-,\quad 
B\to K^*(K) \ell^+\ell^-,\quad B\to X_s\ell^⁺\ell^-
\ee
\be
B\to X_s\nu\bar\nu,\quad, B\to K^*\nu\bar\nu, \quad B\to K\nu\bar\nu,
\ee
\be
\kpn, \quad \klpn, \quad K_L\to\mu^+\mu^-, \quad K_L\to\pi^0\ell^+\ell^-.
\ee

{
Now, it is known from \cite{Buras:2012jb} that in the case of tree-level 
FCNCs mediated by $Z^0$ generally NP effects in $\Delta F=1$ processes 
are larger than in $\Delta F=2$ transitions, while the opposite is true 
for $Z'$ tree-level exchanges with $M_{Z'}$ of order few $\tev$. 
In the concrete 
model considered here there is an additional suppression factor that makes NP
 effects in $\Delta F=2$ transitions to be much smaller than in $\Delta F=1$ 
transitions.
  As these properties are very 
important for our analysis we explain them  here in explicit terms:
\begin{itemize}
\item
Considering first as in \cite{Buras:2012jb} a
general tree-level neutral gauge boson ($A$) contributions to FCNC processes,
a tree-level $A$ contribution to $\Delta F=2$
observables depends quadratically on $\Delta_{L,R}^{ij}(A)/M_{A}$, where
   $\Delta_{L,R}^{ij}(A)$ are flavour-violating couplings with $i,j$ denoting
quark flavours. For any high value of $M_{A}$, even beyond the reach of the
LHC, it is possible to find couplings  $\Delta_{L,R}^{ij}(A)$  which are not
only  consistent with the existing data but    can even remove certain
tensions found within the SM in $\Delta F=2$ processes. The larger $M_{A}$, the larger
$\Delta_{L,R}^{ij}(A)$  are allowed: $\Delta_{L,R}^{ij}(A)\approx a_{ij}M_{A}$
with $a_{ij}$ adjusted to agree with $\Delta F=2$ data. Once
$\Delta_{L,R}^{ij}(A)$ are fixed in this manner, they can be used to
predict $A$ effects in $\Delta F=1$ observables. However here
NP contributions to the amplitudes are proportional to
$\Delta_{L,R}^{ij}(A)/M^2_{A}$ and with the couplings proportional to $M_{A}$, $A$ contributions to $\Delta F=1$ observables increase
with decreasing $M_{A}$ without changing $\Delta F=2$ 
transitions. Eventually, for sufficiently low values of $M_A$ the bounds 
on $\Delta F=1$ processes become stronger than the constraints on 
 $\Delta F=2$ processes requiring the coefficients $a_{ij}$ to be smaller than 
obtained from  $\Delta F=2$ constraints alone. Therefore in turn for these low masses 
of $M_A$, as is the case of $M_Z$, even if NP effects 
mediated by $Z^0$ in $\Delta F=2$ transitions are small,  
significant effects 
in rare $K$ and $B$ decays can be found.  In this manner
flavour-violating $Z$ couplings turn out to be 
an important portal to NP, in our case the physics of vectorial fermions.
\item
In the specific model considered by us this disparity between $\Delta F=2$ 
and $\Delta F=1$ transitions is enhanced by 
the fact that the coefficients $a_{ij}$ in the couplings $\Delta_{L,R}^{ij}(Z)$ 
are proportional to $v^2/M^2$ and consequently strongly suppressed. As $\Delta F=2$ amplitudes are quadratic 
in these couplings and the interference between SM and NP contributions 
dominating NP effects in rare decay branching ratios is linear in them, 
the disparity in question is in our model much larger than could be expected 
on the basis of the general analysis in \cite{Buras:2012jb}.
\end{itemize}
}

Now, even if NP contributions to $B_{s,d}^0-\bar B_{s,d}^0$ mixings 
are found to be 
very small, the structure 
of NP contributions to $\varepsilon_K$ forces flavour violating 
$\Delta^{sd}_L(Z)$
couplings to be sufficiently large in order to 
compensate the negative contributions of LR operators. But as $\Delta^{sd}_L(Z)$
 couplings 
are bounded from above by $K_L\to\mu^+\mu^-$, this compensation is only 
possible if RH couplings are suppressed. Recall that hadronic matrix elements of LR operators are in the
$K$ system not only enhanced through renormalization group effects but 
also receive a large chiral enhancement. 

Thus at the end NP contributions to rare $K$ decays are governed by 
LH couplings of $Z^0$ to quarks with RH couplings being subleading. Due 
to small number of free parameters in TUM this structure is transferred to 
rare $B$ decays. At the end the predictions of TUM for rare $K$ and $B$ 
decays are forced to differ from SM predictions and the deviations from 
the SM take place in a correlated manner. In order to have a better insight 
in this structure we will now derive approximate expressions for various 
amplitudes and master functions governing various 
observables so that the pattern just discussed will be seen in explicit terms.

\boldmath
\subsection{Structure of $Z^0$ Couplings}
\unboldmath
The starting point  of our discussion are suitable approximations for the $\tilde{A}_{ij}^D$, which are sums of three different terms
(summation over $k=1,2,3$) 
\begin{align}
\tilde{A}_{Lij}^D & = \frac{v^2}{M^2} V^*_{ki} V_{kj} \epsilon^{Q2}_k &\tilde{A}_{Rij}^D & = \frac{m_i m_j}{M^2} \frac{V^*_{ki} V_{kj}
}{\epsilon^{Q2}_k}. 
\end{align}
We find
\begin{align}
 \tilde {A}_{L21}^D  & \approx \frac{v^2}{M^2} \left(  \eps^{Q2}_3  \lambda_t  + \eps^{Q2}_2 \, \re \lambda_c \  \right)  & \tilde
{A}_{R21}^D & \approx \frac{m_d m_s}{M^2} \left( \frac{\re \lambda_u}{\eps^{Q2}_1} + i \frac{\im \lambda_c}{\eps^{Q2}_2}   \right) \\
 \tilde {A}_{L31}^D  & \approx \frac{v^2}{M^2} \epsilon^{Q2}_3 \lambda_t^d & \tilde {A}_{R31}^D  & \approx  \frac{m_b m_d}{M^2} \left(
\frac{\lambda_u^d}{\epsilon^{Q2}_1} +  \frac{\re \lambda_c^d}{\epsilon^{Q2}_2} \right)
\\
 \tilde {A}_{L32}^D  & \approx \frac{v^2}{M^2} \epsilon^{Q2}_3 \lambda_t^s & \tilde {A}_{R32}^D  & \approx  \frac{m_s m_b}{M^2} \left(
\frac{ \lambda_u^s}{\epsilon^{Q2}_1} + \frac{\re \lambda_c^s}{\epsilon^{Q2}_2}  \right),
\end{align}
where $\lambda_i$ and $\lambda_i^{q}$ are defined in (\ref{CKMV}). 

 Already at this stage we observe that the LH couplings in $B^0_d$ and 
    $B_s^0$ systems are of MFV type, while in the $K$ system this 
    property is broken by the term involving $\lambda_c$. The RH couplings 
    are suppressed through quark masses but in order to estimate their size
    $X_{ij}$ have to be constrained through FCNC processes as done below.

Using then (\ref{Xij}) we can express these couplings entirely in terms of 
$\epsilon_3^Q$, $X_{13}$ and $X_{23}$, quark masses and the elements of the 
CKM matrix. We find
\begin{align}
 \tilde {A}_{L21}^D  & \approx \frac{v^2}{M^2} \eps^{Q2}_3  \left(   \lambda_t  + X_{23}^2 |V_{cb}|^2 \, \re \lambda_c \  \right)  & \tilde
{A}_{R21}^D  & \approx \frac{m_d m_s}{M^2} \frac{1}{\eps^{Q2}_3} \left( \frac{\re \lambda_u}{X_{13}^2 |V_{ub}|^2} + i \frac{\im
\lambda_c}{X_{23}^2 |V_{cb}|^2}   \right) \\
 \tilde {A}_{L31}^D  & \approx \frac{v^2}{M^2} \epsilon^{Q2}_3 \lambda_t^d & \tilde {A}_{R31}^D  & \approx  \frac{m_b m_d}{M^2}
\frac{1}{\eps^{Q2}_3} \left( \frac{\lambda_u^d}{X_{13}^2 |V_{ub}|^2} +  \frac{\re \lambda_c^d}{X_{23}^2 |V_{cb}|^2} \right)
\\
 \tilde {A}_{L32}^D  & \approx \frac{v^2}{M^2} \epsilon^{Q2}_3 \lambda_t^s & \tilde {A}_{R32}^D  & \approx  \frac{m_s m_b}{M^2}
\frac{1}{\eps^{Q2}_3} \left( \frac{ \lambda_u^s}{X_{13}^2 |V_{ub}|^2} + \frac{\re \lambda_c^s}{X_{23}^2 |V_{cb}|^2}  \right)
\end{align}

We recall that the masses of quarks entering these expressions are evaluated at 
the high scale $M$ and for $M=3\tev$ take the values
\be\label{qmasses}
m_d(M)= 2.3\mev,\qquad  m_s(M)= 45\mev, \qquad m_b(M)=2.4\gev~.
\ee

\boldmath
\subsection{Structure of $\Delta F=2$ Amplitudes}
\unboldmath
Using these results we find the tree-level $Z$ contributions to mixing amplitudes $M_{12}^K$ and $M_{12}^q$. We list them in Appendix~\ref{app:DeltaF2}. Inserting in the latter formulae our nominal values of the CKM parameters in (\ref{fixed}) and quark masses in (\ref{qmasses}) we obtain the following approximate results for quantities of direct 
interest:
{\allowdisplaybreaks
\begin{align}
\frac{{\rm Im} \left(M_{12}^K\right)_\text{Z}} {{\rm Im} \left(M_{12}^K\right)_\text{SM}} & \approx  4.0
\left( \frac{3\TeV}{M}\right)^4 \left[\eps^{Q4}_3 (1+1.2~X_{23}^2) -0.18\frac{|P_1^{\rm LR}(K)|}{X_{13}^2} \right]10^{-3}\label{RR1} \\
\frac{{\rm Re} \left(M_{12}^K\right)_\text{Z}} {{\rm Re} \left(M_{12}^K\right)_\text{SM}} & \approx 3.0\left( \frac{3\TeV}{M}\right)^4 \left[ \eps^{Q4}_3 X_{23}^4 - 0.30|P_1^{\rm LR}(K)| \left(\frac{X_{23}^2}{X_{13}^2} + 0.86\frac{1}{X_{13}^2}\right) \right] 10^{-5} \label{RR2}\\
\re \frac{ \left(M_{12}^{B_d}\right)_\text{Z}} {\left(M_{12}^{B_d}\right)_\text{SM}} & \approx  2.6\left(\frac{ 3\TeV}{M} \right)^4   \,  \eps^{Q4}_3\label{RR3} 10^{-3} \\
\im \frac{ \left(M_{12}^{B_d}\right)_\text{Z}} {\left(M_{12}^{B_d}\right)_\text{SM}} & \approx - 3.0\left(\frac{3\TeV}{M} \right)^4 
\frac{|P_1^{\rm LR}(B_d)|}{X_{13}^2} 10^{-5}\label{RR4}\\
\re \frac{ \left(M_{12}^{B_s}\right)_\text{Z}} {\left(M_{12}^{B_s}\right)_\text{SM}} & \approx 2.5\left(\frac{3\TeV}{M} \right)^4 \left[\eps^{Q4}_3  -4.2\frac{|P_1^{\rm LR}(B_s)|}{X_{13}^2} 10^{-3} \right]10^{-3}\label{RR5} \\
\im \frac{ \left(M_{12}^{B_s}\right)_\text{Z}} {\left(M_{12}^{B_s}\right)_\text{SM}} & \approx 2.3\left(\frac{3\TeV}{M} \right)^4  
 \frac{|P_1^{\rm LR}(B_s)|}{X_{13}^2} 10^{-5}.\label{RR6}
\end{align} }%

For $M=3\tev$ we find  
\be
|P_1^{\rm LR}(K)|\approx 39, \qquad |P_1^{\rm LR}(B_d)|\approx 4.4, \qquad
|P_1^{\rm LR}(B_s)|\approx 4.5.
\ee
All $P_1^{\rm LR}$ are negative and the minus signs have been included in 
the formulae above.
As seen in Appendix~\ref{app:DeltaF2} the above expressions can 
allow us to estimate the size of 
new contributions to $\eps_K$, $\delta \Delta M_K$, $C_{B_q}$, $S_{\psi K_s}$ and
$S_{\psi \phi}$.

The last four results signal that NP effects in $B_q^0-\bar B_q^0$ systems 
are suppressed. However, in order to prove it we have to know the size of 
$X_{13}$. $\eps^{Q}_3$ is $\ord(1)$. To this end it is useful to look first at 
$\Delta F=1$ transitions.

\boldmath
\subsection{Structure of $\Delta F=1$ Amplitudes}
\unboldmath
Proceeding in the same manner we obtain approximate expressions for the relevant quantities in $\Delta F =1$ observables which we list in  Appendix~\ref{app:DeltaF1}. Inserting the nominal values of CKM parameters and quark masses we 
find
{\allowdisplaybreaks
\begin{align}
\Delta X_L(K) \equiv X_L(K) - X(x_t) & = 0.31 \left( \frac{3\TeV}{M} \right)^2 \eps^{Q2}_3 \left[ 1 + 1.1 X_{23}^2
e^{i(\beta-\beta_s)}\right] \label{RR7} \\
 X_R(K) & =  0.05 \left( \frac{3\TeV}{M} \right)^2 
\frac{e^{i(\beta-\beta_s)}}{\eps^{Q2}_3 X_{13}^2} \label{RR8}\\
 \Delta X_L(B_d) & = 0.31 \left( \frac{3\TeV}{M} \right)^2 \eps^{Q2}_3 \label{RR9}\\
 X_R(B_d) & =  - 1.7 \left( \frac{3\TeV}{M} \right)^2 \frac{1}{\eps^{Q2}_3} 
 \frac{1}{X_{13}^2} e^{-i(\beta+\gamma)}10^{-3} \label{RR10}\\
 \Delta X_L(B_s) & = 0.31 \left( \frac{3\TeV}{M} \right)^2 \eps^{Q2}_3 \label{RR11}\\
 X_R(B_s) & =   0.31   \left( \frac{3\TeV}{M} \right)^2\frac{1 }{\eps^{Q2}_3}\left[ \frac{5.5 e^{-i\gamma}}{X_{13}^2} + 
\frac{2.2}{X_{23}^2}\right]e^{-i\beta_s} 10^{-3}\label{RR12}
\end{align}}%
Moreover, we find
\be
\Delta Y_A(K)= \Delta X_L(K) - X_R(K), \qquad
 \Delta Y_d  = \Delta Y_s = \Delta X_L(B_d) = \Delta X_L(B_s)  
\ee
with other relations listed in Appendix~\ref{app:DeltaF1}. $\Delta Y_q$ are 
NP corrections to $Y_q$ in (\ref{equ:Yq}).

\boldmath
\subsection{The Interplay of $\Delta F=2$ and $\Delta F=1$ Transitions}
\unboldmath
With these formulae at hand we can now understand the pattern of NP effects 
that we have outlined in Subsection~\ref{Anomalies}. Indeed:
\begin{itemize}
\item
As seen in (\ref{RR1}) the last term representing LR contribution to $\varepsilon_K$ and being enhanced through $|P_1^{\rm LR}(K)|\approx 39$ suppresses 
$|\varepsilon_K|$ instead of enhancing it as required by the data. For 
$M\le 3\tev$ this contribution is a problem for TUM.
\item
The solution to this problem is a sufficiently small value of $1/X_{13}$ 
accompanied by sufficiently large values of  $\eps^{Q}_3$ and 
$X_{23}$. However $\eps^{Q}_3\le 1.$ 
\item
The required suppression of $1/X_{13}$  because of $\varepsilon_K$ 
and the small quark masses multiplying 
it in the expressions for corrections to the RH  master functions as given in 
in  Appendix~\ref{app:DeltaF1} imply that $\Delta F=1$ transitions are dominated by LH currents.
This is explicitly seen in (\ref{RR8}), (\ref{RR10}) 
and (\ref{RR12}). 
\item
On the other hand $\eps^{Q}_3$ and $X_{23}$ are bounded from above 
by $\mathcal{B}(K_L\to\mu^+\mu^-)$ as with the suppression of RH currents 
$\Delta Y_A(K)\approx \Delta X_L(K)$ and with increasing  
$\eps^{Q}_3$ and $X_{23}$ the upper bound in (\ref{eq:KLmm-bound}) is 
approached. However, as our numerical analysis in the next section shows, 
this bound still allows for values of $\eps^{Q}_3$ and $X_{23}$ necessary to compensate the LR contribution to $\varepsilon_K$ in (\ref{RR1}). But then the  net 
effect of NP in $\varepsilon_K$ is very small and $\vub$ has to be larger 
than its exclusive determinations in order for $\varepsilon_K$ to be 
within $1\sigma$ from the data.
\item
Remarkable are also results in (\ref{RR7}), (\ref{RR9}) and (\ref{RR11}) which 
imply that  all  rare decay $K$ and $B_q$ branching ratios considered by us are predicted to 
be enhanced over their SM values. Moreover, the enhancements of branching ratios for $B_s$ and $B_d$ 
rare decays satisfy CMFV relations. On the other hand the usual CMFV relation between rare $B_q$
decays and rare $K$ decays is broken by 
non-vanishing $X_{23}$ but this effect, as explained in the next 
section  is only present in $\kpn$.
\end{itemize}

 With this general view in mind we can now enter the numerical analysis.

\begin{table}[!tbh]
\center{\begin{tabular}{|l|l|}
\hline
$G_F = 1.16637(1)\times 10^{-5}\gev^{-2}$\hfill\cite{Nakamura:2010zzi} 	&  $m_{B_d}= 5279.5(3)\mev$\hfill\cite{Nakamura:2010zzi}\\
$M_W = 80.385(15) \gev$\hfill\cite{Nakamura:2010zzi}  								&	$m_{B_s} =
5366.3(6)\mev$\hfill\cite{Nakamura:2010zzi}\\
$\sin^2\theta_W = 0.23116(13)$\hfill\cite{Nakamura:2010zzi} 				& 	$F_{B_d} =
(190.6\pm4.6)\mev$\hfill\cite{Laiho:2009eu}\\
$\alpha(M_Z) = 1/127.9$\hfill\cite{Nakamura:2010zzi}									& 	$F_{B_s} =
(227.7\pm6.2)\mev$\hfill\cite{Laiho:2009eu}\\
$\alpha_s(M_Z)= 0.1184(7) $\hfill\cite{Nakamura:2010zzi}								&  $\hat B_{B_d} =
1.26(11)$\hfill\cite{Laiho:2009eu}\\\cline{1-1}
$m_u(2\gev)=(2.1\pm0.1)\mev $ 	\hfill\cite{Laiho:2009eu}						&  $\hat B_{B_s} =
1.33(6)$\hfill\cite{Laiho:2009eu}\\
$m_d(2\gev)=(4.73\pm0.12)\mev$	\hfill\cite{Laiho:2009eu}							& $\hat B_{B_s}/\hat B_{B_d}
= 1.05(7)$ \hfill \cite{Laiho:2009eu} \\
$m_s(2\gev)=(93.4\pm1.1) \mev$	\hfill\cite{Laiho:2009eu}				&
$F_{B_d} \sqrt{\hat
B_{B_d}} = 226(13)\mev$\hfill\cite{Laiho:2009eu} \\
$m_c(m_c) = (1.279\pm 0.013) \gev$ \hfill\cite{Chetyrkin:2009fv}					&
$F_{B_s} \sqrt{\hat B_{B_s}} =
279(13)\mev$\hfill\cite{Laiho:2009eu} \\
$m_b(m_b)=4.19^{+0.18}_{-0.06}\gev$\hfill\cite{Nakamura:2010zzi} 			& $\xi =
1.237(32)$\hfill\cite{Laiho:2009eu}
\\
$m_t(m_t) = 163(1)\gev$\hfill\cite{Laiho:2009eu,Allison:2008xk} &  $\eta_B=0.55(1)$\hfill\cite{Buras:1990fn,Urban:1997gw}  \\
$M_t=172.9\pm0.6\pm0.9 \gev$\hfill\cite{Nakamura:2010zzi} 						&  $\Delta M_d = 0.507(4)
\,\text{ps}^{-1}$\hfill\cite{Nakamura:2010zzi}\\\cline{1-1}
$m_K= 497.614(24)\mev$	\hfill\cite{Nakamura:2010zzi}								&  $\Delta M_s = 17.73(5)
\,\text{ps}^{-1}$\hfill\cite{Abulencia:2006ze,Aaij:2011qx}
\\	
$F_K = 156.1(11)\mev$\hfill\cite{Laiho:2009eu}												&
$S_{\psi K_S}= 0.679(20)$\hfill\cite{Nakamura:2010zzi}\\
$\hat B_K= 0.767(10)$\hfill\cite{Laiho:2009eu}												&
$S_{\psi\phi}= 0.0002\pm 0.087$\hfill\cite{Clarke:1429149}\\
$\kappa_\epsilon=0.94(2)$\hfill\cite{Buras:2008nn,Buras:2010pza}										&
\\	
$\eta_1=1.87(76)$\hfill\cite{Brod:2011ty}												
	& $\tau(B_s)= 1.471(25)\,\text{ps}$\hfill\cite{Asner:2010qj}\\		
$\eta_2=0.5765(65)$\hfill\cite{Buras:1990fn}												
& $\tau(B_d)= 1.518(7) \,\text{ps}$\hfill\cite{Asner:2010qj}\\
$\eta_3= 0.496(47)$\hfill\cite{Brod:2010mj}												
& \\\cline{2-2}
$\Delta M_K= 0.5292(9)\times 10^{-2} \,\text{ps}^{-1}$\hfill\cite{Nakamura:2010zzi}	&
$|V_{us}|=0.2252(9)$\hfill\cite{Nakamura:2010zzi}\\
$|\eps_K|= 2.228(11)\times 10^{-3}$\hfill\cite{Nakamura:2010zzi}					& $|V_{cb}|=(40.6\pm1.3)\times
10^{-3}$\hfill\cite{Nakamura:2010zzi}\\\cline{1-1}
  $\mathcal{B}(B\to X_s\gamma)=(3.55\pm0.24\pm0.09) \times10^{-4}$\hfill\cite{Nakamura:2010zzi}                                                                &
$|V^\text{incl.}_{ub}|=(4.27\pm0.38)\times10^{-3}$\hfill\cite{Nakamura:2010zzi}\\
$\mathcal{B}(B^+\to\tau^+\nu)=(1.64\pm0.34)\times10^{-4}$\hfill\cite{Nakamura:2010zzi}	&
$|V^\text{excl.}_{ub}|=(3.12\pm0.26)\times10^{-3}$\hfill\cite{Laiho:2009eu}\\\cline{1-1}					
$\tau_{B^\pm}=(1641\pm8)\times10^{-3}\,\text{ps}$\hfill\cite{Nakamura:2010zzi}
														&
\\

\hline
\end{tabular}  }
\caption {\textit{Values of the experimental and theoretical
    quantities used as input parameters.}}
\label{tab:input}
\end{table}

\section{Numerical Analysis}\label{sec:numerics}
\setcounter{equation}{0}
\subsection{Procedure in the Trivially Unitary Model}
\label{subsec:5.1}
It is not the goal of this section to present a full-fledged numerical 
analysis of the TUM including present theoretical and experimental 
uncertainties as this would only wash out the effects we want to emphasize.
Therefore, in our numerical analysis we will  choose the values in (\ref{fixed})
as nominal values for four CKM parameters. The remaining input parameters 
are collected in Table~\ref{tab:input}. In any case as NP effects 
in $B_{d,s}^0-\bar B_{d,s}^0$ are very small, the hadronic uncertainties 
in this sector are identical to the ones in the SM. They are more important in the 
case of $\varepsilon_K$, as they play the role in the compensation of LR 
contributions by LL ones and consequently have an impact on the allowed 
values of $X_{13}$ and $X_{23}$ in (\ref{Xij}). Also the uncertainties 
in $F_{B_q}$ are relevant for the predictions of 
$\mathcal{B}(B_{s,d}\to \mu^+\mu^-)$ but lattice calculations made significant 
progress in the last years \cite{Laiho:2009eu}.

The observables analysed numerically in this section depend on thirteen real parameters and one complex phase that is equal to the angle $\gamma$ in the 
unitarity triangle. In the TUM, once the six quark masses and the 
CKM parameters have been determined as explained above, the model contains 
four {\it positive} definite real parameters 
\be\label{4par}
M, \quad \varepsilon_3^Q,\quad s^d_{13}, \quad s^d_{23}
\ee
with  $s_{13}^d$ and 
$s_{23}^d$ smaller than unity and $0.80\le \varepsilon_3^Q\le 1.0$ because 
of the value of the top quark mass.  
 For fixed $M$ eliminating the 10 parameters $\eps^Q_{1,2}, \eps^D_{1,2,3}, \eps^U_{1,2,3}, \delta^d, s_{12}^d$ in favour of 3 CKM mixing
angles, 1 phase and 6 masses, one can find the couplings $\Delta^{ij}_{L,R}$ for $W^\pm$,
$Z^0$ and $H^0$ as 
functions of   $s_{23}^d$, $s_{13}^d$ and  $\varepsilon_3^Q$. 
In what follows we will set $M=3.0~\tev$ which on one hand is still in the 
reach of the LHC and on the other hand is sufficiently large so that the 
upper bound on  $\mathcal{B}(K_L\to\mu^+\mu^-)$  and also 
the LHCb bounds on
$b\to s\ell^+\ell^-$ transitions are satisfied.
As shown in \cite{Buras:2011ph} for such values of $M$ the model is also 
consistent with electroweak constraints. We can 
then vary   $s_{13}^d$, $s_{23}^d$ and  $\varepsilon_3^Q$. All observables considered by us 
depend now on only three real and positive definite free parameters 
 implying various correlations that we are 
going to expose below.

\boldmath
 \subsection{Phenomenology of $\Delta F = 2$ observables}
\begin{figure}[!tb]
\centering
\includegraphics[width = 0.45\textwidth]{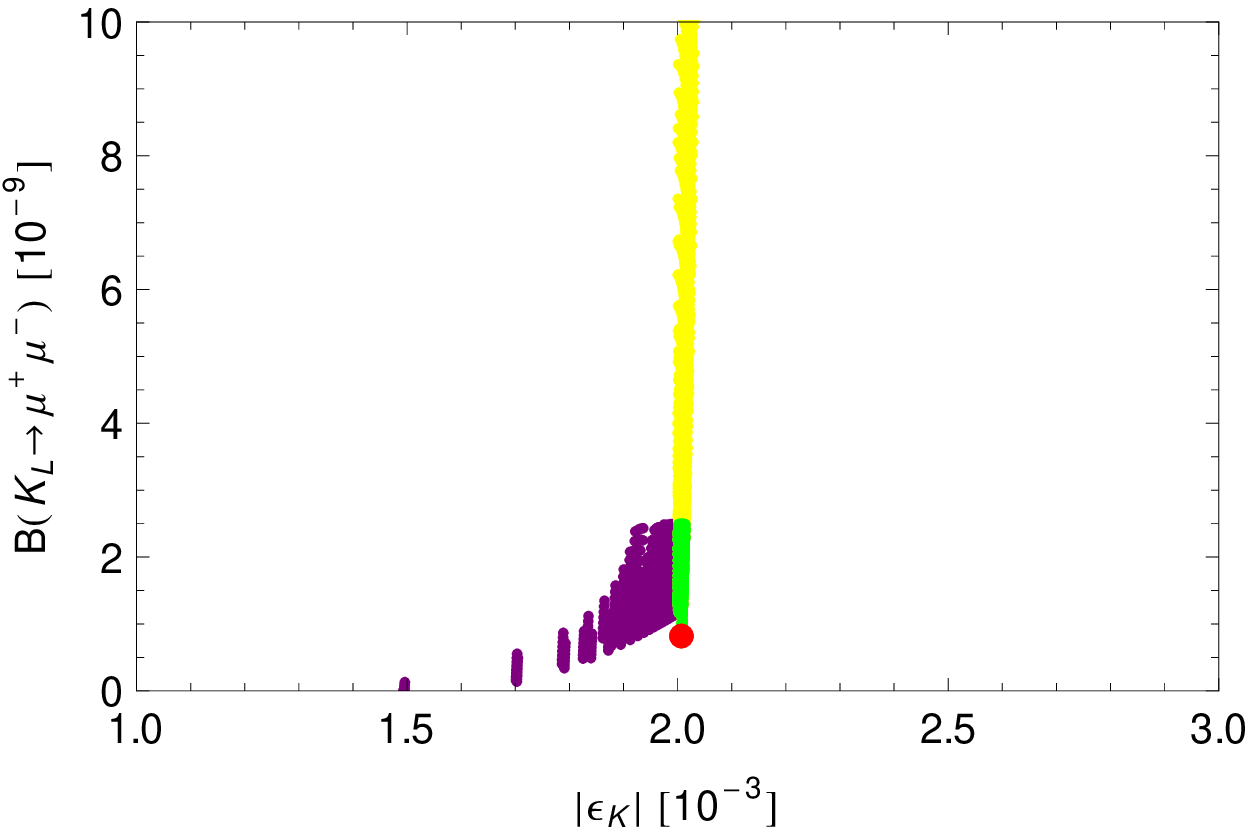}
\includegraphics[width = 0.45\textwidth]{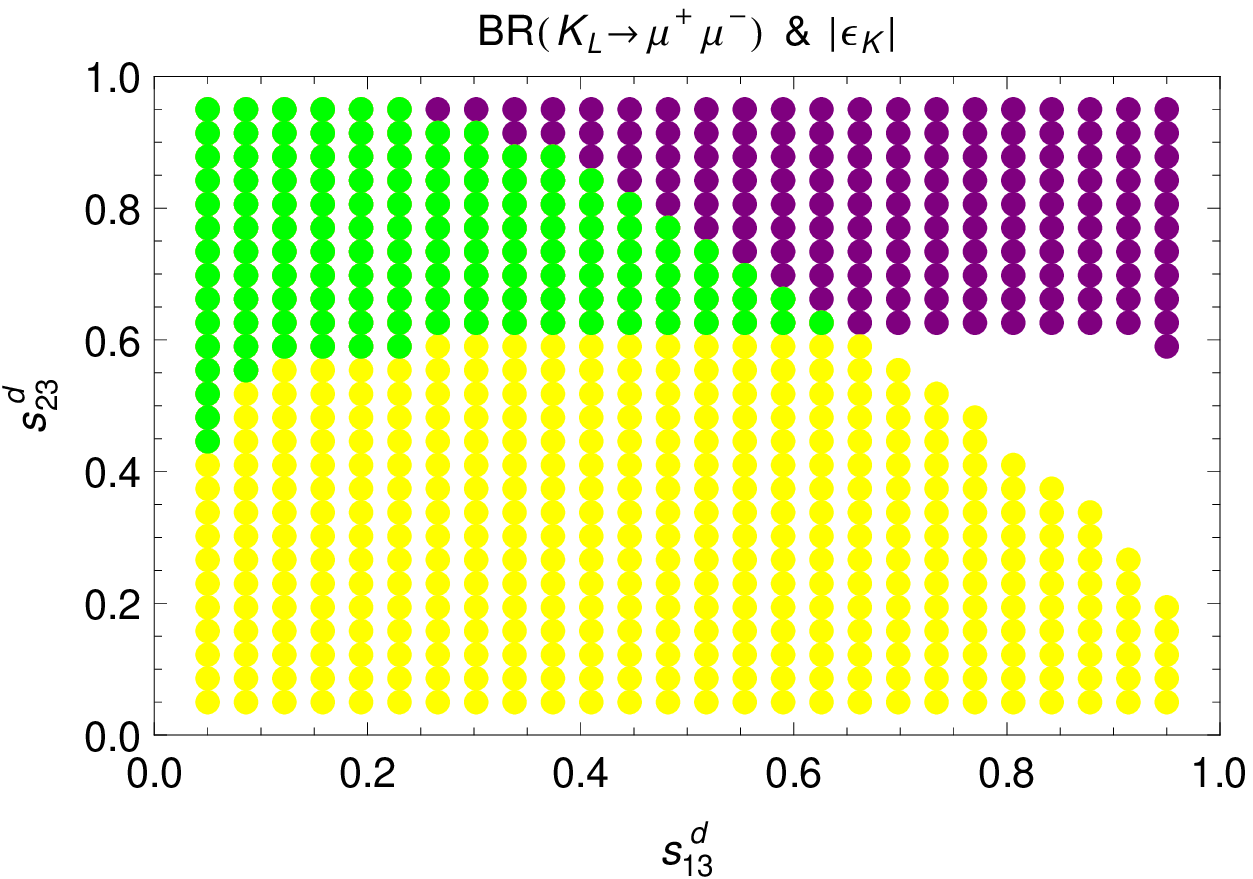}
\caption{\it Left: $\mathcal{B}(K_L\to\mu^+\mu^-)$ versus $|\varepsilon_K|$ for $M = 3~$TeV and $|V_{ub}| = 0.0037$. Green points are
compatible with both bounds for $|\varepsilon_K|$ (\ref{C3}) and $\mathcal{B}(K_L\to\mu^+\mu^-)$ (\ref{eq:KLmm-bound}), yellow is
only compatible with  $|\varepsilon_K|$ and purple only with $\mathcal{B}(K_L\to\mu^+\mu^-)$. The red
point corresponds to the SM central value. Right: Allowed region in the parameter space $(s^d_{13},s^d_{23})$ for $\epsilon_3^Q = 0.9$ due
to $|\varepsilon_K|$ and $\mathcal{B}(K_L\to\mu^+\mu^-)$ bounds (same colour coding as left). }
 \label{fig:epsKKLmumu}~\\[-2mm]\hrule
\end{figure}

\unboldmath
As we already discussed in Section~\ref{sec:anatomy} NP effects in 
$B^0_{s,d}-\bar B^0_{s,d}$ mixings are negligible. However they can be in 
principle large in $\varepsilon_K$. Yet requiring that\footnote{The ranges 
chosen in (\ref{C3}) indicate theoretical and parametric uncertainties in 
both observables.}
\be\label{C3}
0.75\le \frac{\Delta M_K}{(\Delta M_K)_{\rm SM}}\le 1.25,\qquad
2.0\times 10^{-3}\le |\varepsilon_K|\le 2.5 \times 10^{-3}
\ee
and imposing the bound on $\mathcal{B}(K_L\to\mu^+\mu^-)_{\rm SD}$ in
(\ref{eq:KLmm-bound}) we find, as seen in  Fig.~\ref{fig:epsKKLmumu} (left), that $|\varepsilon_K|$ is forced to be at the lower end of
the range in (\ref{C3}) and very close to the SM expectation for the CKM parameters chosen by us.

The colour coding in this plot and the following plots is as follows:
\begin{itemize}
\item
{\it Green} range is allowed by $\varepsilon_K$ through (\ref{C3}) and 
$K_L\to \mu^+\mu^-$ through (\ref{eq:KLmm-bound}).
\item
{\it Yellow} range  is allowed by  $\varepsilon_K$  but not $K_L\to \mu^+\mu^-$. \item
{\it Purple} range  is forbidden by  $\varepsilon_K$ but allowed by 
$K_L\to \mu^+\mu^-$.
 \item
 Points that are forbidden by  both  $\varepsilon_K$ and $K_L\to \mu^+\mu^-$ are not shown.
\end{itemize}

The corresponding 
regions in the space  $(s^d_{13},s^d_{23})$ are shown in  Fig.~\ref{fig:epsKKLmumu} (right) for $\epsilon_3^Q = 0.9$.  
We observe that in the allowed region $s_{23}^d\ge 0.5$, and $s_{13}^d\le 0.6$.

\begin{figure}[!tb]
\centering
\includegraphics[width = 0.65\textwidth]{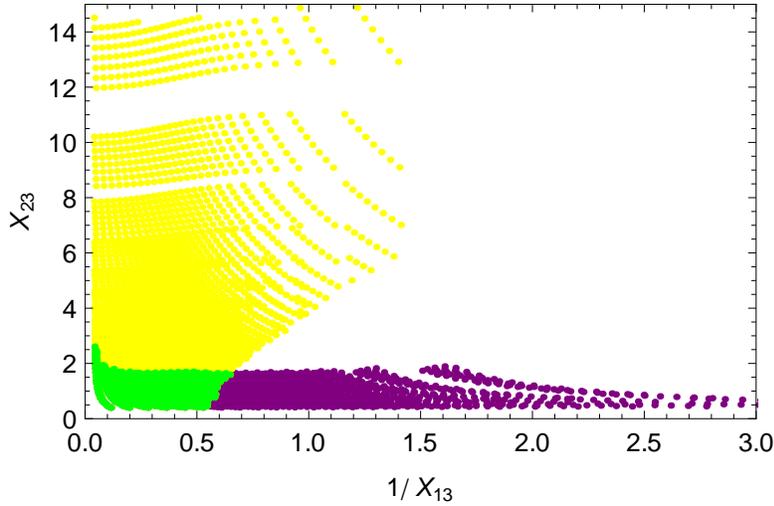}
\caption{\it $X_{23}$ versus $1/X_{13}$. Colour coding as in Fig.~\ref{fig:epsKKLmumu}. }
 \label{fig:X23X13}~\\[-2mm]\hrule
\end{figure}

The fact that NP effects in $\varepsilon_K$ are very small implies the following approximate 
relation between $X_{13}$, $X_{23}$  and $\eps_3^{Q}$
\be\label{RRR1}
\eps^{Q4}_3 (1+1.2~X_{23}^2) -0.18\frac{|P_1^{\rm LR}(K)|}{X_{13}^2}\approx 0
\ee
In  Fig.~\ref{fig:X23X13} we show the four regions in the space $(X_{23},1/X_{13})$. 
We observe that the allowed ranges represented by the green region are 
   roughly:
\be
0.4\le X_{23}\le 2.0, \qquad   0.03 \le \frac{1}{X_{13}}\le 0.70.
\ee

We also recall that for $M=3~\TeV$ in order to reproduce the top-quark mass 
we have $0.80\le \eps_3^Q\le 1$.
Consequently we conclude that the effects of right-handed 
couplings in $\Delta F=1$ decays are negligible and NP contributions to these decays are dominated by left-handed $Z$ couplings to quarks. This is in contrast 
to rare decays in RS model with custodial protection \cite{Blanke:2008yr} where NP effects in these decays were governed by flavour-violating RH couplings of $Z^0$ to quarks.

\boldmath
\subsection{$\mathcal{B}(B_{d,s}\to\mu^+\mu^-)$ and $\Delta M_d/\Delta M_s$
}\label{sec:golden}
\unboldmath

In models with CMFV these observables are related through a
theoretically very clean relation
\cite{Buras:2003td} that in the MTFM and generally in models with non-MFV 
sources and new operators
gets modified as
follows:
 \be\label{eq:golden}
\frac{\mathcal{B}(B_s\to\mu^+\mu^-)}{\mathcal{B}(B_d\to\mu^+\mu^-)}= \frac{\hat
B_{B_d}}{\hat B_{B_s}} \frac{\tau(B_s)}{\tau(B_d)} \frac{\Delta
M_s}{\Delta M_d}\,r\,,
\ee
where
\be\label{eq:r}
r= \left|\frac{Y_A(B_s)}{Y_A(B_d)}\right|^2
\frac{C_{B_d}}{C_{B_s}}\,,\qquad 
{C_{B_{d,s}}= \frac{\Delta M_{d,s}}{(\Delta M_{d,s})_\text{SM}}\,,}
\ee
with $r=1$ in {CMFV} models but generally different from unity.

\begin{figure}[!tb]
\centering
\includegraphics[width = 0.65\textwidth]{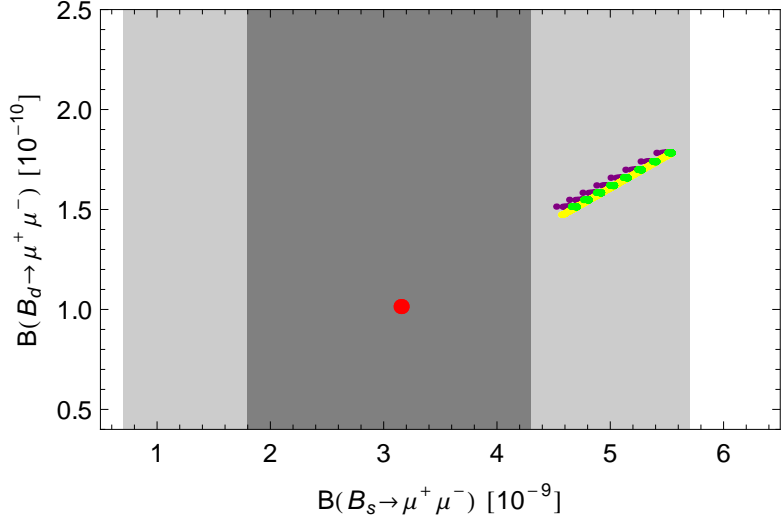}
\caption{\it $\mathcal{B}(B_d\to\mu^+\mu^-)$  versus $\mathcal{B}(B_s\to\mu^+\mu^-)$. Colour coding as in Fig.~\ref{fig:epsKKLmumu}. Gray
region: exp. 1 and $2\sigma$ range of $\mathcal{B}(B_s\to\mu^+\mu^-)$ (see Eq.~(\ref{LHCb2corr})). }
 \label{fig:BdmuvsBsmu}~\\[-2mm]\hrule
\end{figure}

However in TUM when all constraints are taken into account we find $r=1$ and the relation (\ref{eq:golden}) is satisfied very well. 
Moreover $C_{B_{d,s}}=1$. Yet, the values of $\mathcal{B}(B_q\to\mu^+\mu^-)$ 
differ significantly from SM prediction. This we show in   Fig.~\ref{fig:BdmuvsBsmu} indicating the experimental $1\sigma$  and
$2\sigma$ ranges for 
$\mathcal{B}(B_s\to\mu^+\mu^-)$.
The striking 
prediction of TUM are uniquely enhanced values of both branching ratios, 
moreover in the allowed green region these enhancements take place in a correlated manner: the slope 
of the straight line in   Fig.~\ref{fig:BdmuvsBsmu} is given by the formula 
(\ref{eq:golden}) with $r=1$.  We find the ranges:
\be
 4.2 \times 10^{-9}\le\mathcal{B}(B_s\to\mu^+\mu^-)\le 6.0 \times 10^{-9}, \qquad 
1.3 \times 10^{-10}\le\mathcal{B}(B_d\to\mu^+\mu^-)\le 2.0 \times 10^{-10}
\ee
where we added parametric uncertainties not shown in the plot.

\boldmath
\subsection{The $\kpn$ and $\klpn$ Decays}
\unboldmath
In order to understand the pattern of NP contributions to these decays we 
calculate $X_{\rm eff}$ in (\ref{Xeff}) in the TUM. Neglecting 
the small RH contributions we find 
\be\label{Xeff1}
X_{\rm eff}=-\vts\vtd\left[e^{-(\beta-\beta_s)}\tilde X(x_t)+0.37 \left(\frac{3\tev}{M}\right)^2 \eps^{Q2}_3 X_{23}^2\right],
\ee
where
\be
\tilde X(x_t)\equiv X(x_t)+0.31 \left(\frac{3\tev}{M}\right)^2\eps^{Q2}_3.
\ee
The first term in (\ref{Xeff1}) describes a typical CMFV contribution with a modified basic function $\tilde X$ that is uniquely larger than $X$ in the SM. This increase is governed by $\eps^{Q2}_3/M^2$. The second term does not carry any new phases and 
goes beyond CMFV. It contributes only to $\kpn$ and modifies the usual CMFV
correlation between the branching ratios for these two decays.

\begin{figure}[!tb]
\centering
\includegraphics[width = 0.65\textwidth]{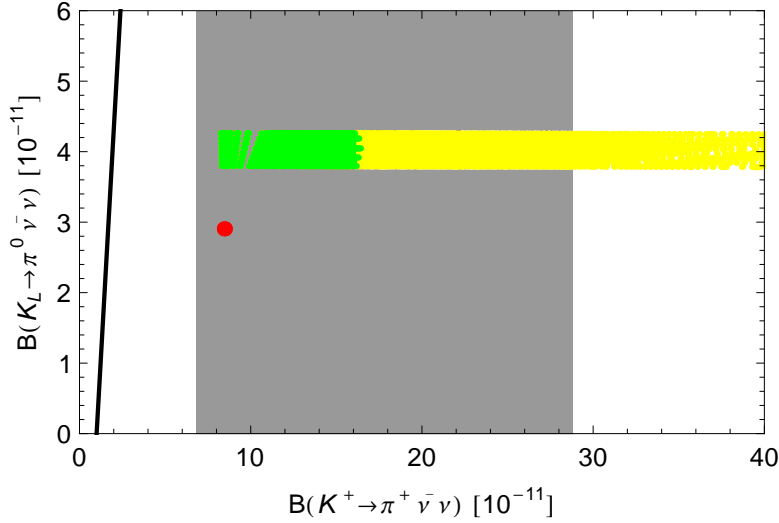}
\caption{\it $\mathcal{B}(\klpn)$  versus $\mathcal{B}(\kpn)$. The black line on the left is the GN bound. Gray region: experimental range
of $\mathcal{B}(\kpn)$ (see Eq.~(\ref{EXP1})). Colour coding as in
Fig.~\ref{fig:epsKKLmumu}. }
 \label{fig:KLpinuvsKppinu}~\\[-2mm]\hrule
\end{figure}

In  Fig.~\ref{fig:KLpinuvsKppinu} we show the correlation between $\mathcal{B}(\kpn)$ and 
$\mathcal{B}(\klpn)$ in TUM. The experimental
$1\sigma$-range for $\mathcal{B}(\kpn)$ \cite{Artamonov:2008qb} and the
model-independent Grossman-Nir (GN) bound \cite{Grossman:1997sk} are also
shown. We observe that  
$\mathcal{B}(\klpn)$ can be as large as  $4.4\cdot 10^{-11}$, that is roughly by a factor of
1.5 larger than its SM value $(3.0\pm 0.6)\cdot 10^{-11}$,  while being still consistent with the measured 
value for $\mathcal{B}(\kpn)$. The latter branching ratio can be enhanced by at
most a factor of 2 but this is sufficient to reach {the} central experimental 
value \cite{Artamonov:2008qb} in (\ref{EXP1}). 

The plot has a shape which differs from the one encountered 
in LHT model or $Z'$ models. The expression in (\ref{Xeff1}) 
explains what is going on:
\begin{itemize}
\item
For a fixed value of $\eps_3^Q$ the branching ratio $\mathcal{B}(\klpn)$ 
is fixed, while $\mathcal{B}(\kpn)$ depending in addition on $X_{23}$ can 
take a significant range of values bounded by $K_L\to\mu^+\mu^-$ and $\varepsilon_K$.  We have then a straight horizontal line. This line 
moves up with increasing $\eps_3^Q$.
\item
For fixed $M$ both branching ratios increase with increasing  $\eps_3^Q$.
\end{itemize}

Thus if $X_{23}$ would vanish, we would just have the case of CMFV.  
Varying  $\eps_3^Q$ we would get a straight line on which 
both branching ratios would increase with increasing  $\eps_3^Q$. This is 
the line which fully describes $\mathcal{B}(\klpn)$ in the TUM.
However,
$X_{23}$ cannot vanish and 
is bounded from below in order to balance negative contributions 
from LR operators to $\varepsilon_K$. This implies the shape in   the Fig.~\ref{fig:KLpinuvsKppinu}.

\boldmath
\subsection{Correlation between $K_L\to \mu^+\mu^-$ and $\kpn$}
\unboldmath

\begin{figure}[!tb]
\centering
\includegraphics[width = 0.65\textwidth]{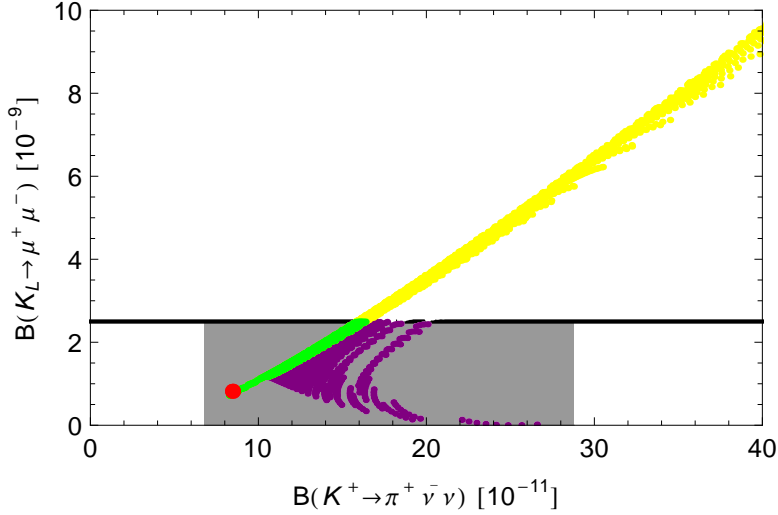}
\caption{\it $\mathcal{B}(K_L\to \mu^+\mu^-)$  versus $\mathcal{B}(\kpn)$. Colour coding as in Fig.~\ref{fig:epsKKLmumu}.  Gray region:
experimental range of $\mathcal{B}(\kpn)$ (see Eq.~(\ref{EXP1})). Black horizontal line: upper bound of $\mathcal{B}(K_L\to \mu^+\mu^-)$
 (see Eq.~(\ref{eq:KLmm-bound})).  }
 \label{fig:KLmuvsKppinu}~\\[-2mm]\hrule
\end{figure}

Next in  Fig.~\ref{fig:KLmuvsKppinu}  we show the correlation between  $\mathcal{B}(K_L\to \mu^+\mu^-)_{\rm SD}$ and
$\mathcal{B}(\kpn)$. As both decays are CP-conserving, a non-trivial correlation is generally expected. The following observations should be made:
\begin{itemize}
\item
Without the upper bound on $\mathcal{B}(K_L\to \mu^+\mu^-)$ the branching 
ratio $\mathcal{B}(\kpn)$ could be much larger.
\item
The fact that the increase of one of the two branching ratios implies uniquely 
the increase of the other one signals the dominance of left-handed currents 
in NP contributions. 
\end{itemize}

Concerning the latter point,
if right-handed couplings were dominating as found 
in RSc scenario  \cite{Blanke:2008yr}  and some $Z'$ scenarios in \cite{Buras:2012jb}, we would find 
an anti-correlation 
i.\,e. an enhancement of $\mathcal{B}(K_L\to \mu^+\mu^-)$ relative to the SM 
would imply suppression of $\mathcal{B}(\kpn)$ and vice versa. 
This different behaviour originates in the fact that the $K^+\to\pi^+\nu\bar\nu$ transition is sensitive to the vector component of the flavour violating $Z$ coupling, while the $K_L\to\mu^+\mu^-$ decay measures its axial component. 
In other words, the correlation between $K^+\to\pi^+\nu\bar\nu$ and $K_L\to\mu^+\mu^-$ offers a clear test of the handedness of NP flavour
violating interactions and  Fig.~\ref{fig:KLmuvsKppinu} shows transparently that in TUM the left-handed 
couplings are at work. A more general discussion of these points can be found
in \cite{Blanke:2009pq}.

\boldmath
\subsection{$K_L\to\pi^0 \ell^+\ell^-$}
\begin{figure}[!tb]
\centering
\includegraphics[width = 0.45\textwidth]{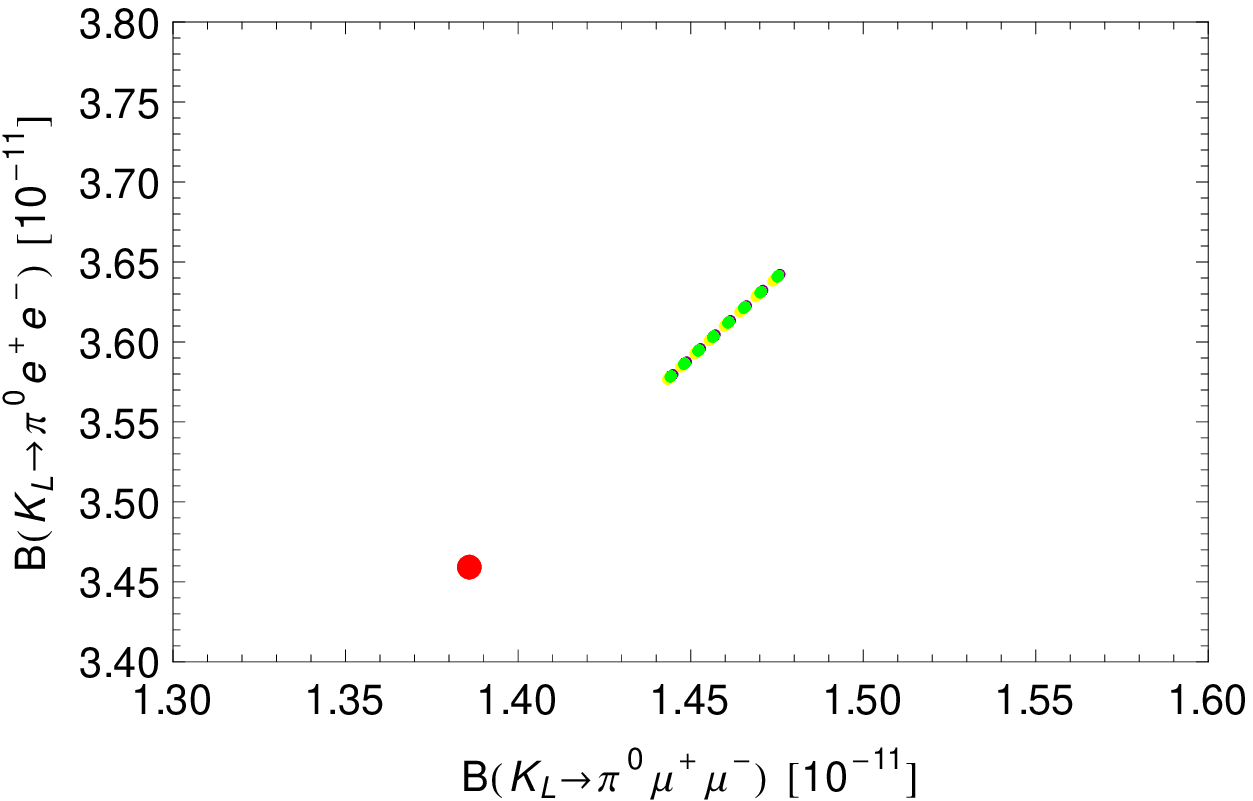}
\includegraphics[width = 0.45\textwidth]{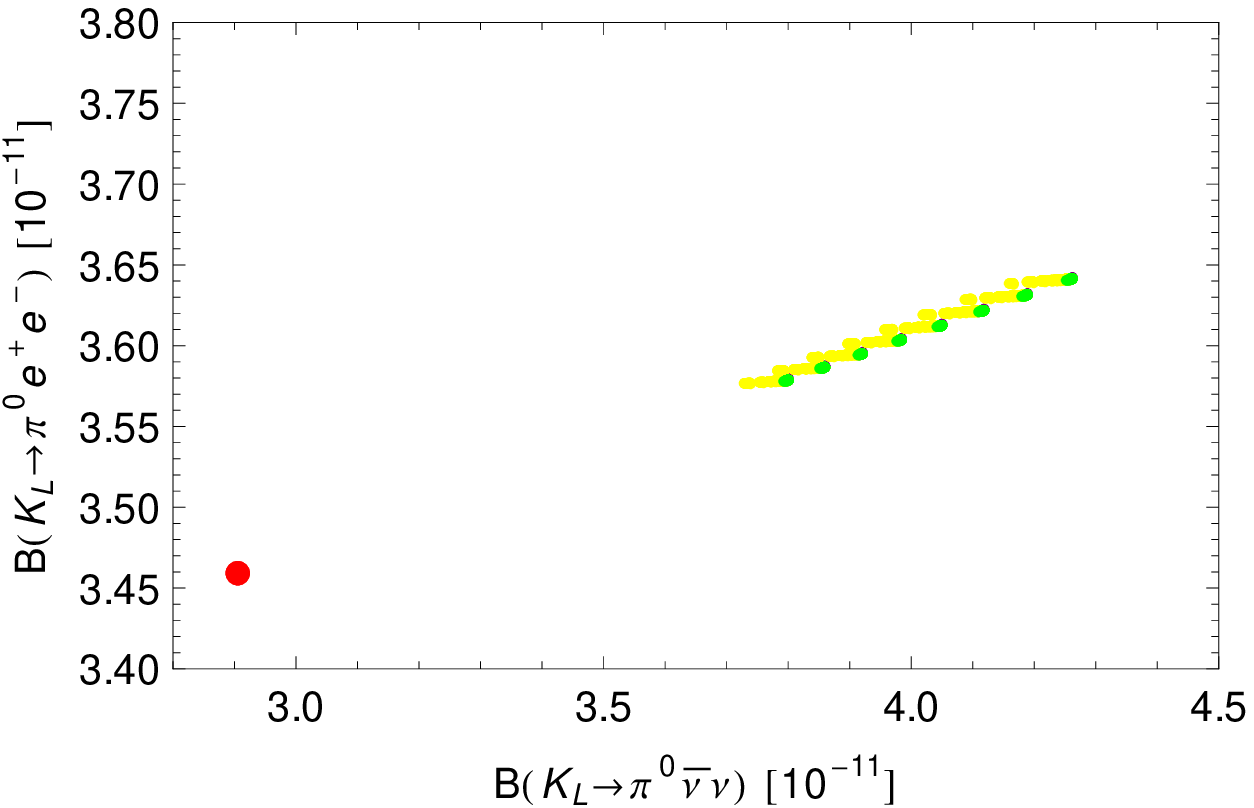}

\caption{\it $\mathcal{B}(K_L\to\pi^0 e^+e^-)$  versus $\mathcal{B}(K_L\to\pi^0 \mu^+\mu-)$ (left) and $\mathcal{B}(\klpn)$ (right). Colour
coding as in Fig.~\ref{fig:epsKKLmumu}. }
 \label{fig:eevsmumu}~\\[-2mm]\hrule
\end{figure}

\unboldmath
In the left panel of Fig.~\ref{fig:eevsmumu}
 we show $\mathcal{B}(K_L\to\pi^0 e^+e^-)$  vs $\mathcal{B}(K_L\to\pi^0 \mu^+\mu-)$.  In the right panel the correlation between  $\mathcal{B}(K_L\to\pi^0 e^+e^-)$ and  $\mathcal{B}(\klpn)$ is shown. We observe that the correlations between all 
three branching ratios are very strong and all branching ratios are enhanced 
relative to the SM values but NP effects in $\mathcal{B}(K_L\to\pi^0 \ell^+\ell^-)$ are as expected much smaller than in 
$\mathcal{B}(\klpn)$. 
 We recall that we only show the results for the constructive interference 
between SM and NP contributions.

\boldmath
\subsection{$B_{s,d}\to \mu^+\mu^-$ versus $\kpn$ and $\klpn$}
\begin{figure}[!tb]
\centering
\includegraphics[width = 0.45\textwidth]{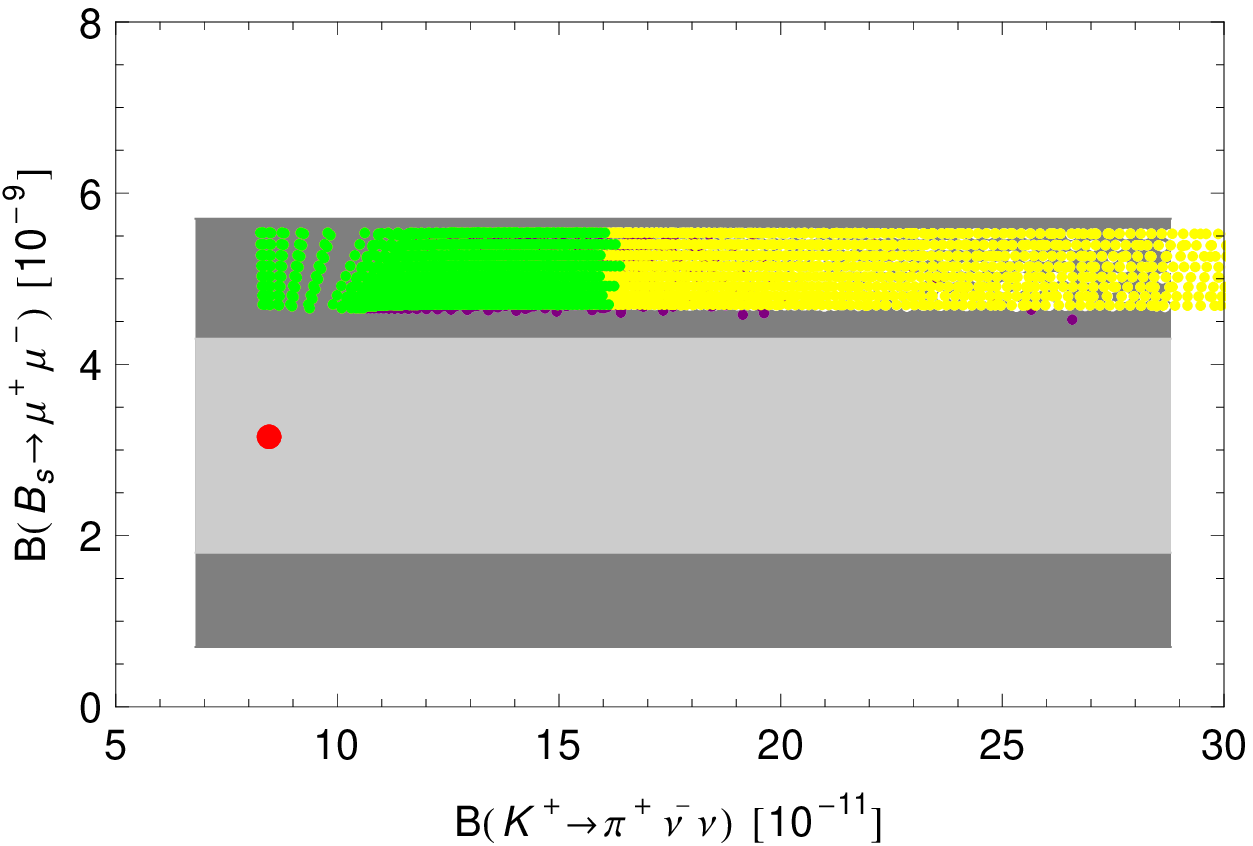}
\includegraphics[width = 0.45\textwidth]{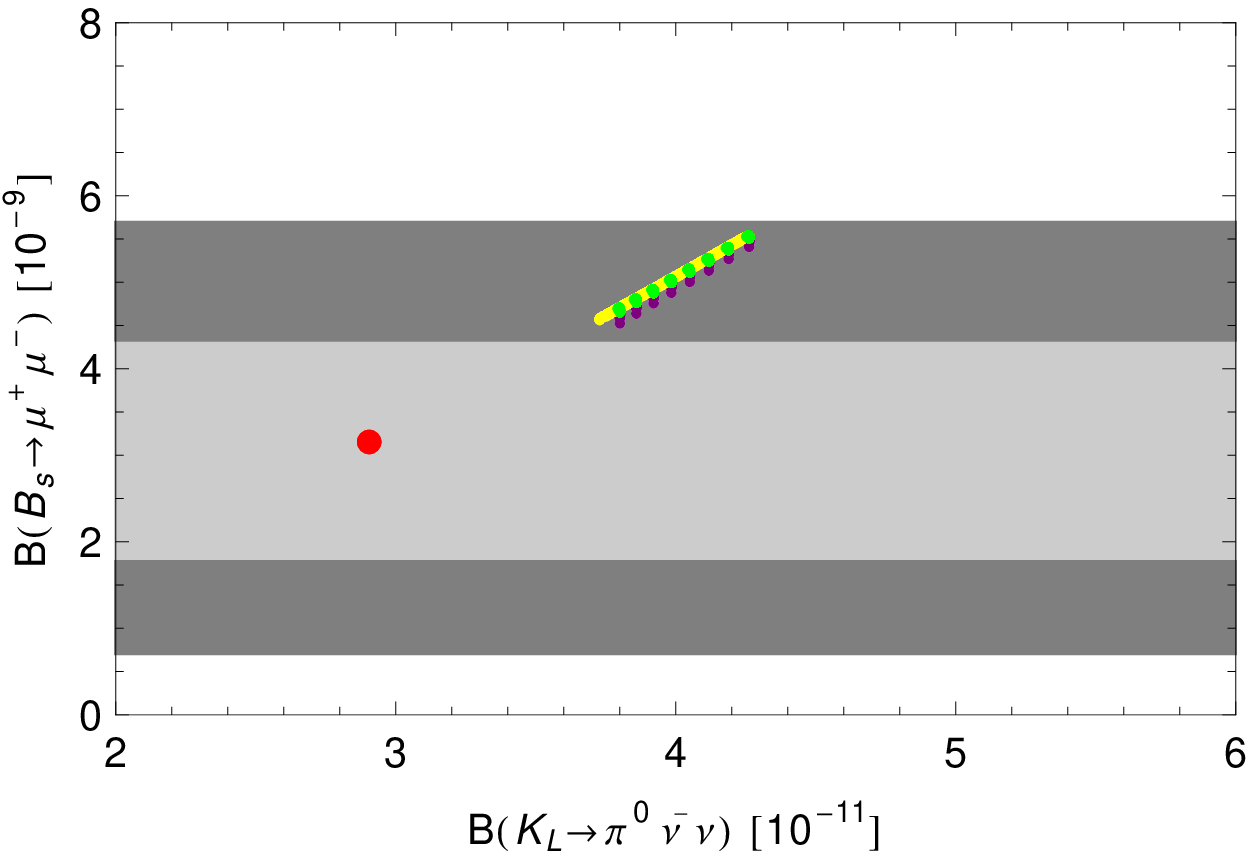}
\caption{\it$B_{s,d}\to \mu^+\mu^-$ versus $\mathcal{B}(\kpn)$ and $\mathcal{B}(\klpn)$. Gray region: exp. range. Colour
coding as in Fig.~\ref{fig:epsKKLmumu}. }
 \label{fig:BsmuvsKpinu}~\\[-2mm]\hrule
\end{figure}

\unboldmath
In view of the small number of parameters in the TUM these decays are correlated 
with each other. In the left panel of  Fig.~\ref{fig:BsmuvsKpinu} we show the 
correlation 
between
$\mathcal{B}(B_{s}\to \mu^+\mu^-)$ and $\mathcal{B}(K^+\to\pi^+\nu\bar\nu)$. This correlation is similar to the one between $\mathcal{B}(\klpn)$
and  $\mathcal{B}(\kpn)$ as $\mathcal{B}(B_{s}\to \mu^+\mu^-)$ similarly to 
$\mathcal{B}(\klpn)$ depends primarily on $\eps_3^Q$. Not surprisingly 
the correlation between $\mathcal{B}(B_{s}\to \mu^+\mu^-)$ and 
$\mathcal{B}(\klpn)$ is very simple. We show it in the right panel of 
  Fig.~\ref{fig:BsmuvsKpinu}.

\boldmath
\subsection{$B \to \{X_s,K, K^*\} \nu\bar \nu$}
\begin{figure}[!tb]
\centering
\includegraphics[width = 0.65\textwidth]{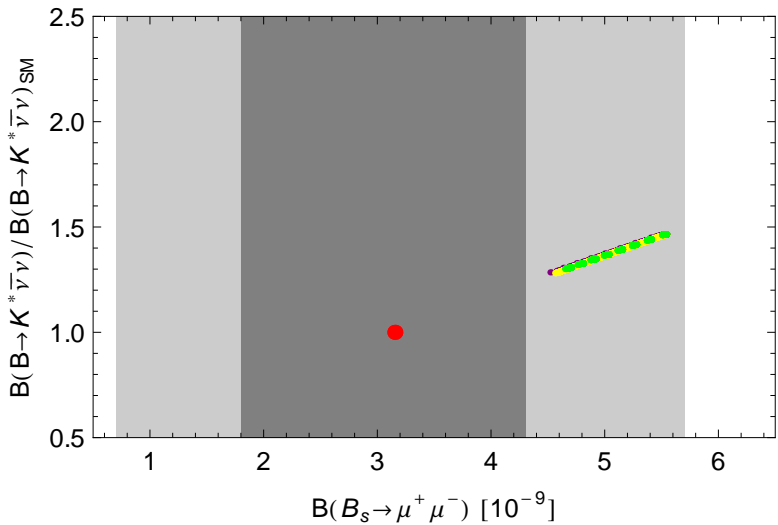}
\caption{\it $\mathcal{B}(B\to K^\star\bar\nu\nu)$ versus $\mathcal{B}(B_{s}\to \mu^+\mu^-)$.  Gray
region: exp. 1 and $2\sigma$ range of $\mathcal{B}(B_s\to\mu^+\mu^-)$ (see Eq.~(\ref{LHCb2corr})).
Colour coding as in Fig.~\ref{fig:epsKKLmumu}. }
 \label{fig:BKstarnuvsBsmu}~\\[-2mm]\hrule
\end{figure}

\unboldmath
As the right-handed currents are suppressed in TUM and $\eta\approx 0$, 
the three branching ratios in question are strongly correlated with each other 
and knowing one of them gives the information about the other two. Moreover, 
as left-handed currents dominate, there is a strong correlation between these 
decays and $B_{s,d}\to\mu^+\mu^-$.  In  Fig.~\ref{fig:BKstarnuvsBsmu} we demonstrate this 
by showing    $\mathcal{B}(B\to K^*\nu\bar\nu)$ versus  $\mathcal{B}(B_{s}\to \mu^+\mu^-)$. 

\subsection{Implications of $b\to s \ell^+\ell^-$ Constraints}\label{bsllc}
Presently the NP effects found by us are consistent with the experimental
data on $B_{s,d}\to\mu^+\mu^-$, although with improved upper bound on 
$B_{s,d}\to\mu^+\mu^-$, the TUM could have problems with describing the data.
 However,
also the data on $B\to X_s \ell^+\ell^-$, $B\to   K^*\ell^+\ell^-$  and
 $B\to   K\ell^+\ell^-$ improved
recently
 by much and it is of interest to see whether they have
an impact on our results. A very extensive model independent
analysis of the impact of the recent LHCb data on the Wilson coefficients
$C_9^{(\prime)}$ and  $C_{10}^{(\prime)}$ has been performed in
\cite{Altmannshofer:2012ir} and we can use these results in our case. Other 
recent analyses of $b\to s\ell^+\ell^-$ can be found in  
\cite{Beaujean:2012uj,DescotesGenon:2012zf}.

In view of suppressed vectorial couplings of $Z$ to muons the bounds on 
$C_9^{(\prime)}$ are easily satisfied.
Therefore we will
only check whether for the ranges of parameters considered
by us  the resulting coefficients $C_{10}^{(\prime)}$  satisfy the model
independent bounds in \cite{Altmannshofer:2012ir}. As these coefficients
are scale independent we can use the formulae in (\ref{C10}) and (\ref{C10P})
and compare the resulting coefficients
with those in the latter paper. 

We find first
that in TUM
\be\label{C101}
\sin^2\theta_W C_{10}^{\rm NP}= -\Delta X_L(B_s)  = -0.31 \left( \frac{3\TeV}{M} \right)^2 \eps^{Q2}_3 ,
\ee
\be\label{C10P1}
\sin^2\theta_W C_{10}^{\prime}=
-X_R(B_s)  =   -0.31  \left( \frac{3\TeV}{M} \right)^2\frac{1 }{\eps^{Q2}_3}\left[ \frac{2.0-5.1i}{X_{13}^2} +  \frac{2.2}{X_{23}^2}\right] 10^{-3}.
\ee

On the other hand 
the allowed $2\sigma$ ranges of $C_{10}^{(\prime)}$ are shown in Figs.~1 and 2 of \cite{Altmannshofer:2012ir}. They are given
approximately as follows:
\begin{subequations}\label{equ:ASconstraint}
\begin{align}
 &-2\leq \re(C_{10}^\prime)\leq 0\,, \quad-2.5\leq \im(C_{10}^\prime)\leq 2.5\,,\\
&-0.8\leq \re(C_{10}^\text{NP})\leq 1.8\,,\quad -3\leq \im(C_{10})\leq 3\,.
\end{align}
\end{subequations}

Especially, 
the new data on
$B\to K^*\mu^+\mu^-$ allow only for  {\it negative} values of the real part
of $C^\prime_{10}$
\be \label{C10C}
\re( C^\prime_{10}) \le 0.
\ee
As seen in (\ref{C10P1}) in TUM this condition is satisfied as 
$\re( C^\prime_{10})$ is predicted to be negative. Moreover,
in view of the suppression of right-handed currents in TUM, these 
data have no impact on our results.

On the other hand the lower bound on the real part of 
$C_{10}^{\rm NP}$ gives an upper bound on the ratio
$\eps_3^Q/M$:
\be\label{e3bound}
\left(\frac{3~\tev}{M}\right) \eps_3^Q\le 0.78.
\ee
 Therefore for $M=3\tev$ these data prefer  $\eps_3^Q$ close to its lower 
bound of $0.8$. Conversely this bound implies 
\be
M\ge 3~ \TeV.
\ee

 If this behaviour will be confirmed in the future by more 
accurate data and improved form factors that enter the analysis of 
 \cite{Altmannshofer:2012ir}, the TUM will favour this low value 
implying rather precise values for various branching ratios for 
$M=3~\TeV$. In particular 
we predict then:
\be\label{P1}
\mathcal{B}(B_{s}\to\mu^+\mu^-)=(4.6\pm 0.4)\times 10^{-9}, \qquad 
\mathcal{B}(B_{d}\to\mu^+\mu^-)=(1.5\pm 0.1)\times 10^{-10}
\ee
and
\be\label{P2}
\mathcal{B}(\klpn)=(3.8\pm0.4) \times  10^{-11}.
\ee
 As these data do not fix $X_{23}$ the branching ratio $\mathcal{B}(\kpn)$ 
is still consistent with all the data for
\be\label{P3}
9\times 10^{-11}\le \mathcal{B}(\kpn)\le 16\times 10^{-11}.
\ee

 We should remark that all the branching ratios depend to an excellent 
approximation on the ratio $\eps^Q_3/M$. Therefore the predictions given 
above can be kept, while satisfying (\ref{e3bound}), for 
$3~\TeV\le M\le 3.8~\TeV$ by increasing $\eps^Q_3$. As  $\eps^Q_3\le 1$, for 
$M\ge 3.8~\TeV$ the values of the branching ratios will decrease with 
increasing $M$.

\section{Summary and Outlook}\label{sec:summary}
\setcounter{equation}{0}
In the present paper we have performed  a detailed analysis of particle-antiparticle mixing and of the most
interesting rare decays of $K$ and $B$ mesons in the MTFM  concentrating 
on its simplest version, the trivially unitary model (TUM),
in which the Yukawa matrix in the heavy down quark sector is 
unitary and the corresponding matrix in the heavy up quark sector is 
a unit matrix. The
new contributions to FCNC processes 
are dominated by  tree-level flavour violating 
$Z$ couplings to quarks. The modifications of the $W$-boson couplings and 
the generated flavour violating Higgs couplings have negligible impact on 
observables considered by us.
Our analysis includes complete renormalization group QCD effects in NP 
contributions at NLO in the case of $\Delta F=2$ and in rare $K$ and $B$ 
decays.

The TUM can correctly reproduce the masses of quarks and the CKM matrix 
leaving for fixed heavy quark mass $M$ three real and positive definite parameters which together 
with CKM couplings and quark masses govern NP contributions to FCNC processes. 
The paucity of free 
parameters in the TUM implies
a number of 
correlations between various observables within the $K$ system, within the $B_{s,d}$ system and  between $K$ and $B_{s,d}$ systems, and also between
$\Delta F=2$ and $\Delta F=1$ observables. These correlations allow for a clear
distinction between this model and other NP scenarios.

The main messages of our paper are as follows:
\begin{itemize}
\item 
The simplest version of the MTFM, the TUM, is capable of describing the 
known quark mass spectrum and the elements of the CKM matrix favouring 
$\vub\approx 0.0037$ and $\gamma\approx 68^\circ$. The  masses of vectorial 
fermions are bounded to be larger than 
$M\approx 3.0\tev$ implying that these fermions are still in the 
reach of the LHC. Precise lower bound would require reduction of 
theoretical uncertainties in FCNC processes.
\item
NP effects in $B_{s,d}^0-\bar B_{s,d}$ observables are very small so that 
$\Delta M_{s,d}$, $S_{\psi\phi}$ and $S_{\psi K_S}$ are basically identical to SM 
predictions. The optimal 
values for $\vub$ and $\gamma$ imply very good  agreement of $\Delta M_{s,d}$ 
and  $S_{\psi\phi}$ with the present data, while  $S_{\psi K_S}\approx 0.72$ is
 by $2\sigma$ higher than its present experimental central value. It 
will be interesting to see how future LHCb results compare with this prediction.
\item
NP effects in $\varepsilon_K$ could in principle be sizable but the 
interplay of NP contributions from LL and LR operators to $\varepsilon_K$, 
the data for the later and the upper bound on $\mathcal{B}(K_L\to\mu^+\mu^-)_{\rm SD}$
makes also NP effects in $\varepsilon_K$ very small. Simultaneously 
right-handed flavour violating $Z$ couplings to quarks are forced to be 
suppressed leaving the corresponding left-handed couplings as the dominant 
source of NP contributions to rare $K$ and $B_{s,d}$ decays.
\item
The pattern of deviations from SM predictions in rare $B$ decays is 
CMFV-like with an important prediction not common to all CMFV models: 
$\mathcal{B}(B_{s,d}\to\mu^+\mu^-)$ are uniquely enhanced by at least $35\%$ relative 
to SM values and can be by almost a factor of two larger than in the SM. While still 
consistent with LHCb results, $\mathcal{B}(B_{s}\to\mu^+\mu^-)$ may turn 
out to be too high to agree with the future improved data. Finding this 
branching ratio to be enhanced would be good news for the TUM.
Also 
$b\to s\nu\bar\nu$ transitions are enhanced by a similar amount.
\item
The model predicts uniquely the enhancements of the branching ratios 
for $\kpn$ and $\klpn$ by similar amount 
as the rare $B$ decay branching ratios. In particular the correlation 
between $\klpn$ and $B_{s,d}\to\mu^+\mu^-$ is CMFV-like but the 
correlation between $\klpn$ and $\kpn$ shows a non-CMFV behaviour. 
NP effects in $K_L\to \pi^0\ell^+\ell^-$ are found to be significantly 
smaller.
\item
The implications of the recent data on other $b\to s\ell^+\ell^-$ 
transitions from LHCb is to suppress some of the effects listed above
so that eventually TUM makes rather sharp predictions for all FCNC 
observables that in the case of rare decays differ from the SM: see  
(\ref{P1})--(\ref{P3}).
\end{itemize}

In view of these very definite predictions we are looking forward to 
improved experimental data  and improved lattice
calculations. The correlations identified in the TUM will allow to
monitor how  this simple NP scenario discussed by us
face the future precision flavour data. In case of difficulties  a possible
way out would be to  increase the value of $M$ or  take $\lambda^U\not=\mathds{1}$ which in turn would allow to 
introduce two CP phases in  $\lambda^D$ and use the parametrization in~(\ref{lambdaumatrix}) for it.
This would have an impact on 
CP-violating observables with smaller effects on CP-conserving ones, 
although it could allow in principle suppressions of various rare decay 
branching ratios which is not possible in the TUM.
It would also introduce NP effects in the $D$ system.
Finally the Yukawa couplings of  vectorial fermions
could be non-unitary matrices. But these generalizations are not yet 
required.

\subsubsection*{Acknowledgements}
We thank Minoru Nagai for useful discussions on $H^0$ contributions to 
 $B\to X_s\gamma$ decay.
This research was done in the context of the ERC Advanced Grant project ``FLAVOUR'' (267104). It was partially supported  by the TUM
Institute for 
Advanced Study (R.Z.) and by the DFG cluster
of excellence ``Origin and Structure of the Universe'' (J.G.).

\begin{appendix}

\section{Rotation matrices from mass diagonalization}\label{app:notation}

In Section~\ref{sec:Diag} we use the following shorthand notation:

\begin{align}
\tilde{\lambda}^U_{12} & \equiv \lambda^U_{13} \lambda^U_{32} - \lambda^U_{12} \lambda^U_{33} & \tilde{\lambda}^U_{21} & \equiv
\lambda^U_{23} \lambda^U_{31} - \lambda^U_{21} \lambda^U_{33} \\
\tilde{\lambda}^U_{22} & \equiv \lambda^U_{23} \lambda^U_{32} - \lambda^U_{22} \lambda^U_{33} & \tilde{\lambda}^U_{31} & \equiv
\lambda^U_{22} \lambda^U_{31} - \lambda^U_{21} \lambda^U_{32} \\
\tilde{\lambda}^U_{13} & \equiv \lambda^U_{13} \lambda^U_{22} - \lambda^U_{12} \lambda^U_{23}
\end{align}
\begin{equation}
\hat{\lambda}^U_{33}   \equiv |\lambda^U_{33}|^2 
\end{equation}
\begin{align}
\tilde{\lambda}^D_{12} & \equiv \lambda^D_{13} \lambda^D_{32} - \lambda^D_{12} \lambda^D_{33} & \tilde{\lambda}^D_{21} & \equiv
\lambda^D_{23} \lambda^D_{31} - \lambda^D_{21} \lambda^D_{33} \\
\tilde{\lambda}^D_{22} & \equiv \lambda^D_{23} \lambda^D_{32} - \lambda^D_{22} \lambda^D_{33} & \tilde{\lambda}^D_{31} & \equiv
\lambda^D_{22} \lambda^D_{31} - \lambda^D_{21} \lambda^D_{32} \\
\tilde{\lambda}^D_{13} & \equiv \lambda^D_{13} \lambda^D_{22} - \lambda^D_{12} \lambda^D_{23}
\end{align}
\begin{align}
\hat{\lambda}^D_{13} & \equiv \lambda^D_{13} \lambda^{D*}_{33} + \eps^{D2}_{23} \lambda^D_{12} \lambda^{D*}_{32} & \hat{\lambda}^D_{23} &
\equiv \lambda^D_{23} \lambda^{D*}_{33} + \eps^{D2}_{23} \lambda^D_{22} \lambda^{D*}_{32} 
\end{align}
\begin{equation}
\hat{\lambda}^D_{33}   \equiv |\lambda^D_{33}|^2 + \eps^{D2}_{23} |\lambda^{D}_{32}|^2 
\end{equation}

Here we give explicit expressions for the entries of the rotation matrices $V_{L,R}^{U,D}$ in Eq.~(\ref{equ:VUL})--(\ref{equ:VDR}): 
\begin{align}
 &u_1^L = \frac{\tilde{\lambda}^U_{12} }{\tilde{\lambda}^U_{22} }\,,\qquad u_2^L  = \frac{\lambda_{13}^U  }{\lambda_{33}^{U}}\,,\qquad
u_3^L =\frac{\lambda_{23}^U }{\lambda_{33}^{U} }\,,\qquad u_4^L  =  \frac{\tilde{\lambda}^{U*}_{13} }{\tilde{\lambda}^{U*}_{22} } \,,\\
 &u_1^R  = \frac{\tilde{\lambda}^{U*}_{21} }{\tilde{\lambda}^{U*}_{22} }\,,\qquad u_2^R  = \frac{\lambda_{31}^{U*} }{\lambda_{33}^{U*}
}\,,\qquad u_3^R  = \frac{\lambda_{32}^{U*} }{\lambda_{33}^{U*} }\,,\qquad  u_4^R =  \frac{\tilde{\lambda}^U_{31} }{\tilde{\lambda}^U_{22}
} \,,
\end{align}
\begin{align}
& d_1^L  = \frac{\tilde{\lambda}^D_{12}}{\tilde{\lambda}^D_{22}}\,,\qquad d_2^L  =
\frac{\hat{\lambda}^D_{13}}{\hat{\lambda}^D_{33}}\,,\qquad d_3^L = \frac{\hat{\lambda}^D_{23}}{\hat{\lambda}^D_{33}}\,,\qquad d_4^L  = 
\frac{\tilde{\lambda}^{D*}_{13}}{\tilde{\lambda}^{D*}_{22}} \,,\\
& d_1^R  =   \left(
\frac{\tilde{\lambda}^{D*}_{21}}{\tilde{\lambda}^{D*}_{22}} + \eps^{D2}_{23} \frac{\lambda_{32}^D  }{\lambda^D_{33}}
\frac{\tilde{\lambda}^{D*}_{31}}{\tilde{\lambda}^{D*}_{22}} \right) \frac{|\lambda_{33}^D |}{\sqrt{\hat{\lambda}^D_{33}}}\,,\qquad d_2^R  =
\frac{\lambda_{31}^{D*}  }{\lambda^{D*}_{33}} \frac{|\lambda_{33}^D|}{\sqrt{\hat{\lambda}^D_{33} }}   \,,\\
&d_3^R  = - 
\frac{\tilde{\lambda}^D_{21}}{\tilde{\lambda}^D_{22}} \,,\qquad d_4^R  =  \frac{|\lambda_{33}^D |}{\sqrt{\hat{\lambda}^D_{33} }}\,,\qquad
d_5^R  = \frac{ \lambda_{32}^{D*}}{\lambda^{D*}_{33}} \frac{|\lambda_{33}^D |}{\sqrt{\hat{\lambda}^D_{33} }}\,,\qquad d_6^R  =
\frac{\tilde{\lambda}^{D}_{31}}{\tilde{\lambda}^{D}_{22}}\,.
\end{align}
To make a comparison with the RS scenario in \cite{Blanke:2008zb} easier we give here the translation between the different
notation used for the rotation matrices (up to phases): 
\begin{align}\begin{split}
 &\omega_{12}^d = d_1^L\,,\qquad \omega_{13}^d = d_2^L\,,\qquad \omega_{23}^d = d_3^L\,,\\
&\omega_{21}^d = - d_1^{L\star}\,,\qquad \omega_{31}^d = d_4^L\,,\qquad \omega_{32}^d = - d_3^{L\star}\,,\end{split}
\end{align}
and similarly with $d\leftrightarrow u$, and
\begin{align}
\begin{split}
 &\rho^u_{12} = u_1^R\,,\qquad \rho_{13}^u = u_2^R\,,\qquad \rho_{23}^u = u_3^R\,,\\
&\rho_{21}^u = - u_1^{R\star}\,,\qquad \rho_{31}^u = u_4^R\,,\qquad \rho_{32}^u = - u_3^{R\star}\,,\end{split}\\
\begin{split}
&\rho_{12}^d = d_1^R\,,\qquad \rho_{13}^d = d_2^R\,,\qquad \rho_{21}^d = d_3^R\,,\qquad \rho_{23}^d = d_5^R\,,\\
&\rho_{22}^d = \rho_{33}^d = d_4^R\,,\qquad \rho_{31}^d = d_6^R\,,\qquad \rho_{32}^d = -d_5^{R\star}\,.\end{split}
\end{align}

\boldmath
\section{Flavour violating  $Z$ and $W$ couplings}\label{app:Deltas}
\unboldmath

 Due to the mixing of SU(2) doublets and singlets flavour violating $Z$ and Higgs couplings are induced and the $W$--couplings are modified.
These effects 
are parametrized by
hermitian matrices $A_{L,R}^{U,D}$ and $A_R^{UD}$ defined already in \cite{Buras:2011ph} and readdressed  in Section~\ref{sec:Diag}. Here we
give analytic formulae for these couplings in the mass eigenstate basis (defined in Section~\ref{sec:Diag}): 
\begin{align}
\tilde{A}^U_L & = 
\frac{1}{M_U^2}\begin{pmatrix}
\frac{m_u^2}{\eps_1^U \eps_1^U}& e^{i b_U} \frac{m_u m_c}{\eps_1^U \eps_2^U} u_1^R & e^{i c_U}
\frac{m_u m_t}{\eps_1^U
\eps_3^U} u_2^R  \\
cc. & \frac{m_c^2}{\eps_2^U \eps_2^U} \left( 1+ |u^R_1|^2 \right)& e^{i (c_U-b_U)} \frac{m_c m_t}{\eps_2^U
\eps_3^U} \left( u_3^R
+ u_1^{R*} u_2^R \right)\\
cc. & cc. & \frac{m_t^2}{\eps_3^U \eps_3^U} \left( 1+ |u^R_2|^2 + |u^R_3|^2 \right)
\end{pmatrix} 
\end{align}
\begin{align}
\tilde{A}^D_L & = 
\frac{1}{M_D^2}\begin{pmatrix}
\frac{m_d^2}{\eps_1^D \eps_1^D}& e^{i b_D} \frac{m_d m_s}{\eps_1^D \eps_2^D} d_1^R & e^{i c_D}
\frac{m_d m_b}{\eps_1^D
\eps_3^D} d_2^R  \\
cc. & \frac{m_s^2}{\eps_2^D \eps_2^D} \left(|d^R_1|^2 + |d^R_4|^2 + \eps^{D4}_{23} |d^R_5|^2 \right)& e^{i (c_D-b_D)}
\frac{m_s
m_b}{\eps_2^D \eps_3^D} \left( d_4^R d_5^R + d_1^{R*} d_2^R - \eps_{23}^{D2} d_4^R d_5^R \right)\\
cc. & cc. & \frac{m_b^2}{\eps_3^D \eps_3^D} \left(|d^R_2|^2 + |d^R_4|^2 + |d^R_5|^2 \right)
\end{pmatrix} 
\end{align}
\begin{align}
\tilde{A}^U_R & = 
\frac{1}{M_Q^2}\begin{pmatrix}
\frac{m_u^2}{\eps_1^Q \eps_1^Q}& e^{i b_U} \frac{m_um_c}{\eps_1^Q \eps_2^Q} u_1^L & e^{i c_U}
\frac{m_um_t}{\eps_1^Q
\eps_3^Q} u_2^L  \\
cc. & \frac{m_c^2}{\eps_2^Q \eps_2^Q} \left( 1+ |u^L_1|^2 \right)& e^{i (c_U-b_U)} \frac{m_cm_t}{\eps_2^Q
\eps_3^Q} \left( u_3^L
+ u_1^{L*} u_2^L \right)\\
cc. & cc. & \frac{m_t^2}{\eps_3^Q \eps_3^Q} \left( 1+ |u^L_2|^2 + |u^L_3|^2 \right)
\end{pmatrix} 
\end{align}
\begin{align}
\tilde{A}^D_R & = 
\frac{1}{M_Q^2}\begin{pmatrix}
\frac{m_d^2}{\eps_1^Q \eps_1^Q}& e^{i b_D} \frac{m_dm_s}{\eps_1^Q \eps_2^Q} d_1^L & e^{i c_D}
\frac{m_dm_b}{\eps_1^Q
\eps_3^Q} d_2^L  \\
cc. & \frac{m_s^2}{\eps_2^Q \eps_2^Q} \left( 1+ |d^L_1|^2 \right)& e^{i (c_D-b_D)} \frac{m_sm_b}{\eps_2^Q
\eps_3^Q} \left( d_3^L
+ d_1^{L*} d_2^L \right)\\
cc. & cc. & \frac{m_b^2}{\eps_3^Q \eps_3^Q} \left( 1+ |d^L_2|^2 + |d^L_3|^2 \right)
\end{pmatrix} 
\end{align}
where "$cc.$" denotes the complex conjugate of the related entry. 
\begin{align}
 \tilde A_R^{UD} & = \frac{1}{M_Q^2}
\begin{pmatrix}
     \frac{m_um_d}{\eps_1^Q \eps_1^Q}  &  e^{i b_D} \frac{m_um_s}{\eps_1^Q \eps_2^Q} d_1^L &  e^{i c_D}
\frac{m_um_b}{\eps_1^Q\eps_3^Q} d_2^L  \\ 
  e^{-i b_U} \frac{m_cm_d}{\eps_1^Q \eps_2^Q} u_1^{L\star} &  \frac{m_cm_s}{\eps_2^Q
\eps_2^Q} \left( 1+ d^L_1 u_1^{L\star} \right) & e^{i (c_D-b_U)} \frac{m_cm_b}{\eps_2^Q\eps_3^Q} \left( d_3^L
+ u_1^{L*} d_2^L \right)\\
e^{-i c_U}\frac{m_tm_d}{\eps_1^Q\eps_3^Q} u_2^{L\star} & e^{-i (c_U-b_D)} \frac{m_tm_s}{\eps_2^Q\eps_3^Q}
\left( u_3^{L\star}
+ u_2^{L*} d_1^L \right)& \frac{m_tm_b}{\eps_3^Q \eps_3^Q} \left( 1+ d^L_2 u_2^{L\star} + d^L_3
u_3^{L\star} \right)
\end{pmatrix}
\end{align}

In the unitary model 
  without the constraint in $\lambda^U$ we get:
  {\allowdisplaybreaks
  \begin{align}
   \tilde A_L^U &=\frac{v^2}{M_U^2}\begin{pmatrix}
                                    \eps_1^{Q2} \left(1+|u_1^L|^2 + |u_4^L|^2\right)& -e^{ib_U} \eps_1^Q\eps_2^Q\left(u_1^L +
  u_3^{L\star} u_4^{L\star}\right) & e^{ic_U} \eps_1^Q \eps_3^Q u_4^{L\star}\\
  				  cc. & \eps_2^{Q2}\left(1 + |u_3^L|^2\right) & -e^{-i(b_U-c_U)} \eps_2^Q\eps_3^Q u_3^L\\
  				  cc. & cc. & \eps_3^{Q2}
                                   \end{pmatrix}\,,\\
   \tilde A_L^D &=\frac{v^2}{M_D^2}\begin{pmatrix}
                                    \eps_1^{Q2}  \left(1+|d_1^L|^2 + |d_4^L|^2\right)&- e^{ib_D} \eps_1^Q\eps_2^Q \left(d_1^L +
  d_3^{L\star} d_4^{L\star}\right) & e^{ic_D} \eps_1^Q
  \eps_3^Q d_4^{L\star}\\
  				  cc. & \eps_2^{Q2}\left(1 + |d_3^L|^2\right) & -e^{-i(b_D-c_D)} \eps_2^Q\eps_3^Q d_3^L\\
  				  cc. & cc. & \eps_3^{Q2}
                                   \end{pmatrix}\,,\\
   \tilde A_R^U &=\frac{v^2}{M_Q^2}\begin{pmatrix}
                                    \eps_1^{U2}\left(1+|u_1^R|^2 + |u_4^R|^2\right) & -e^{ib_U} \eps_1^U\eps_2^U\left( u_1^R +
  u_3^{R\star}u_4^{R\star}\right) & e^{ic_U} \eps_1^U \eps_3^U u_4^{R\star}\\
  				  cc. & \eps_2^{U2}\left(1 + |u_3^R|^2\right) & -e^{-i(b_U-c_U)} \eps_2^U\eps_3^U u_3^R\\
  				  cc. & cc. & \eps_3^{U2}
                                   \end{pmatrix}\,,\\
   \tilde A_R^D &=\frac{v^2}{M_Q^2}\begin{pmatrix}
                                    \eps_1^{D2} \left(1+|d_3^R|^2 + |d_6^R|^2\right) & e^{ib_D} \eps_1^D\eps_2^D \left(d_3^R d_4^R -
  d_5^{R\star}d_6^{R\star}\right) & e^{ic_D} \eps_1^D
  \eps_3^D \left(d_6^{R\star}d_4^R+\eps_{23}^{D2} d_3^R d_5^R\right)\\
  				  cc. & \eps_2^{D2}\left(|d_4^R|^2+|d_5^R|^2\right) & -e^{-i(b_D-c_D)}
  \eps_2^D\eps_3^Dd_5^R\left(d_4^R-\eps_{23}^{D2}d_4^{R\star}\right) \\
  				  cc. & cc. & \eps_3^{D2}\left(|d_4^R|^2+\eps_{23}^{D4}|d_5^R|^2\right)
                                   \end{pmatrix}\,,
  \end{align}}%
  where of course the $u_i^{L,R},\,d_i^{L,R}$ are not all independent because they are functions of the unitary $\lambda^{U,D}$. 
 Inserting the  angle-parametrization from Sec~\ref{sec:unitarymodel} we get for $\tilde A_{L,R}^D$ the same as in TUM but very lengthy
expressions for $\tilde A_{L,R}^U$ which we do not list here.

\boldmath
\section{Approximate Expressions for $Z$ contributions to 
$\Delta F=2$ Amplitudes in TUM}
\label{app:DeltaF2}
\unboldmath
The tree-level $Z$ contributions to mixing amplitudes $M_{12}^K$ and $M_{12}^q$ 
are given approximately as follows
{\allowdisplaybreaks
\begin{align}
\frac{{\rm Im} \left(M_{12}^K\right)_\text{Z}} {{\rm Im} \left(M_{12}^K\right)_\text{SM}} & \approx  0.34
\left( \frac{\TeV}{M}\right)^4 \left[\eps^{Q4}_3 (1+X_{23}^2 |V_{cb}|^2 \frac{\re \lambda_c}{\re \lambda_t}) + \frac{m_s m_d}{v^2}  \frac{\re \lambda_u}{\re \lambda_t}  \frac{|P_1^{\rm LR}(K)|}{X_{13}^2 |V_{ub}|^2 }  \right]\\
\frac{{\rm Re} \left(M_{12}^K\right)_\text{Z}} {{\rm Re} \left(M_{12}^K\right)_\text{SM}} & \approx \left( \frac{\TeV}{M}\right)^4 \left[
\eps^{Q4}_3 X_{23}^4 |V_{cb}|^4 \phantom{ \left( \frac{\TeV}{M}\right)^4}\right. \\ \nonumber
&\qquad \qquad\left. + 2 |P_1^{\rm LR}(K)| \frac{m_s m_d}{v^2} \left( \frac{\re \lambda_u}{\re
\lambda_c} \frac{X_{23}^2 |V_{cb}|^2}{X_{13}^2 |V_{ub}|^2 } + \frac{\re \lambda_u \re \lambda_t}{( \re \lambda_c)^2}  \frac{1}{X_{13}^2
|V_{ub}|^2 } \right) \right] 10^{3} \\
\re \frac{ \left(M_{12}^{B_d}\right)_\text{Z}} {\left(M_{12}^{B_d}\right)_\text{SM}} & \approx  \left(\frac{ \TeV}{M} \right)^4  0.21 \,  \eps^{Q4}_3  \\
\im \frac{ \left(M_{12}^{B_d}\right)_\text{Z}} {\left(M_{12}^{B_d}\right)_\text{SM}} & \approx - 0.42\left(\frac{\TeV}{M} \right)^4 \frac{m_b m_d}{v^2} \im \frac{\lambda^d_u}{\lambda_t^d} \frac{|P_1^{\rm LR}(B_d)|}{X_{13}^2 |V_{ub}|^2}\\
\re \frac{ \left(M_{12}^{B_s}\right)_\text{Z}} {\left(M_{12}^{B_s}\right)_\text{SM}} & \approx 0.20\left(\frac{\TeV}{M} \right)^4 \left[\eps^{Q4}_3  + 2\frac{m_b m_s}{v^2}  \re \frac{\lambda^s_u}{\lambda_t^s} \frac{|P_1^{\rm LR}(B_s)|}{X_{13}^2 |V_{ub}|^2}  \right] \\
\im \frac{ \left(M_{12}^{B_s}\right)_\text{Z}} {\left(M_{12}^{B_s}\right)_\text{SM}} & \approx - 0.40\left(\frac{\TeV}{M} \right)^4  \frac{m_b m_s}{v^2}  \im \frac{\lambda^s_u}{\lambda_t^s} \frac{|P_1^{\rm LR}(B_s)|}{X_{13}^2 |V_{ub}|^2}  .
\end{align} }%
In the $K$ sector the above expressions directly give the new contribution $\delta \eps_K$ and  $\delta \Delta M_K$, i.e.
\begin{align}
\frac{\eps_K^{\rm tot}}{\eps_K^{\rm SM}} & = 1 + \delta \eps_K = 1 + \frac{{\rm Im} \left(M_{12}^K\right)_\text{Z}} {{\rm Im} \left(M_{12}^K\right)_\text{SM}} \\
\frac{\Delta M_K^{\rm tot}}{\Delta M_K^{\rm SM}} & = 1 + \delta \Delta M_K = 1 + \frac{{\rm Re} \left(M_{12}^K\right)_\text{Z}} {{\rm Re} \left(M_{12}^K\right)_\text{SM}}.
\end{align}
In the B-sector one can obtain the relevant quantities 
\begin{align}
C_{B_q} &  = |1 + \frac{(M_{12}^{B_q})_{\rm Z} } {(M_{12}^{B_q})_{\rm SM }}| \\
 \varphi_{B_q} & = \frac{1}{2} \arg \left( {1 + \frac{(M_{12}^{B_q})_{\rm Z} } {(M_{12}^{B_q})_{\rm SM }}} \right) \\
 \frac{S_{\psi K_s}^{\rm tot}}{S_{\psi K_s}^{\rm SM}} & \approx 1 + \frac{2 \varphi_{B_d}}{\tan 2 \beta} \\ 
\frac{S_{\psi \phi}^{\rm tot}}{S_{\psi \phi}^{\rm SM}} & \approx 1 - \frac{2 \varphi_{B_s}}{\tan 2 |\beta_s|}. 
\end{align}
\boldmath
\section{Approximate Expressions for $Z$ contributions to 
$\Delta F=1$ Amplitudes in TUM}
\label{app:DeltaF1}
\unboldmath
Proceeding in the same manner we obtain approximate expressions for the relevant quantities in $\Delta F =1$ observables
{\allowdisplaybreaks
\begin{align}
\Delta X_L(K) \equiv X_L(K) - X(x_t) & = 2.75 \left( \frac{\TeV}{M} \right)^2 \eps^{Q2}_3 \left[ 1 + X_{23}^2 |V_{cb}|^2 \frac{\re \lambda_c}{\lambda_t}\right] \\
 X_R(K) & =  - 2.75   \left( \frac{\TeV}{M} \right)^2 \frac{m_d m_s}{v^2}\frac{ \re \lambda_u }{\lambda_t} \frac{1}{\eps^{Q2}_3 X_{13}^2\vub^2} \\
 \Delta X_L(B_d) & = 2.75 \left( \frac{\TeV}{M} \right)^2 \eps^{Q2}_3 \\
 X_R(B_d) & =  - 2.75   \left( \frac{\TeV}{M} \right)^2 \frac{m_b m_d}{v^2}\frac{1}{\eps^{Q2}_3} \left[ \frac{\lambda_u^d}{\lambda_t^d} \frac{1}{X_{13}^2 |V_{ub}|^2} +  \frac{\re \lambda_c^d}{\lambda_t^d} \frac{1}{X_{23}^2 |V_{cb}|^2}   \right]^* \\
 \Delta X_L(B_s) & = 2.75 \left( \frac{\TeV}{M} \right)^2 \eps^{Q2}_3 \\
 X_R(B_s) & =  - 2.75   \left( \frac{\TeV}{M} \right)^2 \frac{m_b m_s}{v^2}\frac{1 }{\eps^{Q2}_3}\left[ \frac{\lambda_u^s}{\lambda_t^s} \frac{1}{X_{13}^2 |V_{ub}|^2} +  \frac{\re \lambda_c^s}{\lambda_t^s} \frac{1}{X_{23}^2 |V_{cb}|^2}   \right]^*.
\end{align}}%
Moreover one has
\begin{align}
 \Delta Y_d & = \Delta Y_s = \Delta X_L(B_d) = \Delta X_L(B_s)  & Y_d^\prime & = X_R (B_d) & Y_s^\prime & = X_R (B_s) \nonumber \\
\Delta Y_A(K) & = \Delta X_L(K) - X_R(K)& \Delta Z_q^\prime & = \Delta Y_q^\prime&  \Delta Z_q =&  \Delta Y_q 
\end{align}

\end{appendix}

\bibliographystyle{JHEP}
\bibliography{allrefs}

\providecommand{\href}[2]{#2}\begingroup\raggedright\begin{thebibliography}{10}

\bibitem{Buras:2011ph}
A.~J. Buras, C.~Grojean, S.~Pokorski, and R.~Ziegler, {\it {FCNC Effects in a
  Minimal Theory of Fermion Masses}},  {\em JHEP} {\bf 1108} (2011) 028,
  [\href{http://xxx.lanl.gov/abs/1105.3725}{{\tt arXiv:1105.3725}}].

\bibitem{LHCbBsmumu}
{\bf LHCb collaboration} Collaboration, R.~Aaij {\em et.~al.}, {\it {First
  evidence for the decay $B_s\to\mu^+ \mu^-$}},
  \href{http://xxx.lanl.gov/abs/1211.2674}{{\tt arXiv:1211.2674}}.

\bibitem{Buras:2012jb}
A.~J. Buras, F.~De~Fazio, and J.~Girrbach, {\it {The Anatomy of Z' and Z with
  Flavour Changing Neutral Currents in the Flavour Precision Era}},
  \href{http://xxx.lanl.gov/abs/1211.1896}{{\tt arXiv:1211.1896}}.

\bibitem{Blanke:2006sb}
M.~Blanke {\em et.~al.}, {\it {Particle antiparticle mixing, $\varepsilon_K$,
  $\Delta\Gamma_q$, $A_\text{SL}^q$, $A_\text{CP}(B_d \to \psi K_S)$,
  $A_\text{CP}(B_s \to \psi \phi)$ and $B \to X_{s,d} \gamma$ in the Littlest
  Higgs model with T- parity}},  {\em JHEP} {\bf 12} (2006) 003,
  [\href{http://xxx.lanl.gov/abs/hep-ph/0605214}{{\tt hep-ph/0605214}}].

\bibitem{Agashe:2004cp}
K.~Agashe, G.~Perez, and A.~Soni, {\it Flavor structure of warped extra
  dimension models},  {\em Phys. Rev.} {\bf D71} (2005) 016002,
  [\href{http://xxx.lanl.gov/abs/hep-ph/0408134}{{\tt hep-ph/0408134}}].

\bibitem{Blanke:2008zb}
M.~Blanke, A.~J. Buras, B.~Duling, S.~Gori, and A.~Weiler, {\it {$\Delta F=2$
  Observables and Fine-Tuning in a Warped Extra Dimension with Custodial
  Protection}},  {\em JHEP} {\bf 03} (2009) 001,
  [\href{http://xxx.lanl.gov/abs/0809.1073}{{\tt arXiv:0809.1073}}].

\bibitem{Lalak:2010bk}
Z.~Lalak, S.~Pokorski, and G.~G. Ross, {\it {Beyond MFV in family symmetry
  theories of fermion masses}},  {\em JHEP} {\bf 1008} (2010) 129,
  [\href{http://xxx.lanl.gov/abs/1006.2375}{{\tt arXiv:1006.2375}}].

\bibitem{delAguila:2000kb}
F.~del Aguila and J.~Santiago, {\it {Universality limits on bulk fermions}},
  {\em Phys.Lett.} {\bf B493} (2000) 175--181,
  [\href{http://xxx.lanl.gov/abs/hep-ph/0008143}{{\tt hep-ph/0008143}}].

\bibitem{delAguila:2000aa}
F.~del Aguila, M.~Perez-Victoria, and J.~Santiago, {\it {Effective description
  of quark mixing}},  {\em Phys.Lett.} {\bf B492} (2000) 98--106,
  [\href{http://xxx.lanl.gov/abs/hep-ph/0007160}{{\tt hep-ph/0007160}}].

\bibitem{delAguila:2000rc}
F.~del Aguila, M.~Perez-Victoria, and J.~Santiago, {\it {Observable
  contributions of new exotic quarks to quark mixing}},  {\em JHEP} {\bf 0009}
  (2000) 011, [\href{http://xxx.lanl.gov/abs/hep-ph/0007316}{{\tt
  hep-ph/0007316}}].

\bibitem{Buras:2009ka}
A.~J. Buras, B.~Duling, and S.~Gori, {\it {The Impact of Kaluza-Klein Fermions
  on Standard Model Fermion Couplings in a RS Model with Custodial
  Protection}},  {\em JHEP} {\bf 0909} (2009) 076,
  [\href{http://xxx.lanl.gov/abs/0905.2318}{{\tt arXiv:0905.2318}}].

\bibitem{Crivellin:2009sd}
A.~Crivellin, {\it {Effects of right-handed charged currents on the
  determinations of |V(ub)| and |V(cb)|}},  {\em Phys.Rev.} {\bf D81} (2010)
  031301, [\href{http://xxx.lanl.gov/abs/0907.2461}{{\tt arXiv:0907.2461}}].

\bibitem{Chen:2008se}
C.-H. Chen and S.-h. Nam, {\it {Left-right mixing on leptonic and semileptonic
  $b\to u$ decays}},  {\em Phys.Lett.} {\bf B666} (2008) 462--466,
  [\href{http://xxx.lanl.gov/abs/0807.0896}{{\tt arXiv:0807.0896}}].

\bibitem{Buras:2010pz}
A.~J. Buras, K.~Gemmler, and G.~Isidori, {\it {Quark flavour mixing with
  right-handed currents: an effective theory approach}},  {\em Nucl.Phys.} {\bf
  B843} (2011) 107--142, [\href{http://xxx.lanl.gov/abs/1007.1993}{{\tt
  arXiv:1007.1993}}].

\bibitem{Crivellin:2011ba}
A.~Crivellin and L.~Mercolli, {\it {$B\to X_d \gamma$ and constraints on new
  physics}},  {\em Phys.Rev.} {\bf D84} (2011) 114005,
  [\href{http://xxx.lanl.gov/abs/1106.5499}{{\tt arXiv:1106.5499}}].

\bibitem{BelleICHEP}
{\bf Belle Collaboration} Collaboration, I.~Adachi {\em et.~al.}, {\it
  Measurement of $b\to\tau\nu$ with a hadronic tagging method using the full
  data sample of belle},  \href{http://xxx.lanl.gov/abs/1208.4678}{{\tt
  arXiv:1208.4678}}.

\bibitem{Tarantino:2012mq}
C.~Tarantino, {\it {Flavor Lattice QCD in the Precision Era}},
  \href{http://xxx.lanl.gov/abs/1210.0474}{{\tt arXiv:1210.0474}}.

\bibitem{Herrlich:1993yv}
S.~Herrlich and U.~Nierste, {\it {Enhancement of the $K_L - K_S$ mass
  difference by short distance QCD corrections beyond leading logarithms}},
  {\em Nucl. Phys.} {\bf B419} (1994) 292--322,
  [\href{http://xxx.lanl.gov/abs/hep-ph/9310311}{{\tt hep-ph/9310311}}].

\bibitem{Herrlich:1995hh}
S.~Herrlich and U.~Nierste, {\it {Indirect CP violation in the neutral kaon
  system beyond leading logarithms}},  {\em Phys. Rev.} {\bf D52} (1995)
  6505--6518, [\href{http://xxx.lanl.gov/abs/hep-ph/9507262}{{\tt
  hep-ph/9507262}}].

\bibitem{Herrlich:1996vf}
S.~Herrlich and U.~Nierste, {\it {The Complete $|\Delta S|=2$ Hamiltonian in
  the Next-To-Leading Order}},  {\em Nucl. Phys.} {\bf B476} (1996) 27--88,
  [\href{http://xxx.lanl.gov/abs/hep-ph/9604330}{{\tt hep-ph/9604330}}].

\bibitem{Buras:1990fn}
A.~J. Buras, M.~Jamin, and P.~H. Weisz, {\it {Leading and next-to-leading QCD
  corrections to $\varepsilon$ parameter and $B^0-\bar{B}^0$ mixing in the
  presence of a heavy top quark}},  {\em Nucl. Phys.} {\bf B347} (1990)
  491--536.

\bibitem{Urban:1997gw}
J.~Urban, F.~Krauss, U.~Jentschura, and G.~Soff, {\it {Next-to-leading order
  QCD corrections for the $B^0 - \bar B^0$ mixing with an extended Higgs
  sector}},  {\em Nucl. Phys.} {\bf B523} (1998) 40--58,
  [\href{http://xxx.lanl.gov/abs/hep-ph/9710245}{{\tt hep-ph/9710245}}].

\bibitem{Brod:2010mj}
J.~Brod and M.~Gorbahn, {\it {$\epsilon_K$ at Next-to-Next-to-Leading Order:
  The Charm-Top-Quark Contribution}},  {\em Phys.Rev.} {\bf D82} (2010) 094026,
  [\href{http://xxx.lanl.gov/abs/1007.0684}{{\tt arXiv:1007.0684}}].

\bibitem{Brod:2011ty}
J.~Brod and M.~Gorbahn, {\it {Next-to-Next-to-Leading-Order Charm-Quark
  Contribution to the CP Violation Parameter $\varepsilon_K$ and $\Delta
  M_K$}},  {\em Phys.Rev.Lett.} {\bf 108} (2012) 121801,
  [\href{http://xxx.lanl.gov/abs/1108.2036}{{\tt arXiv:1108.2036}}].

\bibitem{Ciuchini:1997bw}
M.~Ciuchini, E.~Franco, V.~Lubicz, G.~Martinelli, I.~Scimemi, {\em et.~al.},
  {\it {Next-to-leading order QCD corrections to $\Delta F = 2$ effective
  Hamiltonians}},  {\em Nucl.Phys.} {\bf B523} (1998) 501--525,
  [\href{http://xxx.lanl.gov/abs/hep-ph/9711402}{{\tt hep-ph/9711402}}].

\bibitem{Buras:2000if}
A.~J. Buras, M.~Misiak, and J.~Urban, {\it {Two loop QCD anomalous dimensions
  of flavor changing four quark operators within and beyond the standard
  model}},  {\em Nucl.Phys.} {\bf B586} (2000) 397--426,
  [\href{http://xxx.lanl.gov/abs/hep-ph/0005183}{{\tt hep-ph/0005183}}].

\bibitem{Buras:2001ra}
A.~J. Buras, S.~Jager, and J.~Urban, {\it {Master formulae for $\Delta F=2$ NLO
  QCD factors in the standard model and beyond}},  {\em Nucl.Phys.} {\bf B605}
  (2001) 600--624, [\href{http://xxx.lanl.gov/abs/hep-ph/0102316}{{\tt
  hep-ph/0102316}}].

\bibitem{Buras:2012fs}
A.~J. Buras and J.~Girrbach, {\it {Complete NLO QCD Corrections for Tree Level
  Delta F = 2 FCNC Processes}},  {\em JHEP} {\bf 1203} (2012) 052,
  [\href{http://xxx.lanl.gov/abs/1201.1302}{{\tt arXiv:1201.1302}}].

\bibitem{Boyle:2012qb}
{\bf RBC and UKQCD Collaborations} Collaboration, P.~Boyle, N.~Garron, and
  R.~Hudspith, {\it {Neutral kaon mixing beyond the standard model with $n_f =
  2+1$ chiral fermions}},  {\em Phys.Rev.} {\bf D86} (2012) 054028,
  [\href{http://xxx.lanl.gov/abs/1206.5737}{{\tt arXiv:1206.5737}}].

\bibitem{Bertone:2012cu}
V.~Bertone, N.~Carrasco, M.~Ciuchini, P.~Dimopoulos, R.~Frezzotti, {\em
  et.~al.}, {\it {Kaon Mixing Beyond the SM from Nf=2 tmQCD and model
  independent constraints from the UTA}},
  \href{http://xxx.lanl.gov/abs/1207.1287}{{\tt arXiv:1207.1287}}.

\bibitem{Bouchard:2011xj}
C.~Bouchard, E.~Freeland, C.~Bernard, A.~El-Khadra, E.~Gamiz, {\em et.~al.},
  {\it {Neutral $B$ mixing from $2+1$ flavor lattice-QCD: the Standard Model
  and beyond}},  \href{http://xxx.lanl.gov/abs/1112.5642}{{\tt
  arXiv:1112.5642}}.

\bibitem{Buras:2008nn}
A.~J. Buras and D.~Guadagnoli, {\it {Correlations among new CP violating
  effects in $\Delta F = 2$ observables}},  {\em Phys. Rev.} {\bf D78} (2008)
  033005, [\href{http://xxx.lanl.gov/abs/0805.3887}{{\tt arXiv:0805.3887}}].

\bibitem{Buras:2010pza}
A.~J. Buras, D.~Guadagnoli, and G.~Isidori, {\it {On $\epsilon_K$ beyond lowest
  order in the Operator Product Expansion}},  {\em Phys.Lett.} {\bf B688}
  (2010) 309--313, [\href{http://xxx.lanl.gov/abs/1002.3612}{{\tt
  arXiv:1002.3612}}].

\bibitem{Bona:2005eu}
{\bf UTfit} Collaboration, M.~Bona {\em et.~al.}, {\it {The UTfit collaboration
  report on the status of the unitarity triangle beyond the standard model. I:
  Model- independent analysis and minimal flavour violation}},  {\em JHEP} {\bf
  03} (2006) 080, [\href{http://xxx.lanl.gov/abs/hep-ph/0509219}{{\tt
  hep-ph/0509219}}].

\bibitem{Buras:2011zb}
A.~J. Buras, L.~Merlo, and E.~Stamou, {\it {The Impact of Flavour Changing
  Neutral Gauge Bosons on $\bar{B}\to X_s \gamma$}},  {\em JHEP} {\bf 1108}
  (2011) 124, [\href{http://xxx.lanl.gov/abs/1105.5146}{{\tt
  arXiv:1105.5146}}].

\bibitem{Blanke:2011ry}
M.~Blanke, A.~J. Buras, K.~Gemmler, and T.~Heidsieck, {\it {$\Delta F = 2$
  observables and $B\to X_q\gamma$; in the Left-Right Asymmetric Model: Higgs
  particles striking back}},  {\em JHEP} {\bf 1203} (2012) 024,
  [\href{http://xxx.lanl.gov/abs/1111.5014}{{\tt arXiv:1111.5014}}].

\bibitem{Blanke:2008yr}
M.~Blanke, A.~J. Buras, B.~Duling, K.~Gemmler, and S.~Gori, {\it {Rare K and B
  Decays in a Warped Extra Dimension with Custodial Protection}},  {\em JHEP}
  {\bf 03} (2009) 108, [\href{http://xxx.lanl.gov/abs/0812.3803}{{\tt
  arXiv:0812.3803}}].

\bibitem{Buchalla:1998ba}
G.~Buchalla and A.~J. Buras, {\it The rare decays $k \to\pi \nu\bar\nu$, $b \to
  x \nu \bar\nu$ and $b\to\ell^+\ell^-$: An update},  {\em Nucl. Phys.} {\bf
  B548} (1999) 309--327, [\href{http://xxx.lanl.gov/abs/hep-ph/9901288}{{\tt
  hep-ph/9901288}}].

\bibitem{Buras:2005gr}
A.~J. Buras, M.~Gorbahn, U.~Haisch, and U.~Nierste, {\it {The rare decay $K^+
  \to \pi^+ \nu \bar\nu$ at the next-to-next-to-leading order in QCD}},  {\em
  Phys. Rev. Lett.} {\bf 95} (2005) 261805,
  [\href{http://xxx.lanl.gov/abs/hep-ph/0508165}{{\tt hep-ph/0508165}}].

\bibitem{Buras:2006gb}
A.~J. Buras, M.~Gorbahn, U.~Haisch, and U.~Nierste, {\it {Charm quark
  contribution to $K^+ \to \pi^+ \nu \bar\nu$ at next-to-next-to-leading
  order}},  {\em JHEP} {\bf 11} (2006) 002,
  [\href{http://xxx.lanl.gov/abs/hep-ph/0603079}{{\tt hep-ph/0603079}}].

\bibitem{Brod:2008ss}
J.~Brod and M.~Gorbahn, {\it {Electroweak Corrections to the Charm Quark
  Contribution to $K^+ \to \pi^+ \nu \bar\nu$}},  {\em Phys. Rev.} {\bf D78}
  (2008) 034006, [\href{http://xxx.lanl.gov/abs/0805.4119}{{\tt
  arXiv:0805.4119}}].

\bibitem{Brod:2010hi}
J.~Brod, M.~Gorbahn, and E.~Stamou, {\it {Two-Loop Electroweak Corrections for
  the $K \to \pi \nu \bar{nu}$ Decays}},  {\em Phys.Rev.} {\bf D83} (2011)
  034030, [\href{http://xxx.lanl.gov/abs/1009.0947}{{\tt arXiv:1009.0947}}].

\bibitem{Misiak:1999yg}
M.~Misiak and J.~Urban, {\it {QCD corrections to FCNC decays mediated by Z
  penguins and W boxes}},  {\em Phys.Lett.} {\bf B451} (1999) 161--169,
  [\href{http://xxx.lanl.gov/abs/hep-ph/9901278}{{\tt hep-ph/9901278}}].

\bibitem{Buras:2004uu}
A.~J. Buras, F.~Schwab, and S.~Uhlig, {\it {Waiting for precise measurements of
  $K^{+} \to \pi^{+} \nu \bar{\nu}$ and $K_{L} \to \pi^0 \nu \bar{\nu}$}},
  {\em Rev. Mod. Phys.} {\bf 80} (2008) 965--1007,
  [\href{http://xxx.lanl.gov/abs/hep-ph/0405132}{{\tt hep-ph/0405132}}].

\bibitem{Isidori:2006yx}
G.~Isidori, {\it {Flavor Physics with light quarks and leptons}},  {\em eConf}
  {\bf C060409} (2006) 035, [\href{http://xxx.lanl.gov/abs/hep-ph/0606047}{{\tt
  hep-ph/0606047}}].

\bibitem{Smith:2006qg}
C.~Smith, {\it {Theory review on rare K decays: Standard model and beyond}},
  \href{http://xxx.lanl.gov/abs/hep-ph/0608343}{{\tt hep-ph/0608343}}.

\bibitem{Mescia:2007kn}
F.~Mescia and C.~Smith, {\it {Improved estimates of rare K decay
  matrix-elements from $K_{\ell3}$ decays}},  {\em Phys. Rev.} {\bf D76} (2007)
  034017, [\href{http://xxx.lanl.gov/abs/0705.2025}{{\tt arXiv:0705.2025}}].

\bibitem{Isidori:2005xm}
G.~Isidori, F.~Mescia, and C.~Smith, {\it {Light-quark loops in $K
  \to\pi\nu\bar\nu$}},  {\em Nucl. Phys.} {\bf B718} (2005) 319--338,
  [\href{http://xxx.lanl.gov/abs/hep-ph/0503107}{{\tt hep-ph/0503107}}].

\bibitem{Artamonov:2008qb}
{\bf E949} Collaboration, A.~V. Artamonov {\em et.~al.}, {\it {New measurement
  of the $K^{+} \to \pi^{+} \nu \bar{\nu}$ branching ratio}},  {\em Phys. Rev.
  Lett.} {\bf 101} (2008) 191802,
  [\href{http://xxx.lanl.gov/abs/0808.2459}{{\tt arXiv:0808.2459}}].

\bibitem{Ahn:2009gb}
{\bf E391a Collaboration} Collaboration, J.~Ahn {\em et.~al.}, {\it
  {Experimental study of the decay $K^0_L\to\pi^0\nu \bar\nu$}},  {\em
  Phys.Rev.} {\bf D81} (2010) 072004,
  [\href{http://xxx.lanl.gov/abs/0911.4789}{{\tt arXiv:0911.4789}}].

\bibitem{Mescia:2006jd}
F.~Mescia, C.~Smith, and S.~Trine, {\it {$K_L\to\pi^0 e^+e^-$ and
  $K_L\to\pi^0\mu^+\mu^-$: A binary star on the stage of flavor physics}},
  {\em JHEP} {\bf 08} (2006) 088,
  [\href{http://xxx.lanl.gov/abs/hep-ph/0606081}{{\tt hep-ph/0606081}}].

\bibitem{Prades:2007ud}
J.~Prades, {\it {ChPT Progress on Non-Leptonic and Radiative Kaon Decays}},
  {\em PoS} {\bf KAON} (2008) 022,
  [\href{http://xxx.lanl.gov/abs/0707.1789}{{\tt arXiv:0707.1789}}].

\bibitem{Isidori:2004rb}
G.~Isidori, C.~Smith, and R.~Unterdorfer, {\it {The rare decay $K_L \to \pi^0
  \mu^+\mu^-$ within the SM}},  {\em Eur. Phys. J.} {\bf C36} (2004) 57--66,
  [\href{http://xxx.lanl.gov/abs/hep-ph/0404127}{{\tt hep-ph/0404127}}].

\bibitem{Friot:2004yr}
S.~Friot, D.~Greynat, and E.~De~Rafael, {\it {Rare kaon decays revisited}},
  {\em Phys. Lett.} {\bf B595} (2004) 301--308,
  [\href{http://xxx.lanl.gov/abs/hep-ph/0404136}{{\tt hep-ph/0404136}}].

\bibitem{Bruno:1992za}
C.~Bruno and J.~Prades, {\it {Rare Kaon Decays in the $1/N_c$-Expansion}},
  {\em Z. Phys.} {\bf C57} (1993) 585--594,
  [\href{http://xxx.lanl.gov/abs/hep-ph/9209231}{{\tt hep-ph/9209231}}].

\bibitem{AlaviHarati:2003mr}
{\bf KTeV} Collaboration, A.~Alavi-Harati {\em et.~al.}, {\it {Search for the
  Rare Decay $K_L \to \pi^0 e^+e^-$}},  {\em Phys. Rev. Lett.} {\bf 93} (2004)
  021805, [\href{http://xxx.lanl.gov/abs/hep-ex/0309072}{{\tt
  hep-ex/0309072}}].

\bibitem{AlaviHarati:2000hs}
{\bf KTEV} Collaboration, A.~Alavi-Harati {\em et.~al.}, {\it {Search for the
  Decay $K_L \to \pi^0 \mu^+ \mu^-$}},  {\em Phys. Rev. Lett.} {\bf 84} (2000)
  5279--5282, [\href{http://xxx.lanl.gov/abs/hep-ex/0001006}{{\tt
  hep-ex/0001006}}].

\bibitem{Blanke:2006eb}
M.~Blanke {\em et.~al.}, {\it {Rare and CP-violating $K$ and $B$ decays in the
  Littlest Higgs model with T-parity}},  {\em JHEP} {\bf 01} (2007) 066,
  [\href{http://xxx.lanl.gov/abs/hep-ph/0610298}{{\tt hep-ph/0610298}}].

\bibitem{Blanke:2008ac}
M.~Blanke, A.~J. Buras, S.~Recksiegel, and C.~Tarantino, {\it {The Littlest
  Higgs Model with T-Parity Facing CP-Violation in $B_s - \bar B_s$ Mixing}},
  \href{http://xxx.lanl.gov/abs/0805.4393}{{\tt arXiv:0805.4393}}.

\bibitem{Buchalla:2003sj}
G.~Buchalla, G.~D'Ambrosio, and G.~Isidori, {\it {Extracting short-distance
  physics from $K_{L,S} \to \pi^0 e^+ e^-$ decays}},  {\em Nucl. Phys.} {\bf
  B672} (2003) 387--408, [\href{http://xxx.lanl.gov/abs/hep-ph/0308008}{{\tt
  hep-ph/0308008}}].

\bibitem{Buras:1994qa}
A.~J. Buras, M.~E. Lautenbacher, M.~Misiak, and M.~Munz, {\it {Direct CP
  violation in $K_L \to\pi^0 e^+ e^-$ beyond leading logarithms}},  {\em Nucl.
  Phys.} {\bf B423} (1994) 349--383,
  [\href{http://xxx.lanl.gov/abs/hep-ph/9402347}{{\tt hep-ph/9402347}}].

\bibitem{Buras:2004ub}
A.~J. Buras, R.~Fleischer, S.~Recksiegel, and F.~Schwab, {\it {Anatomy of
  prominent $B$ and $K$ decays and signatures of CP-violating new physics in
  the electroweak penguin sector}},  {\em Nucl. Phys.} {\bf B697} (2004)
  133--206, [\href{http://xxx.lanl.gov/abs/hep-ph/0402112}{{\tt
  hep-ph/0402112}}].

\bibitem{Gorbahn:2006bm}
M.~Gorbahn and U.~Haisch, {\it {Charm quark contribution to $K_L \to \mu^+
  \mu^-$ at next-to-next-to-leading order}},  {\em Phys. Rev. Lett.} {\bf 97}
  (2006) 122002, [\href{http://xxx.lanl.gov/abs/hep-ph/0605203}{{\tt
  hep-ph/0605203}}].

\bibitem{Isidori:2003ts}
G.~Isidori and R.~Unterdorfer, {\it {On the short-distance constraints from
  $K_{L,S} \to \mu^+ \mu^-$ }},  {\em JHEP} {\bf 01} (2004) 009,
  [\href{http://xxx.lanl.gov/abs/hep-ph/0311084}{{\tt hep-ph/0311084}}].

\bibitem{DescotesGenon:2011pb}
S.~Descotes-Genon, J.~Matias, and J.~Virto, {\it {An analysis of $B_{d,s}$
  mixing angles in presence of New Physics and an update of $B_s \to K^{0*}
  \bar K^{0*}$}},  {\em Phys.Rev.} {\bf D85} (2012) 034010,
  [\href{http://xxx.lanl.gov/abs/1111.4882}{{\tt arXiv:1111.4882}}].

\bibitem{deBruyn:2012wj}
K.~De~Bruyn, R.~Fleischer, R.~Knegjens, P.~Koppenburg, M.~Merk, {\em et.~al.},
  {\it {Branching Ratio Measurements of $B_s$ Decays}},  {\em Phys.Rev.} {\bf
  D86} (2012) 014027, [\href{http://xxx.lanl.gov/abs/1204.1735}{{\tt
  arXiv:1204.1735}}].

\bibitem{deBruyn:2012wk}
K.~De~Bruyn, R.~Fleischer, R.~Knegjens, P.~Koppenburg, M.~Merk, {\em et.~al.},
  {\it {Probing New Physics via the $B^0_s\to \mu^+\mu^-$ Effective Lifetime}},
   {\em Phys.Rev.Lett.} {\bf 109} (2012) 041801,
  [\href{http://xxx.lanl.gov/abs/1204.1737}{{\tt arXiv:1204.1737}}].

\bibitem{Fleischer:2012fy}
R.~Fleischer, {\it {On Branching Ratios of $B_s$ Decays and the Search for New
  Physics in $B^0_s\to \mu^+\mu^-$}},
  \href{http://xxx.lanl.gov/abs/1208.2843}{{\tt arXiv:1208.2843}}.

\bibitem{Aaij:2012ac}
{\bf LHCb collaboration} Collaboration, R.~Aaij {\em et.~al.}, {\it {Strong
  constraints on the rare decays $B_s \to \mu^+ \mu^-$ and $B^0 \to \mu^+
  \mu^-$}},  \href{http://xxx.lanl.gov/abs/1203.4493}{{\tt arXiv:1203.4493}}.

\bibitem{Buras:2012ru}
A.~J. Buras, J.~Girrbach, D.~Guadagnoli, and G.~Isidori, {\it {On the Standard
  Model prediction for BR(B{s,d} to mu+ mu-)}},  {\em Eur.Phys.J.} {\bf C72}
  (2012) 2172, [\href{http://xxx.lanl.gov/abs/1208.0934}{{\tt
  arXiv:1208.0934}}].

\bibitem{Altmannshofer:2009ma}
W.~Altmannshofer, A.~J. Buras, D.~M. Straub, and M.~Wick, {\it {New strategies
  for New Physics search in $B \to K^{*} \nu \bar{\nu}$, $B \to K \nu
  \bar{\nu}$ and $B \to X_{s} \nu \bar{\nu}$ decays}},  {\em JHEP} {\bf 04}
  (2009) 022, [\href{http://xxx.lanl.gov/abs/0902.0160}{{\tt
  arXiv:0902.0160}}].

\bibitem{Kamenik:2009kc}
J.~F. Kamenik and C.~Smith, {\it {Tree-level contributions to the rare decays
  $B^+ \to \pi^+ \nu \bar\nu, B^+\to K^+ \nu\bar\nu$, and $B^+\to
  K^{*+}\nu\bar\nu$ in the Standard Model}},  {\em Phys.Lett.} {\bf B680}
  (2009) 471--475, [\href{http://xxx.lanl.gov/abs/0908.1174}{{\tt
  arXiv:0908.1174}}].

\bibitem{Bartsch:2009qp}
M.~Bartsch, M.~Beylich, G.~Buchalla, and D.-N. Gao, {\it {Precision Flavour
  Physics with $B\to K \nu \bar\nu$ and $B\to K l^+ l^-$}},  {\em JHEP} {\bf
  0911} (2009) 011, [\href{http://xxx.lanl.gov/abs/0909.1512}{{\tt
  arXiv:0909.1512}}].

\bibitem{Barate:2000rc}
{\bf ALEPH} Collaboration, R.~Barate {\em et.~al.}, {\it {Measurements of $BR(b
  \to \tau^- \bar\nu_\tau X)$ and $BR(b \to \tau^- \bar\nu_\tau D^{*\pm} X)$
  and upper limits on $BR(B^- \to \tau^- \bar\nu_\tau)$ and $BR(b \to \nu
  \bar\nu)$}},  {\em Eur. Phys. J.} {\bf C19} (2001) 213--227,
  [\href{http://xxx.lanl.gov/abs/hep-ex/0010022}{{\tt hep-ex/0010022}}].

\bibitem{:2007zk}
{\bf BELLE} Collaboration, K.~F. Chen {\em et.~al.}, {\it {Search for $B\to
  h^{(*)} \nu \bar\nu$ Decays at Belle}},  {\em Phys. Rev. Lett.} {\bf 99}
  (2007) 221802, [\href{http://xxx.lanl.gov/abs/0707.0138}{{\tt
  arXiv:0707.0138}}].

\bibitem{:2008fr}
{\bf BABAR} Collaboration, B.~Aubert {\em et.~al.}, {\it {Search for $B\to K^*
  \nu \bar\nu$ decays}},  {\em Phys. Rev.} {\bf D78} (2008) 072007,
  [\href{http://xxx.lanl.gov/abs/0808.1338}{{\tt arXiv:0808.1338}}].

\bibitem{Bona:2009cj}
{\bf UTfit Collaboration} Collaboration, M.~Bona {\em et.~al.}, {\it {An
  Improved Standard Model Prediction Of $BR(B\to \tau \nu)$ And Its
  Implications For New Physics}},  {\em Phys.Lett.} {\bf B687} (2010) 61--69,
  [\href{http://xxx.lanl.gov/abs/0908.3470}{{\tt arXiv:0908.3470}}].

\bibitem{Altmannshofer:2009ne}
W.~Altmannshofer, A.~J. Buras, S.~Gori, P.~Paradisi, and D.~M. Straub, {\it
  {Anatomy and Phenomenology of FCNC and CPV Effects in SUSY Theories}},  {\em
  Nucl.Phys.} {\bf B830} (2010) 17--94,
  [\href{http://xxx.lanl.gov/abs/0909.1333}{{\tt arXiv:0909.1333}}].

\bibitem{Lunghi:2008aa}
E.~Lunghi and A.~Soni, {\it {Possible Indications of New Physics in
  $B_d$-mixing and in $\sin(2 \beta)$ Determinations}},  {\em Phys. Lett.} {\bf
  B666} (2008) 162--165, [\href{http://xxx.lanl.gov/abs/0803.4340}{{\tt
  arXiv:0803.4340}}].

\bibitem{Buras:2010mh}
A.~J. Buras, M.~V. Carlucci, S.~Gori, and G.~Isidori, {\it {Higgs-mediated
  FCNCs: Natural Flavour Conservation vs. Minimal Flavour Violation}},  {\em
  JHEP} {\bf 1010} (2010) 009, [\href{http://xxx.lanl.gov/abs/1005.5310}{{\tt
  arXiv:1005.5310}}].

\bibitem{Buras:2012sd}
A.~J. Buras and J.~Girrbach, {\it {On the Correlations between Flavour
  Observables in Minimal $U(2)^3$ Models}},  {\em JHEP} {\bf 1301} (2013) 007,
  [\href{http://xxx.lanl.gov/abs/1206.3878}{{\tt arXiv:1206.3878}}].

\bibitem{Nakamura:2010zzi}
{\bf Particle Data Group} Collaboration, K.~Nakamura {\em et.~al.}, {\it
  {Review of particle physics}},  {\em J.Phys.G} {\bf G37} (2010) 075021.

\bibitem{Laiho:2009eu}
J.~Laiho, E.~Lunghi, and R.~S. Van~de Water, {\it {Lattice QCD inputs to the
  CKM unitarity triangle analysis}},  {\em Phys. Rev.} {\bf D81} (2010) 034503,
  [\href{http://xxx.lanl.gov/abs/0910.2928}{{\tt arXiv:0910.2928}}]. Updates
  available on {\tt http://latticeaverages.org/}.

\bibitem{Chetyrkin:2009fv}
K.~Chetyrkin, J.~Kuhn, A.~Maier, P.~Maierhofer, P.~Marquard, {\em et.~al.},
  {\it {Charm and Bottom Quark Masses: An Update}},  {\em Phys.Rev.} {\bf D80}
  (2009) 074010, [\href{http://xxx.lanl.gov/abs/0907.2110}{{\tt
  arXiv:0907.2110}}].

\bibitem{Allison:2008xk}
{\bf HPQCD Collaboration} Collaboration, I.~Allison {\em et.~al.}, {\it
  {High-Precision Charm-Quark Mass from Current-Current Correlators in Lattice
  and Continuum QCD}},  {\em Phys.Rev.} {\bf D78} (2008) 054513,
  [\href{http://xxx.lanl.gov/abs/0805.2999}{{\tt arXiv:0805.2999}}].

\bibitem{Abulencia:2006ze}
{\bf CDF Collaboration} Collaboration, A.~Abulencia {\em et.~al.}, {\it
  {Observation of $B^0_s - \bar{B}^0_s$ Oscillations}},  {\em Phys.Rev.Lett.}
  {\bf 97} (2006) 242003, [\href{http://xxx.lanl.gov/abs/hep-ex/0609040}{{\tt
  hep-ex/0609040}}].

\bibitem{Aaij:2011qx}
{\bf LHCb Collaboration} Collaboration, R.~Aaij {\em et.~al.}, {\it
  {Measurement of the $B^0_s - \bar{B}^0_s$ oscillation frequency $\Delta M_s$
  in $B^0_s \to D_s^-(3) \pi$ decays}},  {\em Phys.Lett.} {\bf B709} (2012)
  177--184, [\href{http://xxx.lanl.gov/abs/1112.4311}{{\tt arXiv:1112.4311}}].

\bibitem{Clarke:1429149}
P.~Clarke, {\it Results on cp violation in $b_s$ mixing}, .
  http://cdsweb.cern.ch/record/1429149/files/LHCb-TALK-2012-029.pdf.

\bibitem{Asner:2010qj}
{\bf Heavy Flavor Averaging Group} Collaboration, D.~Asner {\em et.~al.}, {\it
  {Averages of $b$-hadron, $c$-hadron, and $\tau$-lepton Properties}},
  \href{http://xxx.lanl.gov/abs/1010.1589}{{\tt arXiv:1010.1589}}. Long author
  list - awaiting processing.

\bibitem{Buras:2003td}
A.~J. Buras, {\it {Relations between $\Delta M_{s,d}$ and $B_{s,d} \to \mu^+
  \mu^-$ in models with minimal flavour violation}},  {\em Phys. Lett.} {\bf
  B566} (2003) 115--119, [\href{http://xxx.lanl.gov/abs/hep-ph/0303060}{{\tt
  hep-ph/0303060}}].

\bibitem{Grossman:1997sk}
Y.~Grossman and Y.~Nir, {\it {$K_L\to\pi^0\nu\bar\nu$ beyond the standard
  model}},  {\em Phys. Lett.} {\bf B398} (1997) 163--168,
  [\href{http://xxx.lanl.gov/abs/hep-ph/9701313}{{\tt hep-ph/9701313}}].

\bibitem{Blanke:2009pq}
M.~Blanke, {\it {Insights from the Interplay of $K\rightarrow \pi
  \nu\overline{\nu}$ and $\epsilon_K$ on the New Physics Flavour Structure}},
  {\em Acta Phys.Polon.} {\bf B41} (2010) 127,
  [\href{http://xxx.lanl.gov/abs/0904.2528}{{\tt arXiv:0904.2528}}].

\bibitem{Altmannshofer:2012ir}
W.~Altmannshofer and D.~M. Straub, {\it {Cornering New Physics in $b\to
  s\gamma$ Transitions}},  {\em JHEP} {\bf 1208} (2012) 121,
  [\href{http://xxx.lanl.gov/abs/1206.0273}{{\tt arXiv:1206.0273}}].

\bibitem{Beaujean:2012uj}
F.~Beaujean, C.~Bobeth, D.~van Dyk, and C.~Wacker, {\it {Bayesian Fit of
  Exclusive $b \to s \bar\ell\ell$ Decays: The Standard Model Operator Basis}},
   {\em JHEP} {\bf 1208} (2012) 030,
  [\href{http://xxx.lanl.gov/abs/1205.1838}{{\tt arXiv:1205.1838}}].

\bibitem{DescotesGenon:2012zf}
S.~Descotes-Genon, J.~Matias, M.~Ramon, and J.~Virto, {\it {Implications from
  clean observables for the binned analysis of $B\to K*\mu^+\mu^-$ at large
  recoil}},  \href{http://xxx.lanl.gov/abs/1207.2753}{{\tt arXiv:1207.2753}}.

\end{thebibliography}\endgroup

\end{document}